\documentclass[11pt,a4paper]{article}
\usepackage[margin=23mm]{geometry}
\usepackage[T1]{fontenc}
\usepackage{lmodern}
\usepackage{amsmath,amssymb,amsthm,mathtools}
\usepackage{booktabs,enumitem,needspace,longtable,array}
\usepackage[hidelinks]{hyperref}
\newtheorem{theorem}{Theorem}
\newtheorem{lemma}[theorem]{Lemma}
\newtheorem{corollary}[theorem]{Corollary}
\newtheorem{proposition}[theorem]{Proposition}
\newtheorem{remark}[theorem]{Remark}
\newcommand{\Tr}{\operatorname{Tr}}

\newcommand{\Herm}{\operatorname{Herm}}
\newcommand{\cH}{\mathcal H}
\newcommand{\cV}{\mathcal V}
\newcolumntype{L}[1]{>{\raggedright\arraybackslash}p{#1}}
\newcommand{\JS}{J_{\infty,\mathrm{pt}}^{\mathrm S}}
\newcommand{\JR}{J^{\mathrm R}}
\title{Asymptotic Attainability of the Channel RLD Fisher-Information Bound}
\author{Masahito Hayashi\thanks{Email: \texttt{hmasahito@cuhk.edu.cn}}\\
School of Data Science, The Chinese University of Hong Kong, Shenzhen,\\
Longgang District, Shenzhen, 518172, China\\
International Quantum Academy, Futian District, Shenzhen 518048, China\\
Graduate School of Mathematics, Nagoya University, Nagoya, 464-8602, Japan}
\date{}
\begin{document}
\maketitle
\begin{abstract}
The channel right-logarithmic-derivative (RLD) Fisher information is a computable upper bound on the optimal asymptotic symmetric-logarithmic-derivative (SLD) Fisher information per channel use over causally ordered, parameter-independent strategies. The bound is not always attained, even if the Choi matrix is positive definite. We determine exactly when equality holds using a semidefinite feasibility condition involving only the Choi matrix and its first derivative. The condition allows a degenerate largest eigenvalue of the RLD matrix and rank-deficient Choi matrices with derivatives supported on their range. The original channel family may change rank nearby. For generalized dephasing channels with a positive-definite correlation matrix at the evaluation point, the condition becomes a system of linear equations for nonnegative certificate weights that sum to one. At a completely dephasing point we also calculate the optimal SLD information for every finite number of uses and show that a repeated, ancilla-free probe attains it, even against adaptive strategies. Explicit examples give an unbounded ratio of the channel RLD information to the optimal asymptotic SLD information per use, either as the dimension grows or at fixed qubit dimension as the smallest Choi eigenvalue tends to zero. We derive the equality criterion and its consequences from the asymptotic optimization of Zhou and Jiang.

\end{abstract}

\section{Introduction}
For a quantum channel depending on an unknown parameter, an input probe produces an output state whose sensitivity can be measured by symmetric-logarithmic-derivative (SLD) Fisher information. A single-use experiment allows an input and an ancillary system. With repeated uses, the experimenter may entangle the inputs and perform parameter-independent adaptive operations in a definite causal order. The quantity of interest is the largest SLD information per channel use when all these strategies are allowed.
For the family $\Lambda=(\Lambda_\theta)_\theta$, let $J_n^{\mathrm S}(\Lambda;0)$ be the supremum of the output SLD Fisher information at $\theta=0$ over these strategies with $n$ channel uses.
Write $\JR(\Lambda)$ for the channel right-logarithmic-derivative (RLD) information. When it is finite, Eq.~\eqref{eq:finite-rld-bound} gives $J_n^{\mathrm S}(\Lambda;0)\le n\JR(\Lambda)$, so the optimized SLD information grows at most linearly with $n$. We study the coefficient of this growth. With $\JS$ denoting the optimal pointwise asymptotic SLD coefficient, the bound reads
\[
 \JS\le\JR.
\]
We call the equality $\JS=\JR$ \emph{RLD saturation}. This compares the asymptotic coefficient with the channel bound. If a specified input attains the channel bound in one use, independent repetitions attain the same information per use and hence give asymptotic saturation. If the asymptotic inequality is strict, the RLD formula overestimates the information obtainable even by asymptotically optimized, adaptive experiments.

The asymptotic optimization itself is known. Fujiwara and Imai~\cite{FI2008} derived a Kraus-representation minimization for one-use information. Channel-extension bounds developed this approach for repeated uses~\cite{DDKG2012,KD2013}. For channels in the regime of at most linear information growth, Zhou and Jiang~\cite[Theorem~2]{ZJ} identified the optimal asymptotic coefficient and proved that parallel strategies attain the same coefficient as sequential strategies. In contrast, the channel RLD bound has a direct expression in the Choi matrix and its derivative~\cite{Hayashi}. It corresponds to one feasible choice in the Kraus optimization~\cite[Appendix~C]{KD2013}. Whether that particular choice is optimal is the equality question studied here. Related RLD bounds for adaptive, multiparameter settings appear in~\cite{KW,AD2022}. Recent work establishes RLD-based additive bounds even for quantum processes without definite causal order~\cite{HayashiOuyang2026,HuangWilde2026}.

Zhou and Jiang also characterize stationarity at an optimal
Kraus gauge~\cite[Appendix~F, Eq.~(F4)]{ZJ}: a single density
operator supported on the largest-eigenvalue space of the Kraus
objective matrix must have zero trace pairing with the linear change of that matrix in every feasible gauge direction. We test the particular gauge that gives the RLD
value. To do so, we express both its feasible variations and their
annihilator using only the evaluation-point Choi data. For rank-deficient
Choi matrices the resulting operator equation is imposed on their
support.

We answer this question using the Choi matrix $C$ at the evaluation point and its derivative $D$. There are several Kraus representations with these same channel data, and their derivatives can differ. The known asymptotic optimization minimizes over that freedom. We describe it by a Hermitian Choi-space variable $Y$, chosen so that $Y=0$ gives the RLD expression. Consequently, $\JS=\JR$ exactly when no admissible $Y$ improves upon $Y=0$. Our criterion replaces this minimization by a feasibility test. The density operator $\sigma$ must be supported on the largest-eigenvalue space of the RLD matrix $R(C,D)$, and $\sigma$ and a Hermitian $H$ must satisfy a linear operator equation. The test allows a degenerate largest eigenvalue and applies also to rank-deficient $C$ provided $D$ is supported on its range. The density operator $\sigma$ serves as an optimality certificate; its role is distinct from that of a physical input probe.

The examples address three separate questions: whether the asymptotic RLD bound is saturated, whether a prescribed one-use input attains the channel RLD value, and whether the optimized finite-use information is additive. For generalized dephasing channels with a positive-definite
correlation matrix at the evaluation point, a directed graph gives
a saturation test. At complete dephasing we also obtain an exact
formula for every number of uses and show that repeated ancilla-free
probes are optimal. Other qubit dephasing channels can exhibit
asymptotic saturation with strictly smaller one-use information. The replacer-channel examples give unbounded RLD-to-SLD ratios, and the rank-changing example shows why the information must be evaluated at the specified parameter point.

\begin{table}[t]
\centering
\footnotesize
\renewcommand{\arraystretch}{1.14}
\caption{Questions addressed by the examples. The rows distinguish asymptotic saturation, finite-use optimality, and attainment by a specified probe.}\label{tab:example-guide}
\begin{tabular}{@{}L{.27\linewidth}L{.62\linewidth}@{}}
\toprule
Question & Examples and conclusion\\
\midrule
Is the RLD bound tight? & Generalized Pauli channels attain it; measurement--preparation and correlated-qutrit examples show both equality and strict inequality.\\
How does $J_1^{\mathrm S}$ compare with $\JS$? & Some qubit dephasing channels have $J_1^{\mathrm S}<\JS=\JR$. At complete dephasing, repeating an ancilla-free input is optimal for every use count.\\
Can $\JR/\JS$ diverge? & Yes. The dimension grows for the star graph. The smallest Choi eigenvalue tends to zero at fixed dimension for the qubit replacer channel.\\
Does a specified probe suffice? & A maximally entangled one-use input requires conditions stronger than channel-level asymptotic equality.\\
What if the rank changes? & A rank-changing replacer channel shows that pointwise SLD information need not equal its limit at nearby parameter values.\\
\bottomrule
\end{tabular}
\end{table}

Section~\ref{sec:preliminaries} introduces the Choi variational formula, and Section~\ref{sec:positive} states the equality criteria. Sections~\ref{sec:pauli}, \ref{sec:full-support-examples}, and
\ref{sec:applications} study generalized Pauli, measurement--preparation,
and generalized dephasing channels, respectively. The measurement--preparation
section includes a replacer channel whose Choi rank changes at the evaluation
point. Section~\ref{sec:prescribed-probes} determines when a maximally
entangled input attains the channel RLD value in one use. Section~\ref{sec:positive-proof} proves the general and positive-definite saturation criteria, derives a directional-derivative characterization of saturation, and gives finite-dimensional tests. Appendices~\ref{app:tangent} and~\ref{app:choi-translation} justify
the affine replacement and the Choi variational formula.
Appendix~\ref{app:mp-kraus} constructs the Kraus representation
used for the upper bound in Section~\ref{sec:measure-prepare}; Appendix~\ref{app:comparisons} compares earlier results (Table~\ref{tab:prior-work}) and provides an expanded inventory of the examples (Table~\ref{tab:examples}).

\section{Setting and the Choi variational formula}\label{sec:preliminaries}
\subsection{Channels, Choi matrices, and the evaluation point}
This section defines the information quantities to be compared and
derives the Choi variational formula for the optimal asymptotic
SLD coefficient. We begin with the local channel data used in
both calculations.

We describe the channel at the evaluation point and its first
derivative by two Choi matrices, $C$ and $D$.
Let $\cH_A$ and $\cH_B$ be finite-dimensional input and output Hilbert spaces. We consider a family of completely positive trace-preserving (CPTP) maps
\[
 \Lambda_\theta:\mathcal B(\cH_A)\longrightarrow\mathcal B(\cH_B)
\]
on an open interval containing $0$, differentiable at $0$. The parameter value at which information is evaluated is always $\theta=0$. Thus a dot denotes a derivative at zero unless another point is specified.

Choose an orthonormal input basis $\{|j\rangle\}$. We use the unnormalized Choi matrix with the output factor first:
\begin{equation}\label{eq:choi}
 C_\theta:=\sum_{j,k}\Lambda_\theta(|j\rangle\langle k|)\otimes|j\rangle\langle k|,
 \qquad C:=C_0,\qquad D:=\left.\frac{dC_\theta}{d\theta}\right|_0.
\end{equation}
The matrices $C$ and $D$ act on $\cH_B\otimes\cH_A$ and obey
\[
 C\ge0,\qquad D=D^\dagger,\qquad \Tr_B C=I_A,\qquad \Tr_B D=0.
\]
Here $\Tr_B$ traces out the output system. Conversely, the channel is recovered from its Choi matrix by
\begin{equation}\label{eq:choi-inverse}
 \Lambda_\theta(X)=\Tr_A[C_\theta(I_B\otimes X^{\mathsf T})],
 \qquad X\in\mathcal B(\cH_A),
\end{equation}
where the transpose is taken in the chosen input basis. This inverse correspondence will also identify the channel defined by the affine tangent in Appendix~\ref{app:tangent}.

\subsection{Pointwise SLD information and the RLD upper bound}\label{subsec:pointwise-info}
The channel and its derivative at the evaluation point determine
the output state and its derivative for a fixed protocol.
Each protocol produces an output state whose SLD information is
determined by that state and its derivative. We first define the
state information and then optimize it over channel protocols.
Let $\rho$ and $\dot\rho$ be the state and its derivative at the evaluation point of a differentiable state family $\rho_\theta$ on an open interval. Its SLD information is
\begin{equation}\label{eq:SLD-state}
 F^{\mathrm S}(\rho,\dot\rho)=\Tr(\rho L^2),\qquad
 \dot\rho=\tfrac12(\rho L+L\rho),\qquad L=L^\dagger.
\end{equation}
For any $v\in\ker\rho$, the nonnegative function $\langle v|\rho_\theta|v\rangle$ vanishes at the evaluation point. Its derivative is therefore zero. If $P_0$ projects onto $\ker\rho$, then $\langle v|P_0\dot\rho P_0|v\rangle=0$ for every $v$ in that kernel. A Hermitian operator with zero quadratic form on its whole space is zero, so $P_0\dot\rho P_0=0$. In an eigenbasis $\rho=\sum_i\lambda_i|i\rangle\langle i|$, the SLD equation gives $L_{ij}=2\dot\rho_{ij}/(\lambda_i+\lambda_j)$ when $\lambda_i+\lambda_j>0$. The remaining entries are in the kernel block, where $\dot\rho_{ij}=0$, and their values do not affect the information. Hermiticity gives $|L_{ij}|^2=|L_{ji}|^2$. Exchanging $i,j$ in a copy of the trace sum and averaging gives
\[
 \Tr(\rho L^2)=\sum_{i,j}\lambda_i|L_{ij}|^2
 =\tfrac12\sum_{i,j}(\lambda_i+\lambda_j)|L_{ij}|^2.
\]
Substituting the determined entries of $L$ gives
\begin{equation}\label{eq:SLD-spectral}
 F^{\mathrm S}(\rho,\dot\rho)
 =2\sum_{i,j:\,\lambda_i+\lambda_j>0}
 \frac{|\langle i|\dot\rho|j\rangle|^2}{\lambda_i+\lambda_j}.
\end{equation}
Terms with zero denominator are omitted. We use this expression for the finite-dimensional examples below.

Let $J_n^{\mathrm S}(\Lambda;0)$ be the supremum of the output SLD information at zero over all parameter-independent, causally ordered protocols using $\Lambda_\theta$ $n$ times. Such protocols may use ancillas and adaptive processing between uses. Parallel protocols are included. The input preparation, intermediate operations, and final processing do not depend on the unknown parameter. The symbol $\Lambda$ in an information quantity denotes the entire family $(\Lambda_\theta)_\theta$, not just the channel $\Lambda_0$.

When the limit exists, write
\begin{equation}\label{eq:Jinf}
 \JS(\Lambda):=\lim_{n\to\infty}\frac{J_n^{\mathrm S}(\Lambda;0)}{n}.
\end{equation}
The results below establish existence of this limit under their stated assumptions and show that parallel strategies attain the same coefficient asymptotically. The limit is over the number of uses. The word \emph{pointwise} means that each protocol is evaluated using its output state $\rho_n(0)$ and derivative $\dot\rho_n(0)$; we do not first evaluate information at $\theta\ne0$ and then let $\theta$ tend to zero. These procedures can differ at a change of rank.

\paragraph{SLD information of a positive-definite qubit state.}
We use the following formula for the two-dimensional states in the
replacer and generalized dephasing examples.
Consider a differentiable qubit state family
$\rho_\theta=(I+\boldsymbol s(\theta)\cdot\boldsymbol\sigma)/2$,
with $|\boldsymbol s|<1$ at the evaluation point. Here
$\boldsymbol\sigma=(\sigma_X,\sigma_Y,\sigma_Z)$ denotes the Pauli matrices
\[
 \sigma_X=\begin{pmatrix}0&1\\1&0\end{pmatrix},\qquad
 \sigma_Y=\begin{pmatrix}0&-i\\i&0\end{pmatrix},\qquad
 \sigma_Z=\begin{pmatrix}1&0\\0&-1\end{pmatrix}.
\] Write
$\dot{\boldsymbol s}=\partial_\theta\boldsymbol s$ there and expand
the SLD as $L=\ell_0 I+\boldsymbol\ell\cdot\boldsymbol\sigma$.
Comparing the scalar and Pauli coefficients in
$\dot\rho=(\rho L+L\rho)/2$ gives
\[
 \ell_0+\boldsymbol s\cdot\boldsymbol\ell=0,
 \qquad \boldsymbol\ell+\ell_0\boldsymbol s=\dot{\boldsymbol s}.
\]
Taking the dot product of the second equation with $\boldsymbol s$
and using the first yields
$\ell_0=-(\boldsymbol s\cdot\dot{\boldsymbol s})/(1-|\boldsymbol s|^2)$;
the denominator is positive by full rank. Since the SLD equation
also implies $F^{\mathrm S}=\Tr(\dot\rho L)$, substituting
$\boldsymbol\ell=\dot{\boldsymbol s}-\ell_0\boldsymbol s$ gives
\begin{equation}\label{eq:bloch-sld}
 F^{\mathrm S}=\Tr(\dot\rho L)
   =\dot{\boldsymbol s}\cdot\boldsymbol\ell
   =|\dot{\boldsymbol s}|^2+
      \frac{(\boldsymbol s\cdot\dot{\boldsymbol s})^2}{1-|\boldsymbol s|^2}.
\end{equation}

\subsection{Supported Choi data and the variational formula}\label{subsec:supported-variational}
We now evaluate the optimization over protocols using the Choi
data $C,D$. The two quantities to be compared are the optimal asymptotic SLD
coefficient and its RLD upper bound.
The support of $C$ determines both the condition for a finite RLD
value and the space of variables in the SLD minimization.

Write the positive spectral part of $C$ as
\begin{equation}\label{eq:pseudoinverse}
 C=\sum_{j:\,\lambda_j>0}\lambda_j|u_j\rangle\langle u_j|,\qquad
 C^+:=\sum_{j:\,\lambda_j>0}\lambda_j^{-1}|u_j\rangle\langle u_j|,\qquad
 P:=\sum_{j:\,\lambda_j>0}|u_j\rangle\langle u_j|.
\end{equation}
Thus $P$ projects onto the range, or support, of $C$, and $C^+$ is its Moore--Penrose inverse: the inverse on that range, extended by zero to the kernel. For full support, meaning $C>0$, we have $P=I_{BA}$ and $C^+=C^{-1}$.

For the finite-RLD extension to rank-deficient $C$, assume
\begin{equation}\label{eq:supp}
 D=PDP.
\end{equation}
It requires the derivative to act entirely within the support of $C$ at zero. This condition does not require the original family to keep the same rank nearby. For example, weight proportional to $\theta^2$ can enter an additional output subspace without contributing to $D$. The replacer example in Section~\ref{sec:rank-changing} gives such a family. The support condition is a substantive restriction on the first derivative, rather than a consequence of differentiability alone.

Under Eq.~\eqref{eq:supp}, define the channel RLD matrix and its scalar information by
\begin{equation}\label{eq:R}
 R(C,D):=\Tr_B(DC^+D),\qquad r(C,D):=\lambda_{\max}(R(C,D)),\qquad \JR(\Lambda):=r(C,D).
\end{equation}
The positive semidefinite matrix $R(C,D)$ acts on the full input space $\cH_A$; its largest eigenvalue $r(C,D)$ is the scalar channel RLD information~\cite{Hayashi}. We write $R$ and $r$ when $C,D$ are clear. Since the finite-use RLD bound established in Eq.~\eqref{eq:finite-rld-bound} controls $J_n^{\mathrm S}$ for every $n$, the asymptotic coefficient satisfies
\begin{equation}\label{eq:bound}
 \JS(\Lambda)\le\JR(\Lambda)=r(C,D).
\end{equation}
The existence of the limit on the left under the stated support assumption follows from Theorem~\ref{thm:variational}.

To decide whether the upper bound in Eq.~\eqref{eq:bound} is attained,
we shall test whether $Y=0$ minimizes a Choi-space objective. The
variable $Y$ parametrizes the freedom in the Kraus derivatives
while leaving the Choi data $C,D$ fixed.

Let $P$ be the support projection of $C$. The feasible Hermitian
Choi-space variables satisfy both the support constraint and the
zero output-partial-trace constraint:
\begin{equation}\label{eq:V}
 \cV_P:=\{Y\in\Herm(\cH_B\otimes\cH_A):Y=PYP,\ \Tr_B Y=0\}.
\end{equation}
For any supported Hermitian $Y=PYP$, define the input-space operator
\begin{equation}\label{eq:M}
 M_{C,D}(Y):=\Tr_B[(D-2iY)C^+(D+2iY)].
\end{equation}
In particular, $Y=0$ belongs to $\cV_P$, and substituting it in
Eq.~\eqref{eq:M} yields $M_{C,D}(0)=R(C,D)$, so testing whether
$Y=0$ minimizes the largest eigenvalue tests the RLD value.

The Kraus optimization can be applied to a family of constant Choi
rank that shares the data $C,D$. Let $\widehat\Lambda_\theta$ have
Choi matrix $\widehat C_\theta=C+\theta D$. On $\operatorname{Ran}P$,
the least eigenvalue of $C$ is positive, so $\widehat C_\theta$
is positive definite there whenever $|\theta|\|D\|_\infty$ is
smaller than that eigenvalue. Both $C$ and $D$ vanish on the
orthogonal complement by Eq.~\eqref{eq:supp}. Also
$\Tr_B\widehat C_\theta=I_A$ since $\Tr_B D=0$. Thus
$\widehat\Lambda_\theta$ is a quantum channel with the same
Choi support for all sufficiently small $|\theta|$.

For a fixed $n$-use protocol, write its output as
$\mathcal F(\Lambda_\theta,\ldots,\Lambda_\theta)$, where
$\mathcal F$ is multilinear in the $n$ slots and includes the
fixed input preparation and controls, including any memory.
Differentiating slot by slot gives
\begin{align*}
 \rho_n(0)&=\mathcal F(\Lambda_0,\ldots,\Lambda_0),\\
 \dot\rho_n(0)&=\sum_{j=1}^n
 \mathcal F(\Lambda_0,\ldots,\dot\Lambda_0,\ldots,\Lambda_0)
 \quad\text{with }\dot\Lambda_0\text{ in slot }j\text{ of the }j\text{th term}.
\end{align*}
The affine family shares both $\Lambda_0$ and $\dot\Lambda_0$,
so the pair is the same for every fixed protocol. Taking
suprema gives the same $n$-use information separately for
parallel and adaptive protocols. Lemma~\ref{lem:tangent} proves
the protocol identity and the support property of the output derivative.

To obtain a differentiable Kraus representation, let
$V:\mathbb C^s\to\operatorname{Ran}P$ be the isometry whose columns
form an orthonormal basis of the fixed support, with $s=\operatorname{rank}C$.
Taking the positive square root in that basis gives
$Q_\theta=V(V^\dagger\widehat C_\theta V)^{1/2}$, a differentiable
factor with $s$ columns satisfying
$Q_\theta Q_\theta^\dagger=\widehat C_\theta$.
The columns are the vectorized Kraus operators of
$\widehat\Lambda_\theta$. For a Hermitian $s\times s$ matrix $h$,
the factor $Q_\theta e^{-i\theta h}$ represents the same channel and
has derivative $\dot Q_h=\dot Q-iQh$ at zero, where $Q=Q_0$.
Writing $K_\ell$ and $\dot K_{h,\ell}$ for the operators represented
by the columns of $Q$ and $\dot Q_h$, respectively, define
\[
 \beta_h=i\sum_\ell K_\ell^\dagger\dot K_{h,\ell},\qquad
 \alpha_h=\sum_\ell\dot K_{h,\ell}^\dagger\dot K_{h,\ell}.
\]
The Kraus minimization imposes $\beta_h=0$ and minimizes the
norm of $\alpha_h$. Lemma~\ref{lem:gauge-correspondence}
identifies each $h$ with
a Hermitian $Y=PYP$, with identities
\[
 \beta_h^{\mathsf T}=\Tr_B Y,\qquad
 4\alpha_h^{\mathsf T}=M_{C,D}(Y).
\]
In particular, $Y=0$ provides a gauge with $\beta_h=0$.
For a family $\mathcal N$ in this setting, with a differentiable
Kraus representation admitting such a gauge, Zhou and
Jiang~\cite[Theorem~2, Eq.~(12)]{ZJ} give
\[
 \JS(\mathcal N)=4\min_{h=h^\dagger:\,\beta_h=0}
                         \|\alpha_h\|_\infty
 \quad\text{(Eq.~\eqref{eq:ZJ-SQL}).}
\]
They establish the existence of the asymptotic coefficient and its
equality for parallel and causally ordered adaptive protocols.
Since $\alpha_h$ is positive semidefinite and transposition
preserves eigenvalues, the two identities convert their formula
to the Choi minimization below. The equality for each $n$
transfers the limits to the original family. Attainment of the
Choi minimum follows from the coercivity in
Lemma~\ref{lem:coercivity}.

\begin{theorem}[Choi variational formula for finite RLD]\label{thm:variational}
For every differentiable finite-dimensional CPTP family satisfying $D=PDP$, the limit in Eq.~\eqref{eq:Jinf} exists and
\begin{equation}\label{eq:var}
 \JS(\Lambda)=\min_{Y\in\cV_P}\lambda_{\max}(M_{C,D}(Y)),
 \qquad M_{C,D}(0)=R(C,D).
\end{equation}
The minimum is attained. Parallel strategies achieve the same asymptotic coefficient as causally ordered adaptive strategies. No constancy of the original family's Choi rank is required.
\end{theorem}

Appendix~\ref{app:choi-translation} derives the theorem from the asymptotic optimization of Zhou and Jiang~\cite[Theorem~2]{ZJ}. For the affine replacement $\widehat\Lambda$, Lemma~\ref{lem:gauge-correspondence} identifies the Hermitian $h_0$ corresponding to $Y=0$. Inserting $Y=0$ into its constraint and objective identities gives
\[
 \beta_{h_0}=0,\qquad
 4\alpha_{h_0}^{\mathsf T}=M_{C,D}(0)=R(C,D).
\]
Since $\alpha_{h_0}$ is positive semidefinite and transposition preserves eigenvalues, $4\|\alpha_{h_0}\|_\infty=r(C,D)$. Applying the finite-use bound of Zhou and Jiang~\cite[Eqs.~(9)--(10)]{ZJ} to this representation of $\widehat\Lambda$ gives
\begin{equation}\label{eq:finite-rld-bound}
 J_n^{\mathrm S}(\Lambda;0)
 =J_n^{\mathrm S}(\widehat\Lambda;0)
 \le 4n\|\alpha_{h_0}\|_\infty=n\JR(\Lambda).
\end{equation}
The first equality uses Eq.~\eqref{eq:finite-eq}, which transfers the bound to the original family at each use count.

\subsection{The full-support specialization}
For the positive-definite saturation criterion in the next section,
the variational formula simplifies: $P=I_{BA}$, $C^+=C^{-1}$, and
every Hermitian $Y$ satisfies the support constraint. The RLD matrix in Eq.~\eqref{eq:R} therefore has the ordinary-inverse form
\begin{equation}\label{eq:R-positive}
 R(C,D)=\Tr_B(DC^{-1}D).
\end{equation}
Its largest eigenvalue is the channel RLD information. The feasible set in Eq.~\eqref{eq:V} reduces to Hermitian $Y$ with $\Tr_B Y=0$. On this set, substituting $C^{-1}$ in Eq.~\eqref{eq:M} gives
\begin{equation}\label{eq:M-positive}
 M_{C,D}(Y)=\Tr_B[(D-2iY)C^{-1}(D+2iY)].
\end{equation}
The matrix $M_{C,D}(Y)$ acts on $\cH_A$ and is positive semidefinite. 

\begin{proposition}[Choi form of the asymptotic SLD coefficient when $C>0$]\label{prop:positive-var}
Let $\Lambda_\theta$ be a differentiable finite-dimensional CPTP family with $C=C_0>0$ and $D=\dot C_0$. Under these assumptions, the optimized asymptotic SLD coefficient is given by the following Choi-space minimization:
\begin{equation}\label{eq:var-positive}
 \JS(\Lambda)=\min_{Y=Y^\dagger:\,\Tr_B Y=0}
 \lambda_{\max}(M_{C,D}(Y)),\qquad M_{C,D}(0)=R(C,D).
\end{equation}
The minimum is attained. Parallel strategies attain the same asymptotic coefficient as the full class of causally ordered protocols in the definition of $J_n^{\mathrm S}(\Lambda;0)$ in Subsection~\ref{subsec:pointwise-info}.
\end{proposition}

\begin{proof}[Proof of Proposition~\ref{prop:positive-var}]
For $C>0$ the support projection is $P=I_{BA}$ and
$C^+=C^{-1}$; the supported feasible set becomes
$\cV_P=\{Y=Y^\dagger:\Tr_B Y=0\}$. Therefore the general formula
in Theorem~\ref{thm:variational} becomes precisely
Eq.~\eqref{eq:var-positive}, with $M_{C,D}$ as in
Eq.~\eqref{eq:M-positive} and $M_{C,D}(0)=R(C,D)$.
Attainment and the parallel/adaptive equality are inherited from the
same theorem. This proves every assertion of the proposition.
\end{proof}

\section{Saturation criteria}\label{sec:positive}
\subsection{Full-support criterion}
The main result of this paper is a necessary and sufficient
condition for the optimal asymptotic SLD coefficient to attain the
channel RLD upper bound. The variational formula of the preceding
section reduces this question to whether $Y=0$ is optimal. We
state the condition first for a positive-definite Choi matrix and
then on the support of a possibly rank-deficient Choi matrix.

We first treat full support, $C>0$, so that the criterion can
be read as an equation on the entire Choi space. The criterion tests
optimality at $Y=0$ by
looking for a density operator $\sigma$ supported on the
largest-eigenvalue space of $R(C,D)$. A single such $\sigma$ must
satisfy an operator equation that makes its trace pairing with the
linear variation vanish in every feasible direction.

\begin{theorem}[RLD saturation for a positive-definite Choi matrix]\label{thm:positive}
Let $\Lambda_\theta$ be a differentiable finite-dimensional CPTP family with $C=C_0>0$ and $D=\dot C_0$. Then the limit in Eq.~\eqref{eq:Jinf} exists. Equality $\JS(\Lambda)=\JR(\Lambda)$ holds if and only if there are a density operator $\sigma$ and a Hermitian operator $H$ on $\cH_A$ such that
\begin{equation}\label{eq:positive-state}
 \sigma\ge0,\qquad \Tr\sigma=1,\qquad
 R(C,D)\sigma=r(C,D)\sigma,
\end{equation}
and
\begin{equation}\label{eq:L}
 2i[(I_B\otimes\sigma)DC^{-1}-C^{-1}D(I_B\otimes\sigma)]
 =I_B\otimes H.
\end{equation}
\end{theorem}

Zhou and Jiang~\cite[Appendix~F, Eq.~(F4)]{ZJ} give a stationarity condition at an optimal Kraus representation. Here we test the representation giving the RLD bound~\cite[Appendix~C]{KD2013}. The proof of the criterion is given in Section~\ref{sec:positive-proof}, and Appendix~\ref{app:stationarity} relates the two conventions.

To apply the theorem, first find the largest-eigenvalue space of $R(C,D)$. The equation $R(C,D)\sigma=r(C,D)\sigma$ restricts $\sigma$ to density operators on that space. If this space is one-dimensional, $\sigma$ is already determined: it is the projection onto a normalized largest-eigenvalue eigenvector. In a higher-dimensional space, mixtures as well as pure states are allowed.

Next evaluate the left-hand side of Eq.~\eqref{eq:L} for such a $\sigma$. The equation asks whether it has the form $I_B\otimes H$. In an output basis, every off-diagonal block must vanish and every diagonal block must be the same input-space matrix $H$. For example, when $\dim\cH_B=2$, the required block form is
\[
 I_B\otimes H=\begin{pmatrix}H&0\\0&H\end{pmatrix},
 \qquad H:\cH_A\longrightarrow\cH_A.
\]
Thus $H$ can be read off from a diagonal block once the block conditions hold. With $C,D$ given, both equations of the theorem are linear in $\sigma,H$; the remaining restrictions are $\sigma\ge0$ and $\Tr\sigma=1$.

\subsection{General saturation criterion}\label{sec:general-criterion}
The preceding subsection gives the saturation criterion for a
full-rank Choi matrix. We now formulate the criterion on the support
of the Choi matrix $C$ at the evaluation point, covering both
full-rank and rank-deficient $C$. Let $P$ be the projection onto
that support. As in the variational formula, we assume $D=PDP$ at
the evaluation point.

Every feasible direction satisfies $Y=PYP$, so its trace pairing
with a Choi-space operator depends only on the compression of that
operator to $\operatorname{Ran}P$. The general criterion therefore
imposes the operator equation on $\operatorname{Ran}P$, with
$\sigma,H$ acting on the input space. For full-rank $C$, this support
is the entire Choi space.

To see why saturation requires the two operators below, let $E$ be the
largest-eigenvalue space of $R(C,D)$ and compress the coefficient
of $t$ in $M_{C,D}(tY)$ to $E$. If that compression is positive
definite for a feasible direction $Y$, the right derivative of
$t\mapsto\lambda_{\max}M_{C,D}(-tY)$ at zero is strictly negative
by Lemma~\ref{lem:spectral-certificate}. Hence small positive steps
along $-Y$ strictly lower the objective. Thus, when $Y=0$ is optimal,
the real linear space of compressed variations contains no
positive-definite matrix. Separation then gives one density operator
$\sigma_E$ on $E$ whose trace pairing vanishes with every compressed
variation. Extend $\sigma_E$ by zero to an input operator $\sigma$.
The vanishing pairings say that the operator on the left of
Eq.~\eqref{eq:station} is orthogonal to the kernel of the partial
trace on supported Hermitian operators. The orthogonal complement
consists of operators $P(I_B\otimes H)P$. Section~\ref{sec:general-proof}
proves both implications using Lemmas~\ref{lem:spectral-certificate}
and~\ref{lem:supported-annihilator}.

\Needspace{15\baselineskip}
\begin{theorem}[General finite-RLD saturation theorem]\label{thm:criterion}
For a differentiable finite-dimensional CPTP family satisfying $D=PDP$, the limit in Eq.~\eqref{eq:Jinf} exists. Equality $\JS(\Lambda)=\JR(\Lambda)$ holds if and only if there are Hermitian operators $\sigma,H$ on $\cH_A$ satisfying
\[
 \sigma\ge0,\qquad \Tr\sigma=1,\qquad R(C,D)\sigma=r(C,D)\sigma,
\]
and
\begin{equation}\label{eq:station}
 P\,2i[(I_B\otimes\sigma)DC^+-C^+D(I_B\otimes\sigma)]P
 =P(I_B\otimes H)P.
\end{equation}
The support condition $D=PDP$ is imposed at the evaluation point
$\theta=0$; the Choi rank of $C_\theta$ need not be constant in a
neighborhood of that point.
\end{theorem}

When $C>0$, substituting $P=I_{BA}$ and $C^+=C^{-1}$ reduces
Eq.~\eqref{eq:station} to Eq.~\eqref{eq:L}. Section~\ref{sec:applications}
applies the supported criterion to generalized dephasing channels.

\section{Generalized Pauli channels}\label{sec:pauli}
Our first example is a generalized Pauli channel. It applies a
randomly chosen Weyl operator, with the unknown parameter entering
only through the selection probabilities. We use this structure
to compare the channel RLD bound with the information obtained
by measuring the output.
Let the input and output both have dimension $d$. Choose parameter-independent Weyl operators $W_\alpha$, $1\le\alpha\le d^2$, normalized by
\[
 \Tr(W_\alpha^\dagger W_\beta)=d\delta_{\alpha\beta}.
\]
Consider a differentiable family
\begin{equation}\label{eq:pauli-channel}
 \Lambda_\theta(X):=\sum_{\alpha=1}^{d^2}p_\alpha(\theta)
 W_\alpha XW_\alpha^\dagger,\qquad
 p_\alpha(\theta)\ge0,\quad\sum_\alpha p_\alpha(\theta)=1.
\end{equation}
Assume that every $p_\alpha:=p_\alpha(0)$ is positive, and define $\dot p_\alpha:=p'_\alpha(0)$. The positive probabilities give a positive-definite Choi matrix.
For a fixed input state $\rho$ on the channel input and a reference
system, the output can be prepared from a classical register with
distribution $p_\theta$. The preparation map sends the register
state $|\alpha\rangle\langle\alpha|$ to
$(W_\alpha\otimes I)\rho(W_\alpha^\dagger\otimes I)$.
Measuring the register in its label basis and preparing the
corresponding state defines a parameter-independent CPTP map.
By monotonicity, the output SLD information at zero is at most
the classical Fisher information $I(p):=\sum_\alpha\dot p_\alpha^2/p_\alpha$.
This is the state-preparation argument in
Ref.~\cite[Theorem~6.2 and Exercise~6.51]{HayashiQIT2017}.

For $n$ uses, fix any causally ordered adaptive protocol. Given
a label sequence $(\alpha_1,\ldots,\alpha_n)$, apply $W_{\alpha_j}$
in slot $j$ and run the protocol's fixed controls. The resulting
final state for each label sequence defines a CPTP preparation
map on the label register, independent of $\theta$. Averaging
over $p_\theta^{\otimes n}$ gives the output of the original
protocol by linearity in the channel uses. The product distribution
has Fisher information $nI(p)$, so $J_n^{\mathrm S}(\Lambda;0)\le nI(p)$.

To attain the upper bound, use the normalized maximally entangled
input $|\Phi\rangle=d^{-1/2}\sum_j|j\rangle|j\rangle$.
Define $|W_\alpha\rangle\rangle:=\sum_{b,j}(W_\alpha)_{bj}|b\rangle_B|j\rangle_A$ and $|\Phi_\alpha\rangle:=|W_\alpha\rangle\rangle/\sqrt d$. The inner product of the vectorized Weyl operators is the trace inner product.
\[
 \langle\langle W_\alpha|W_\beta\rangle\rangle
 =\sum_{b,j}\overline{(W_\alpha)_{bj}}(W_\beta)_{bj}
 =\Tr(W_\alpha^\dagger W_\beta)=d\delta_{\alpha\beta}.
\]
Dividing each vector by $\sqrt d$ gives $d^2$ orthonormal vectors, which form a basis of $\cH_B\otimes\cH_A$. The output conditional on $\alpha$ is $|\Phi_\alpha\rangle$,
so the output state is
\[
 \sum_\alpha p_\alpha(\theta)|\Phi_\alpha\rangle\langle\Phi_\alpha|.
\]
Measuring in this orthonormal basis recovers the label with
distribution $p_\theta$ and yields classical Fisher information
$I(p)$. The output SLD information is therefore $I(p)$ by
monotonicity under the measurement and the upper bound above.
Independent repetitions give $J_n^{\mathrm S}(\Lambda;0)=nI(p)$
for every $n\ge1$, and hence $\JS=I(p)$.
See Ref.~\cite[Exercises~6.52--6.54]{HayashiQIT2017} for the
orthogonal-output argument and the generalized Pauli example.
Fujiwara and Imai~\cite[Theorem~1]{FI2003} established the optimality
of maximally entangled inputs for generalized Pauli channels,
including parallel extensions.

We next compute the channel RLD value using the same basis.
Substituting a unitary channel into Eq.~\eqref{eq:choi} gives its Choi matrix $|W_\alpha\rangle\rangle\langle\langle W_\alpha|=d|\Phi_\alpha\rangle\langle\Phi_\alpha|$. In the chosen vectorization, summing over the output index gives
\[
 \Tr_B(|W_\alpha\rangle\rangle\langle\langle W_\alpha|)
 =(W_\alpha^\dagger W_\alpha)^{\mathsf T}=I_A.
\]
Dividing by $d$ gives $\Tr_B|\Phi_\alpha\rangle\langle\Phi_\alpha|=I_A/d$.
The Choi data and their inverse are consequently
\begin{equation}
C=d\sum_\alpha p_\alpha|\Phi_\alpha\rangle\langle\Phi_\alpha|>0,
 D=d\sum_\alpha\dot p_\alpha|\Phi_\alpha\rangle\langle\Phi_\alpha|,\qquad C^{-1}=\frac1d\sum_\alpha\frac1{p_\alpha}|\Phi_\alpha\rangle\langle\Phi_\alpha|. \label{eq:pauli-data}
\end{equation}
In this basis every eigenvalue $dp_\alpha$ of $C$ is positive. The matrices $C,D,C^{-1}$ are diagonal in the same basis. Multiplication and then the output partial trace give
\begin{equation}
DC^{-1}D=d\sum_\alpha\frac{\dot p_\alpha^2}{p_\alpha}
 |\Phi_\alpha\rangle\langle\Phi_\alpha|,\qquad R(C,D)=\left(\sum_\alpha\frac{\dot p_\alpha^2}{p_\alpha}\right)I_A,
 r(C,D)=\sum_\alpha\frac{\dot p_\alpha^2}{p_\alpha}. \label{eq:pauli-R}
\end{equation}
The largest eigenvalue is $I(p)$, so the optimal SLD value
computed above attains the RLD bound.
In particular,
\begin{equation}\label{eq:pauli-values}
 J_1^{\mathrm S}(\Lambda;0)=\JS(\Lambda)=\JR(\Lambda)
 =\sum_\alpha\frac{\dot p_\alpha^2}{p_\alpha}.
\end{equation}

We now verify the saturation criterion of Theorem~\ref{thm:positive}.
Since $R$ is a scalar matrix, its largest-eigenvalue space is the entire input space. Choose $\sigma=I_A/d$. This is a density operator satisfying Eq.~\eqref{eq:positive-state}. For the other condition,
\[
 2i[(I_B\otimes\sigma)DC^{-1}-C^{-1}D(I_B\otimes\sigma)]
 =\frac{2i}{d}(DC^{-1}-C^{-1}D)=0.
\]
Thus $\sigma=I_A/d$ and $H=0$ satisfy the two conditions for saturation.

For the qubit example, relabel the four Weyl operators by indices $0,1,2,3$ rather than the indices $1,\ldots,d^2$ used in the general formula. Take $W_0=I$, $W_1=\sigma_X$, $W_2=\sigma_Y$, $W_3=\sigma_Z$, the identity and the three Pauli matrices. One concrete family is
\[
 p_0(\theta)=1-3(q+\theta),\qquad
 p_1(\theta)=p_2(\theta)=p_3(\theta)=q+\theta,
 \qquad 0<q<1/3,
\]
on a sufficiently small interval around zero. Here $\dot p_0=-3$ and $\dot p_1=\dot p_2=\dot p_3=1$, so
\[
 r(C,D)=\frac9{1-3q}+\frac3q.
\]
At $q=1/4$ this gives $R=48I_2$, $\sigma=I_2/2$, $H=0$, and information $48$ at one use as well as asymptotically. Mixing only a proper subset of the Weyl operators gives a rank-deficient Choi matrix. For such mixtures the general saturation test takes the support-restricted form of Subsection~\ref{sec:general-criterion}.

\section{Measurement--preparation channels}\label{sec:full-support-examples}
\subsection{Basis measurement followed by state preparation}\label{sec:measure-prepare}
Our next examples are measurement--preparation channels. In the
model below, sending a basis state lets us choose which output
state is prepared. This gives a way to compare the channel SLD
and RLD information through the information of the individual
output states.

We consider a projective measurement in a fixed orthonormal input
basis $\{|x\rangle\}_{x=1}^d$, followed by preparation of the state
$\rho_{x,\theta}$ for outcome $x$:
\begin{equation}\label{eq:mp}
 \Lambda_\theta(X):=\sum_{x=1}^d\langle x|X|x\rangle\rho_{x,\theta}.
\end{equation}
These channels are entanglement breaking~\cite{HSR2003}. The
outcome with the largest state SLD information can differ from the
one with the largest state RLD information.

Suppose that $\rho_x=\rho_{x,0}>0$ for every $x$, and set $D_x=\dot\rho_{x,0}$. Let $L_x$ denote the SLD of $(\rho_x,D_x)$. The corresponding scalar state RLD and SLD information values are
\begin{align}\label{eq:state-figures}
 r_x&=\Tr(D_x\rho_x^{-1}D_x),&
 f_x&=\Tr(\rho_x L_x^2),&
 D_x&=\tfrac12(\rho_x L_x+L_x\rho_x),\quad L_x=L_x^\dagger.
\end{align}

The following state-level comparison identifies when the SLD and RLD
information of a prepared state agree. We formulate it also for a rank-deficient
state whose derivative stays on its support, since it is reused for the
Choi state produced by a maximally entangled input in Proposition~\ref{prop:maxent}.
\begin{lemma}[State SLD and RLD information for a derivative supported at the evaluation point]\label{lem:state-gap}
Let $\rho\ge0$ be a density operator, $P_\rho$ its support projection,
and $\dot\rho=\dot\rho^\dagger=P_\rho\dot\rho P_\rho$. In an orthonormal
eigenbasis of $\rho$ on its support, write $\rho|i\rangle=\lambda_i|i\rangle$
with $\lambda_i>0$. Then
\begin{equation}\label{eq:state-gap}
 \Tr(\dot\rho\,\rho^+\dot\rho)-F^{\mathrm S}(\rho,\dot\rho)
 =\sum_{i<j}\frac{(\lambda_i-\lambda_j)^2}
 {\lambda_i\lambda_j(\lambda_i+\lambda_j)}
 |\langle i|\dot\rho|j\rangle|^2\ge0.
\end{equation}
Equality holds if and only if $[\rho,\dot\rho]=0$.
\end{lemma}
\begin{proof}
The support condition allows both expressions to be computed on
$\operatorname{Ran}P_\rho$, where $\rho^+$ has eigenvalues $\lambda_i^{-1}$.
The trace on the left has diagonal terms
$|\dot\rho_{ii}|^2/\lambda_i$ and, for each $i<j$, off-diagonal terms
$(\lambda_i^{-1}+\lambda_j^{-1})|\dot\rho_{ij}|^2$.
By Eq.~\eqref{eq:SLD-spectral}, the SLD expression has the same diagonal
terms and off-diagonal terms
$4|\dot\rho_{ij}|^2/(\lambda_i+\lambda_j)$.
The diagonal terms cancel. For each pair $i<j$ the remaining coefficient is
\[
 \frac{1}{\lambda_i}+\frac{1}{\lambda_j}-\frac{4}{\lambda_i+\lambda_j}
 =\frac{(\lambda_i-\lambda_j)^2}{\lambda_i\lambda_j(\lambda_i+\lambda_j)}.
\]
This gives Eq.~\eqref{eq:state-gap}.
Every summand is nonnegative, and it vanishes exactly when
$(\lambda_i-\lambda_j)\dot\rho_{ij}=0$.
These are precisely the matrix entries of $[\rho,\dot\rho]=0$ on the
support; both operators vanish on the orthogonal complement.
\end{proof}

\begin{proposition}[Information and RLD saturation for fixed-basis measurement and preparation]\label{prop:mp}
For the channel in Eq.~\eqref{eq:mp}, assume $\rho_{x,0}>0$ for every $x$. Then
\begin{equation}\label{eq:mp-value}
 \JR=\max_x r_x,\qquad \JS=\max_x f_x.
\end{equation}
Consequently, $\JS=\JR$ if and only if at least one index $x_*$ satisfies
\begin{equation}\label{eq:mp-test}
 r_{x_*}=\max_x r_x,\qquad [\rho_{x_*},D_{x_*}]=0.
\end{equation}
The conditions impose no commutativity requirement on the other output states.
\end{proposition}
\Needspace{6\baselineskip}
\begin{proof}
\leavevmode\par\nobreak

\Needspace{3\baselineskip}
We compute the RLD value from the diagonal Choi blocks. For the
SLD coefficient, repeating a fixed input gives the lower bound,
and the Kraus construction in Appendix~\ref{app:mp-kraus} gives
the upper bound. Comparing the resulting values yields the
equality condition.

\noindent\textbf{Step 1: Prove that $\JR=\max_x r_x$.}\par\nobreak

Since $\Lambda_\theta(|j\rangle\langle k|)=\delta_{jk}\rho_{j,\theta}$,
Eq.~\eqref{eq:choi} and differentiation give the Choi data
\[
 C=\sum_x\rho_x\otimes|x\rangle\langle x|,\qquad
 D=\sum_x D_x\otimes|x\rangle\langle x|.
\]
Each block $\rho_x$ is positive definite, so $C>0$ and $C^{-1}=\sum_x\rho_x^{-1}\otimes|x\rangle\langle x|$. Proposition~\ref{prop:positive-var} therefore ensures that the asymptotic limit exists. Multiplying the blocks and taking the partial trace gives
\[
 R=\Tr_B(DC^{-1}D)=\sum_x\Tr(D_x\rho_x^{-1}D_x)|x\rangle\langle x|
   =\sum_x r_x|x\rangle\langle x|.
\]
Hence $\JR=\lambda_{\max}(R)=\max_x r_x$.

\Needspace{3\baselineskip}
\noindent\textbf{Step 2: Prove that $\JS\ge\max_x f_x$.}\par\nobreak

For every $x$ and $n$, preparing $|x\rangle$ at each use produces $n$ copies of $\rho_{x,\theta}$. Additivity of the SLD information gives $J_n^{\mathrm S}/n\ge f_x$. Taking the limit and then maximizing over $x$ gives $\JS\ge\max_x f_x$.

\Needspace{3\baselineskip}
\noindent\textbf{Step 3: Prove that $\JS\le\max_x f_x$.}\par\nobreak

To bound the optimum over all causally ordered protocols, use
the Kraus representation in Lemma~\ref{lem:mp-kraus-factor}.
Its two identities give $\beta_0=0$ and
$4\alpha_0=\sum_x f_x|x\rangle\langle x|$ for the gauge $h=0$
relative to that representation. Here $h=0$ leaves the constructed
Kraus derivatives as they are. The Choi point $Y=0$ instead selects
the gauge that gives the RLD value.
Substituting this feasible gauge
into Eq.~\eqref{eq:ZJ-SQL} gives
\[
 \JS\le4\|\alpha_0\|_\infty
 =\Big\|\sum_x f_x|x\rangle\langle x|\Big\|_\infty
 =\max_x f_x.
\]
Together with Step~2 this proves $\JS=\max_x f_x$.

\Needspace{3\baselineskip}
\noindent\textbf{Step 4: Prove that $\JS=\JR$ holds if and only if there is an $x_*$ with $r_{x_*}=\max_xr_x$ and $[\rho_{x_*},D_{x_*}]=0$.}\par\nobreak

Apply Lemma~\ref{lem:state-gap} to each $\rho_x>0$ and
$\dot\rho=D_x$. Its equality condition is
\[
 f_x\le r_x,\qquad f_x=r_x\ \Longleftrightarrow\ [\rho_x,D_x]=0.
\]
We need this equality for an index that actually attains the maximum,
not for every $x$. 

For necessity, assume $\JS=\JR$ and choose $x_*$ maximizing $f_x$. Then
\[
 \max_x f_x=f_{x_*}\le r_{x_*}\le\max_x r_x=\max_x f_x.
\]
Equality throughout shows that $x_*$ maximizes $r_x$ and $[\rho_{x_*},D_{x_*}]=0$. For sufficiency, choose an $x_*$ maximizing $r_x$ with $[\rho_{x_*},D_{x_*}]=0$. Then
\[
 \max_x r_x=r_{x_*}=f_{x_*}\le\max_x f_x\le\max_x r_x.
\]
Thus Eq.~\eqref{eq:mp-value} gives $\JS=\JR$.
\end{proof}
\subsection{Saturation without commuting Choi data}\label{sec:noncommuting-saturation}
By Proposition~\ref{prop:mp}, the RLD bound is attained if some
outcome maximizes the state RLD information and its prepared state
commutes with its derivative. We construct a two-outcome channel
in which the first outcome meets this condition, while the state
for the second outcome and its derivative do not commute.
Let the input basis be $|1\rangle,|2\rangle$, let the output be a
qubit, and define
\begin{equation}\label{eq:elementary-mp}
 \Lambda_\theta(X):=\sum_{x=1}^2\langle x|X|x\rangle\rho_{x,\theta},
 \qquad \rho_{x,\theta}:=\rho_x+\theta D_x.
\end{equation}
For this channel, choose
\begin{equation}\label{eq:saturating-mp-data}
 \rho_1=I_2/2,\quad D_1=\sigma_Z/3,\qquad
 \rho_2=(I_2+\sigma_X/2)/2,\quad D_2=\sigma_Z/4.
\end{equation}
Both $\rho_x$ are positive definite, and $\Tr D_x=0$, so the affine states in Eq.~\eqref{eq:elementary-mp} are density operators for sufficiently small $|\theta|$. In the output--input ordering,
\[
 C=\sum_x\rho_x\otimes|x\rangle\langle x|>0,\qquad
 D=\sum_xD_x\otimes|x\rangle\langle x|,\qquad
 C^{-1}=\sum_x\rho_x^{-1}\otimes|x\rangle\langle x|.
\]
The inverse of the prepared-state block for outcome $x=2$ is
\[
 \rho_2^{-1}=\frac83 I_2-\frac43\sigma_X.
\]
Using $\sigma_Z\sigma_X\sigma_Z=-\sigma_X$, we obtain
\begin{equation}
\Tr(D_1\rho_1^{-1}D_1)=\Tr(2I_2/9)=4/9,\qquad \Tr(D_2\rho_2^{-1}D_2)=\Tr(I_2/6+\sigma_X/12)=1/3, \label{eq:elementary-mp-R}
\end{equation}
and therefore $R=\operatorname{diag}(4/9,1/3)$ in the input basis. The largest eigenvalue is simple, so Eq.~\eqref{eq:positive-state} requires $\sigma=|1\rangle\langle1|$. Substitution in the second condition selects only the first input component:
\[
 2i[(I_B\otimes\sigma)DC^{-1}-C^{-1}D(I_B\otimes\sigma)]
 =2i(D_1\rho_1^{-1}-\rho_1^{-1}D_1)\otimes|1\rangle\langle1|=0.
\]
Take $H=0$. Theorem~\ref{thm:positive} now proves $\JS=\JR=4/9$, even though $[\rho_2,D_2]\ne0$ and hence $[C,D]\ne0$. The input vector $|2\rangle$ is outside the largest-eigenvalue space of $R$, so its output state and derivative do not contribute to Eq.~\eqref{eq:L} for this choice of $\sigma$.

Here a physical probe can again be identified separately: input $|1\rangle$ produces $\rho_{1,\theta}=I_2/2+\theta\sigma_Z/3$, whose SLD at zero is $2\sigma_Z/3$ and whose SLD information is $4/9$. This input attains the bound already at one use.

\subsection{A strict RLD upper bound}\label{sec:strict-full-support}
We change only $D_1$ so that outcome $2$, whose state and derivative
do not commute, becomes the unique maximizer of the state RLD information.
Specifically, keep both states and $D_2$ fixed and decrease the
coefficient of $\sigma_Z$ in $D_1$ from $1/3$ to $1/4$. To calculate the information values in
both examples, write $\rho_x=(I+c_x\sigma_X)/2$ and $D_x=a_x\sigma_Z$, with $|c_x|<1$. The matrices $\sigma_X$ and $\sigma_Z$ anticommute, so $L_x=2a_x\sigma_Z$ satisfies $(\rho_xL_x+L_x\rho_x)/2=D_x$. Also $\rho_x^{-1}=2(I-c_x\sigma_X)/(1-c_x^2)$. Since $L_x^2=4a_x^2 I_2$, its expectation in $\rho_x$ is $4a_x^2$.
Conjugation by $\sigma_Z$ gives
$D_x\rho_x^{-1}D_x=2a_x^2(I_2+c_x\sigma_X)/(1-c_x^2)$,
whose trace uses $\Tr\sigma_X=0$. Thus
\begin{equation}\label{eq:qubit-block-info}
 f_x=4a_x^2,\qquad r_x=\frac{4a_x^2}{1-c_x^2}.
\end{equation}
The changed model has $a_1=a_2=1/4$, $c_1=0$, and $c_2=1/2$. Equation~\eqref{eq:qubit-block-info} gives
\[
 r_1=\frac{1/4}{1-0^2}=\frac14,\qquad
 r_2=\frac{1/4}{1-(1/2)^2}=\frac13.
\]
The diagonal form of $R$ in Proposition~\ref{prop:mp} gives $R=\operatorname{diag}(1/4,1/3)$ and $r=1/3$. The largest-eigenvalue space has therefore changed from $\operatorname{span}\{|1\rangle\}$ to $\operatorname{span}\{|2\rangle\}$.
To see how the general operator criterion gives the same conclusion, note that $R\sigma=r\sigma$ forces $\sigma=|2\rangle\langle2|$. In contrast with the preceding example, the selected output state and derivative do not commute. The left-hand side of Eq.~\eqref{eq:L} is
\begin{align}\label{eq:strict-mp-station}
 2i(D_2\rho_2^{-1}-\rho_2^{-1}D_2)\otimes|2\rangle\langle2|
 &=\frac43\sigma_Y\otimes|2\rangle\langle2|\ne0.
\end{align}
Since $\Tr\sigma_Y=0$, the output partial trace of the displayed operator is $\tfrac43(\Tr\sigma_Y)|2\rangle\langle2|=0$. The partial trace of $I_B\otimes H$ is $2H$, so Eq.~\eqref{eq:L} would force $H=0$ while its left-hand side is nonzero. Equation~\eqref{eq:L} alone is satisfied by $\sigma=|1\rangle\langle1|$ and $H=0$, since $\rho_1$ and $D_1$ commute. But Eq.~\eqref{eq:positive-state} requires $\sigma=|2\rangle\langle2|$, for which the nonzero operator in Eq.~\eqref{eq:strict-mp-station} rules out every $H$. Thus the two conditions cannot hold together, and Theorem~\ref{thm:positive} gives $\JS<\JR=1/3$.

In this example, $a_1=a_2=1/4$, so Eq.~\eqref{eq:qubit-block-info} gives $f_1=f_2=1/4$. Proposition~\ref{prop:mp} therefore gives $\JS=1/4<1/3=\JR$.

\subsection{Replacer channels}
A replacer channel is the special case of the measurement--preparation
model in Subsection~\ref{sec:measure-prepare} in which the same state
$\rho_\theta$ is prepared for every outcome. Summing over the
outcomes gives $\Lambda_\theta(X)=\Tr(X)\rho_\theta$~\cite{CMW2016}.
The preceding example had a strict gap between the RLD bound and
the optimal asymptotic SLD coefficient. We use replacer channels
to ask how large their ratio can be at fixed dimension and to
examine the effect of a change in Choi rank at the evaluation point.

To evaluate the optimum over adaptive strategies, we first reduce
the experiment to one in which the output states are supplied directly.
For an operator $X$ on the input $A$ and a retained memory $M$,
\[
 (\Lambda_\theta\otimes\mathrm{id}_M)(X)
 =\rho_\theta\otimes\Tr_A X.
\]
The identity follows on product operators and extends by linearity.
To simulate a fixed $n$-use adaptive protocol from $\rho_\theta^{\otimes n}$,
prepare its initial input and memory as usual. At each call discard the
current input by partial trace, attach the next copy of $\rho_\theta$
to the remaining memory, and apply the protocol's next fixed control.
The final state is therefore obtained from $\rho_\theta^{\otimes n}$
by a channel independent of $\theta$. Monotonicity and product additivity
of SLD information give the upper bound $nF^{\mathrm S}(\rho_0,\dot\rho_0)$.
Sending fixed inputs and collecting the outputs attains this bound, so
\[
 J_n^{\mathrm S}(\Lambda;0)=nF^{\mathrm S}(\rho_0,\dot\rho_0)
 \qquad(n\ge1).
\]

\paragraph{An unbounded gap at fixed dimension.}
We use the reduction to state information to compare the two
channel values with a qubit output. We evaluate the information
at $\theta=0$ for each fixed $c<1$
before taking $c\to1^-$.

\begin{corollary}[Unbounded RLD overestimate at fixed dimension]\label{cor:full-support-unbounded}
For $0<c<1$ and a fixed real $a\ne0$, consider the qubit
replacer-channel family
\[
 \Lambda_{c,\theta}(X)=\Tr(X)\rho_{c,\theta},\qquad
 \rho_{c,\theta}=\frac{I+c\sigma_X}{2}+\theta a\sigma_Z
\]
for $|\theta|<(1-c)/(2|a|)$.
Its Choi matrix at zero is positive definite, and
\begin{equation}\label{eq:full-support-unbounded}
 \JS(\Lambda_c)=4a^2,\qquad
 \JR(\Lambda_c)=\frac{4a^2}{1-c^2},\qquad
 \frac{\JR(\Lambda_c)}{\JS(\Lambda_c)}=\frac1{1-c^2}.
\end{equation}
Thus, even for a fixed qubit input and output dimension and
positive-definite Choi matrices, the ratio between the finite
RLD upper bound and the optimal asymptotic SLD coefficient has
no universal finite upper bound. The smallest Choi eigenvalue
$(1-c)/2$ tends to zero as $c\to1^-$.
\end{corollary}
\begin{proof}
The smallest eigenvalue of $\rho_{c,0}$ is $(1-c)/2>0$. The perturbation has norm $|\theta||a|$, so $\rho_{c,\theta}$ is positive definite for $|\theta|<(1-c)/(2|a|)$ and has trace one. Thus $C=\rho_{c,0}\otimes I_2>0$ and $D=a\sigma_Z\otimes I_2$. Equation~\eqref{eq:qubit-block-info} gives the state values $4a^2$ and $4a^2/(1-c^2)$ for this output. The replacer identity above gives $\JS$ from the state SLD value.
For the RLD, $\Tr_B(DC^{-1}D)=\Tr(a\sigma_Z\rho_{c,0}^{-1}a\sigma_Z)I_2$,
so the state RLD value is also the channel RLD value. Taking their
ratio gives Eq.~\eqref{eq:full-support-unbounded}.
\end{proof}

\paragraph{A change in Choi rank at the evaluation point.}\label{sec:rank-changing}
Starting with the same two-level state and derivative, we add a third,
orthogonal output level of weight $\kappa\theta^2$. Here $a,c,\kappa$
are fixed and $\theta$ varies. The added level changes the Choi rank
without contributing to the derivative at zero. We compare the SLD
information at zero with the limit of its values at nonzero $\theta$.

Let the input space be $\mathbb C^2$ and the output space $\mathbb C^3$ with orthonormal basis $|0\rangle,|1\rangle,|2\rangle$. Take $a>0$, $0\le c<1$, and $\kappa>0$. Define the output state and its derivative data by
\begin{align}\label{eq:states}
 \rho_0(c)&:=\frac12\begin{pmatrix}1&c&0\\c&1&0\\0&0&0\end{pmatrix},
 &T(a)&:=a\begin{pmatrix}1&0&0\\0&-1&0\\0&0&0\end{pmatrix},\\
 \rho_\theta(a,c,\kappa)&:=(1-\kappa\theta^2)(\rho_0(c)+\theta T(a))+\kappa\theta^2|2\rangle\langle2|.
 \nonumber
\end{align}
The smallest eigenvalue of the upper $2\times2$ block of $\rho_0(c)$ is $(1-c)/2>0$. Its perturbation $\theta T(a)$ has norm $|\theta|a$, so the upper block of $\rho_0(c)+\theta T(a)$ stays positive definite for $|\theta|<(1-c)/(2a)$. This block has trace one. For $|\theta|<1/\sqrt\kappa$, its weight $1-\kappa\theta^2$ is positive, and the lower block of $\rho_\theta$ has weight $\kappa\theta^2$, positive when $\theta\ne0$. The weights sum to one. Thus $\rho_\theta$ is a density operator near zero, with rank two at zero and rank three at nearby nonzero $\theta$.

Set $\Lambda_\theta(X)=\Tr(X)\rho_\theta(a,c,\kappa)$. Its Choi matrix is $C_\theta=\rho_\theta\otimes I_2$, so
$\operatorname{rank}C_\theta=2\operatorname{rank}\rho_\theta$.
The Choi rank is therefore four at zero and six at nearby nonzero $\theta$. Differentiation at zero gives $D=T(a)\otimes I_2$, since the weight $\kappa\theta^2$ has zero derivative there. The support projection of $C$ is $P=P_{\rm out}\otimes I_2$,
where $P_{\rm out}=|0\rangle\langle0|+|1\rangle\langle1|$.
Since $P_{\rm out}T(a)P_{\rm out}=T(a)$, we have $D=PDP$. The SLD information values below are evaluated pointwise at zero, as in Eq.~\eqref{eq:Jinf}. 
\begin{proposition}[Exact equality and strict inequality in rank-changing families]\label{prop:family}
For the replacer channel with output state in Eq.~\eqref{eq:states},
\begin{equation}\label{eq:values}
 \JR=\frac{4a^2}{1-c^2},\qquad
 \JS=4a^2.
\end{equation}
Thus $c=0$ gives saturation and every $0<c<1$ gives strict non-saturation, despite rank change and nonzero $D$ in both cases.
\end{proposition}
\Needspace{6\baselineskip}
\begin{proof}
\leavevmode\par\nobreak

\Needspace{3\baselineskip}
\noindent\textbf{Step 1: Prove that $\JR=4a^2/(1-c^2)$.}\par\nobreak

On the first two output coordinates,
$\rho_0(c)^+=2(I_2-c\sigma_X)/(1-c^2)$. It is zero on the third.
Since $T(a)=a\sigma_Z$ on those coordinates, the supported block of
$T(a)\rho_0(c)^+T(a)$ is
$2a^2(I_2+c\sigma_X)/(1-c^2)$ by $\sigma_Z\sigma_X\sigma_Z=-\sigma_X$.
It follows that
\begin{equation*}
R\overset{(a)}{=}\Tr\!\left[T(a)\rho_0(c)^+T(a)\right] I_2
 \overset{(b)}{=}\frac{4a^2}{1-c^2}I_2,\qquad \JR\overset{(c)}{=}\frac{4a^2}{1-c^2}.
\end{equation*}
Here (a) substitutes $C^+=\rho_0(c)^+\otimes I_2$ and
$D=T(a)\otimes I_2$ into $R=\Tr_B(DC^+D)$. Equality (b)
uses $\Tr\sigma_X=0$ on the two-dimensional block, and (c)
takes the largest eigenvalue of $R$.

\Needspace{3\baselineskip}
\noindent\textbf{Step 2: Compute $F^{\mathrm S}(\rho_0(c),T(a))=4a^2$.}\par\nobreak

At zero the output state is $\rho_0(c)$ and its tangent is $T(a)$. On the upper two coordinates, set $L=2a\sigma_Z$ and extend it by zero on the third. On that block,
\[
 \frac{L\rho_0(c)+\rho_0(c)L}{2}
 =a\sigma_Z+\frac{ac}{2}(\sigma_Z\sigma_X+\sigma_X\sigma_Z)
 =a\sigma_Z.
\]
Both sides of the SLD equation vanish on the third coordinate.
Since $L^2=4a^2P_{\rm out}$,
\[
 F^{\mathrm S}(\rho_0(c),T(a))=\Tr[\rho_0(c)L^2]=4a^2.
\]

\Needspace{4\baselineskip}
\noindent\textbf{Step 3: Obtain the channel SLD coefficient.}\par\nobreak

The replacer identity gives $J_n^{\mathrm S}(\Lambda;0)=4na^2$
from the state SLD value in Step~2. Dividing by $n$ and taking the
limit gives $\JS=4a^2$. Comparing with Step~1 proves saturation for
$c=0$ and strict inequality for $0<c<1$.
\end{proof}

\paragraph{SLD information near the rank change.}
The pointwise SLD value is $4a^2$ at zero. We now show that the state
SLD information evaluated at nonzero $\theta$ tends to $4a^2+4\kappa$. For sufficiently small $\theta$ with $1-c^2-4a^2\theta^2>0$, let $\tau_\theta$ be the restriction of $\rho_0(c)+\theta T(a)$
to the first two output coordinates. This normalized $2\times2$ state
is positive definite, so Eq.~\eqref{eq:bloch-sld} applies. Its Bloch vector is $\boldsymbol s=(c,0,2a\theta)$ and derivative $\dot{\boldsymbol s}=(0,0,2a)$. Thus $|\dot{\boldsymbol s}|^2=4a^2$, $\boldsymbol s\cdot\dot{\boldsymbol s}=4a^2\theta$, and $1-|\boldsymbol s|^2=1-c^2-4a^2\theta^2$. Substituting into Eq.~\eqref{eq:bloch-sld} gives
\[
 F^{\mathrm S}(\tau_\theta,\partial_\theta\tau_\theta)
 =4a^2+\frac{(4a^2\theta)^2}{1-c^2-4a^2\theta^2}
 =\frac{4a^2(1-c^2)}{1-c^2-4a^2\theta^2}.
\]
Write $p=\kappa\theta^2$. For sufficiently small $\theta\ne0$ the state has two orthogonal blocks $(1-p)\tau_\theta$ and $p$, both with positive weight. For a block $s\tau$ with $s>0$ and $\Tr\tau=1$, its derivative is $\dot s\,\tau+s\dot\tau$. If $L_\tau$ solves the SLD equation for $\tau$, then
\[
 \tfrac12\{s\tau,L_\tau+(\dot s/s)I\}
 =s\dot\tau+\dot s\,\tau ,
\]
where $\{U,V\}=UV+VU$. Its contribution to the information is
\[
 \Tr[s\tau(L_\tau+(\dot s/s)I)^2]
 =sF^{\mathrm S}(\tau,\dot\tau)+2\dot s\,\Tr(\tau L_\tau)+\dot s^2/s
 =sF^{\mathrm S}(\tau,\dot\tau)+\dot s^2/s .
\]
The last equality uses $\Tr(\tau L_\tau)=\Tr\dot\tau=0$. The one-dimensional block has a constant normalized state and contributes only its weight derivative. Summing the two blocks gives
\[
 F^{\mathrm S}(\rho_\theta,\partial_\theta\rho_\theta)
 =(1-p)F^{\mathrm S}(\tau_\theta,\partial_\theta\tau_\theta)
       +\frac{\dot p^2}{1-p}+\frac{\dot p^2}{p}.
\]
The last two terms sum to the classical information in the two block probabilities,
\[
 \frac{(2\kappa\theta)^2}
 {\kappa\theta^2(1-\kappa\theta^2)}
 =\frac{4\kappa}{1-\kappa\theta^2}.
\]
Consequently,
\begin{equation}\label{eq:nearby-QFI}
 F^{\mathrm S}(\rho_\theta,\partial_\theta\rho_\theta)
 =(1-\kappa\theta^2)
 \frac{4a^2(1-c^2)}{1-c^2-4a^2\theta^2}
 +\frac{4\kappa}{1-\kappa\theta^2},\qquad\theta\ne0,
\end{equation}
where $\rho_\theta=\rho_\theta(a,c,\kappa)$ is the state in Eq.~\eqref{eq:states}. Thus
\begin{equation}\label{eq:pointwise-nearby}
 F^{\mathrm S}(\rho_0(c),T(a))=4a^2,
 \qquad
 \lim_{\theta\to0,\,\theta\ne0}
 F^{\mathrm S}(\rho_\theta,\partial_\theta\rho_\theta)
 =4a^2+4\kappa.
\end{equation}
This is an instance of the known discontinuity of SLD information at changes of rank~\cite{Safranek2017}. The coefficient in Eq.~\eqref{eq:values} uses the value at zero. For $\theta\ne0$, the new level contributes
$\dot p^2/p=(2\kappa\theta)^2/(\kappa\theta^2)=4\kappa$.
At zero both $p$ and $\dot p$ vanish, and that level contributes
nothing to the pointwise SLD information.

\section{Generalized dephasing channels}\label{sec:applications}
\subsection{When is the RLD bound attained?}\label{sec:correlation}
Our next example is a generalized dephasing channel. In a fixed
basis it preserves the diagonal entries of the input state and
multiplies each off-diagonal entry by a parameter-dependent
coefficient. The effect of the parameter can therefore be studied
entry by entry. Fix an orthonormal basis
$\{|j\rangle\}_{j=1}^d$ of $\mathbb C^d$ and write the channel as
\begin{equation}\label{eq:corr-channel}
 [\Lambda_\theta(X)]_{jk}:=(A_\theta)_{jk}X_{jk},\qquad
 A_\theta\ge0,\quad (A_\theta)_{jj}=1.
\end{equation}
The coefficient matrix $A_\theta$ is a correlation matrix, meaning
that it is positive semidefinite with unit diagonal
(see also Ref.~\cite{SinghDatta2022}).

The Choi definition gives
$C_\theta=\sum_{j,k}(A_\theta)_{jk}|jj\rangle\langle kk|$,
where $|jj\rangle=|j\rangle_B|j\rangle_A$. Positivity of $A_\theta$
therefore gives complete positivity. The unchanged diagonal gives trace
preservation. At the evaluation point write $A=A_0$ and $G=\dot A_0$.
The unit diagonal of $A_\theta$ gives $G_{jj}=0$.
We assume $A>0$ in the saturation criterion below. Even then, for $d>1$
the $d$-dimensional Choi support lies in a space of dimension $d^2$,
so we apply the general criterion in Theorem~\ref{thm:criterion}.
On this support the criterion reduces to equations for weights on
the basis indices. We derive those equations here and use them in
the later examples.

Define the quantities that will enter the saturation test by
\begin{equation}\label{eq:corr-R}
 q_j:=(GA^{-1}G)_{jj},\quad q_*:=\max_j q_j,\quad I_*:=\{j:q_j=q_*\}.
\end{equation}

The following lemma gives the Choi data on their support and the
diagonal form of $R$ used in the saturation criterion. It also
expresses the feasible directions and the variational objective in
$d\times d$ matrices for the calculation at complete dephasing.
\begin{lemma}[Choi compression for generalized dephasing channels]\label{lem:corr-compression}
Let $A>0$ have unit diagonal and let $G=G^\dagger$ have zero diagonal.
Define the isometry $W:\mathbb C^d\to\mathbb C^d\otimes\mathbb C^d$
by $W|j\rangle=|j\rangle_B|j\rangle_A$, and let $C=WAW^\dagger$,
$D=WGW^\dagger$. Then
\[
 \begin{gathered}
 P=WW^\dagger,\quad C^+=WA^{-1}W^\dagger,\quad
 \Tr_B(WZW^\dagger)=\operatorname{diag}(Z),\\
 \cV_P=\{WZW^\dagger:Z=Z^\dagger,\ \operatorname{diag}(Z)=0\},\qquad
 R(C,D)=\operatorname{diag}(GA^{-1}G),
 \end{gathered}
\]
and for every such Hermitian $Z$,
\begin{equation}\label{eq:corr-compression}
 M_{C,D}(WZW^\dagger)
 =\operatorname{diag}[(G-2iZ)A^{-1}(G+2iZ)].
\end{equation}
Here $\operatorname{diag}(B)$ denotes the diagonal matrix formed from
the diagonal entries of $B$. In particular the largest-eigenvalue space
of $R(C,D)$ is spanned by $|j\rangle$ with
$(GA^{-1}G)_{jj}=\max_k(GA^{-1}G)_{kk}$.
\end{lemma}
\begin{proof}
The Choi definition~\eqref{eq:choi} and Eq.~\eqref{eq:corr-channel}
give $C_\theta=WA_\theta W^\dagger$; differentiating gives $D$.
Since $A>0$ and $W^\dagger W=I$, the range of $C$ is $\operatorname{ran}W$,
so $P=WW^\dagger$ and $C^+=WA^{-1}W^\dagger$.
For matrix units,
$\Tr_B(|jj\rangle\langle kk|)=\delta_{jk}|j\rangle\langle k|$;
linearity proves the partial-trace identity. Every supported Hermitian
$Y$ is uniquely $WZW^\dagger$ for Hermitian $Z$, and the partial-trace
condition is exactly $\operatorname{diag}(Z)=0$.
For a feasible $Y=WZW^\dagger$, the objective is
\begin{equation*}
M_{C,D}(WZW^\dagger)
 \overset{(a)}{=}\Tr_B\!\left[W(G-2iZ)A^{-1}(G+2iZ)W^\dagger\right] \overset{(b)}{=}\operatorname{diag}\!\left[(G-2iZ)A^{-1}(G+2iZ)\right].
\end{equation*}
Equality (a) substitutes $D=WGW^\dagger$ and $C^+=WA^{-1}W^\dagger$
in Eq.~\eqref{eq:M} and uses $W^\dagger W=I$. Equality (b) uses
the partial-trace identity proved above. At $Z=0$, this gives
$R=M_{C,D}(0)=\operatorname{diag}(GA^{-1}G)$. Its largest-eigenvalue
space is spanned by the basis vectors with maximizing diagonal entries.
\end{proof}

The RLD matrix is diagonal by Lemma~\ref{lem:corr-compression}.
Its largest eigenvalue is $q_*$ and the corresponding eigenspace is
spanned by the basis vectors indexed by $I_*$. We can then test RLD saturation using only the diagonal entries
$w_j=\sigma_{jj}$ of the auxiliary density operator in
Theorem~\ref{thm:criterion}.

\begin{proposition}[Linear equations for RLD saturation]\label{prop:corr}
Consider the channels in Eq.~\eqref{eq:corr-channel} with $A=A_0>0$
and $G=\dot A_0$. Then $\JS=\JR=q_*$ if and
only if there are real weights $w_j$ satisfying
\begin{equation}\label{eq:corr-test}
 \begin{gathered}
 w_j\ge0,\qquad \sum_jw_j=1,\qquad w_j=0\quad (j\notin I_*),\\
 w_j(GA^{-1})_{jk}=w_k(A^{-1}G)_{jk}
 \quad\text{for all }j\ne k.
 \end{gathered}
\end{equation}
Taking real and imaginary parts of the off-diagonal equations gives real linear constraints on the unknown weights $w_j$.
\end{proposition}
\Needspace{6\baselineskip}
\begin{proof}
\leavevmode\par\nobreak

\Needspace{3\baselineskip}
\noindent\textbf{Step 1: Verify $D=PDP$ and
$R(C,D)=\operatorname{diag}(q_1,\ldots,q_d)$.}\par\nobreak

Lemma~\ref{lem:corr-compression}, applied to the channel in
Eq.~\eqref{eq:corr-channel}, gives $D=PDP$ and
$R(C,D)=\operatorname{diag}(q_1,\ldots,q_d)$.
Thus Theorem~\ref{thm:criterion} applies, and $I_*$ indexes the
largest-eigenvalue space of $R(C,D)$.

\Needspace{3\baselineskip}
\noindent\textbf{Step 2: Derive the weight conditions from RLD saturation.}\par\nobreak

If the RLD bound is saturated, Theorem~\ref{thm:criterion} gives a pair $\sigma,H$ satisfying its operator conditions. For that pair, set $w_j=\langle j|\sigma|j\rangle$. Positivity and $\Tr\sigma=1$ give $w_j\ge0$ and $\sum_jw_j=1$. The $j$th diagonal entry of $R\sigma=q_*\sigma$ gives $(q_j-q_*)w_j=0$, so $w_j=0$ for $j\notin I_*$. For any input operator $X$, the $(j,k)$ entry of $W^\dagger(I_B\otimes X)W$ is $\delta_{jk}X_{jk}$, so this matrix is $\operatorname{diag}(X_{11},\ldots,X_{dd})$. To derive the off-diagonal equations in Eq.~\eqref{eq:corr-test}, multiply Eq.~\eqref{eq:station} by $W^\dagger$ on the left and $W$ on the right. Set $B_\sigma=W^\dagger(I_B\otimes\sigma)W=\operatorname{diag}(w)$. The compressed left-hand side becomes
\begin{align*}
 &2iW^\dagger P[(I_B\otimes\sigma)DC^+-C^+D(I_B\otimes\sigma)]PW\\
 &\qquad\overset{(a)}{=}2i[B_\sigma GA^{-1}-A^{-1}G B_\sigma]
 \overset{(b)}{=}\operatorname{diag}(H_{11},\ldots,H_{dd}).
\end{align*}
In (a), $PW=W$, $DC^+=WGA^{-1}W^\dagger$, and $C^+D=WA^{-1}GW^\dagger$ by Lemma~\ref{lem:corr-compression}. Equality (b) uses Eq.~\eqref{eq:station} and the diagonal identity with $X=H$. At $(j,k)$ with $j\ne k$, the right-hand side is zero, so $w_j(GA^{-1})_{jk}=w_k(A^{-1}G)_{jk}$.

The resulting weights satisfy all the conditions in Eq.~\eqref{eq:corr-test}.

\Needspace{4\baselineskip}
\noindent\textbf{Step 3: Construct a pair $(\sigma,H)$ from the weights.}\par\nobreak

Take $w$ satisfying Eq.~\eqref{eq:corr-test}. The diagonal operator
$\sigma=\operatorname{diag}(w)$ is positive semidefinite with trace one.
Since $(q_j-q_*)w_j=0$ for every $j$, it also satisfies $R\sigma=q_*\sigma$.
Put $B_\sigma=\operatorname{diag}(w)$. The off-diagonal equations make
$2i(B_\sigma GA^{-1}-A^{-1}GB_\sigma)$ diagonal, and
$(GA^{-1})^\dagger=A^{-1}G$ makes it Hermitian. Choose $H$ to be this
diagonal matrix. With $S_\sigma$ denoting the expression
$2i[(I_B\otimes\sigma)DC^+-C^+D(I_B\otimes\sigma)]$, the compression
computed in Step~2 gives
\[
 W^\dagger S_\sigma W=H=W^\dagger(I_B\otimes H)W.
\]
Multiplying on the left by $W$ and on the right by $W^\dagger$ gives
$PS_\sigma P=P(I_B\otimes H)P$ because $P=WW^\dagger$.
The constructed pair satisfies Theorem~\ref{thm:criterion}, so the
RLD bound is attained asymptotically.
\end{proof}

Proposition~\ref{prop:corr} restricts the support of the weights
to the indices that maximize $q_j$. If the largest eigenvalue is
simple, all weight must be on one index. The following corollary
tests that single candidate in the weight equations.

\begin{corollary}[Non-saturation for a simple largest RLD eigenvalue]
\label{cor:dephasing-simple}
For a generalized dephasing channel with $A>0$ and $G\ne0$,
a simple largest eigenvalue of $R(C,D)$ implies $\JS<\JR$.
\end{corollary}
\begin{proof}
Let $j$ be the unique maximizer of $q_j$ in Eq.~\eqref{eq:corr-R}.
The only possible weights in Proposition~\ref{prop:corr} are
$w_j=1$ and $w_k=0$ for $k\ne j$. With $B=GA^{-1}$, the equations
would then give $B_{jk}=0$ for $k\ne j$. Since $G=BA$, the zero
diagonal of $G$ and $A_{jj}=1$ give $0=G_{jj}=B_{jj}$. The entire
$j$th row of $G$ would vanish, giving $q_j=0$. But
$GA^{-1}G=(A^{-1/2}G)^\dagger(A^{-1/2}G)$ is nonzero positive
semidefinite, so its largest diagonal entry is positive.
The equations therefore have no solution.
\end{proof}

\subsection{One-use and asymptotic SLD information for qubit dephasing}\label{subsec:qubit-dephasing}
We apply the weight criterion to a qubit and compare the resulting
asymptotic value with the optimal information from one use.
For a qubit correlation matrix with off-diagonal entry $a$ at the
evaluation point and $|a|<1$, the two diagonal entries of $R$ coincide.
Equal weights also satisfy the off-diagonal equations, so the RLD bound
is attained asymptotically.
The optimal SLD information from one use can nevertheless be smaller.
The formulas below recover the values of Zhou and Jiang~\cite[Sec.~III,
Eq.~(15), and Appendix~B, Eq.~(B9)]{ZJ}, who use the phase and noise
convention $a=(1-2p_{\rm noise})e^{-i\phi}$. We obtain the asymptotic
value from Proposition~\ref{prop:corr} and the one-use value by
calculating the SLD information of the output.

\begin{corollary}[Known information formulas for qubit dephasing]\label{cor:qubit}
Let $|a|<1$, and suppose that the background correlation matrix and its derivative are
\begin{equation}\label{eq:qubit-data}
 A=\begin{pmatrix}1&a\\\overline a&1\end{pmatrix},\qquad
 G=\begin{pmatrix}0&b\\\overline b&0\end{pmatrix},\qquad a,b\in\mathbb C.
\end{equation}
The optimized one-use SLD Fisher information (arbitrary ancilla allowed), the pointwise asymptotic SLD coefficient, and the channel RLD information satisfy
\begin{align}\label{eq:qubit-values}
 J_1^{\mathrm S}
 &=|b|^2+\frac{[\operatorname{Re}(\overline a b)]^2}{1-|a|^2},\\
 \JS&=\JR=\frac{|b|^2}{1-|a|^2}.
\end{align}
In particular,
\begin{equation}\label{eq:qubit-gap}
 \JS-J_1^{\mathrm S}=
 \frac{[\operatorname{Im}(\overline a b)]^2}{1-|a|^2}.
\end{equation}
Thus one use attains the RLD value if and only if $\operatorname{Im}(\overline a b)=0$, whereas asymptotic RLD saturation holds for every pair $(a,b)$ under the stated assumptions.
\end{corollary}
\Needspace{6\baselineskip}
\begin{proof}
\leavevmode\par\nobreak

\Needspace{3\baselineskip}
\noindent\textbf{Step 1: Prove that $\JS=\JR=|b|^2/(1-|a|^2)$.}\par\nobreak

The condition $|a|<1$ gives $A>0$, so Proposition~\ref{prop:corr} applies. We use
\[
 A^{-1}=\frac{1}{1-|a|^2}\begin{pmatrix}1&-a\\-\bar a&1\end{pmatrix}
\]
to calculate the two types of coefficients in Proposition~\ref{prop:corr}.
The diagonal entries of $GA^{-1}G$ are
$q_1=q_2=|b|^2/(1-|a|^2)$, so both indices belong to $I_*$.
For the off-diagonal equations, multiplication gives
$(GA^{-1})_{12}=(A^{-1}G)_{12}=b/(1-|a|^2)$, and $(GA^{-1})^\dagger=A^{-1}G$ gives the equality at $(2,1)$ by conjugation. Thus $w_1=w_2=1/2$ satisfies Eq.~\eqref{eq:corr-test}, so $\JS=\JR=q_*=|b|^2/(1-|a|^2)$.

\Needspace{3\baselineskip}
\noindent\textbf{Step 2: Reduce the one-use optimization to $J_1^{\mathrm S}=\sup_{p\in[0,1]}F^{\mathrm S}(\rho(p),\dot\rho(p))$.}\par\nobreak

To obtain the output matrices $\rho(p)$ and $\dot\rho(p)$ for this optimization, first consider a pure input to the qubit channel with an arbitrary ancillary system. Such an input can be written as
\[
 |\psi\rangle=\sqrt p\,|1\rangle|u\rangle+
 \sqrt{1-p}\,|2\rangle|v\rangle,
 \qquad 0\le p\le1,
\]
where $|u\rangle$ and $|v\rangle$ are normalized, but need not be orthogonal. The values $p$ and $1-p$ are the weights of the two system-basis components in the input; they are not generally eigenvalues of the output state. Because $|1\rangle$ and $|2\rangle$ are orthogonal, the vectors $|1\rangle|u\rangle$ and $|2\rangle|v\rangle$ are orthogonal. Write $a_1=u$, $a_2=v$, $p_1=p$, and $p_2=1-p$. Applying the channel and the identity on the ancilla to each input term gives
\[
 (\Lambda_\theta\otimes\mathrm{id})
 (|j\rangle\langle k|\otimes|a_j\rangle\langle a_k|)
 =(A_\theta)_{jk}(|j\rangle\otimes|a_j\rangle)(\langle k|\otimes\langle a_k|).
\]
Multiplying by $\sqrt{p_jp_k}$ and summing shows that the output and its derivative both act on this span. In the indicated orthonormal basis their matrices are
\[
 \rho(p)=\begin{pmatrix}
 p&\sqrt{p(1-p)}\,a\\
 \sqrt{p(1-p)}\,\overline a&1-p
 \end{pmatrix},\quad
 \dot\rho(p)=\begin{pmatrix}
 0&\sqrt{p(1-p)}\,b\\
 \sqrt{p(1-p)}\,\overline b&0
 \end{pmatrix}.
\]
It is sufficient to optimize over pure inputs: every mixed input can be purified, and tracing out the purifying ancilla cannot increase SLD Fisher information. Every $p\in[0,1]$ is realized without an ancilla by $\sqrt p\,|1\rangle+\sqrt{1-p}\,|2\rangle$, with the same pair of matrices. Thus
\[
 J_1^{\mathrm S}=\sup_{p\in[0,1]}F^{\mathrm S}(\rho(p),\dot\rho(p)).
\]

\Needspace{3\baselineskip}
\noindent\textbf{Step 3: Prove that $J_1^{\mathrm S}=|b|^2+[\operatorname{Re}(\bar a b)]^2/(1-|a|^2)$, attained at $p=1/2$.}\par\nobreak

The output and its derivative are both represented on the two-dimensional
span in Step~2 and vanish on its orthogonal complement. For $0<p<1$,
the output matrix on this span has determinant $p(1-p)(1-|a|^2)>0$. We may therefore apply the full-rank qubit SLD identity in Eq.~\eqref{eq:bloch-sld} to this output state and its derivative.
Set $k=\sqrt{p(1-p)}$. For $\rho(p)$ in Step~2, $\boldsymbol s=(2k\operatorname{Re}a,-2k\operatorname{Im}a,2p-1)$, and the derivative is $\dot{\boldsymbol s}=(2k\operatorname{Re}b,-2k\operatorname{Im}b,0)$. Hence
\begin{equation*}
1-|\boldsymbol s|^2 = 4p(1-p)(1-|a|^2),\qquad |\dot{\boldsymbol s}|^2 = 4p(1-p)|b|^2,\qquad \boldsymbol s\cdot\dot{\boldsymbol s}
   =4p(1-p)\operatorname{Re}(\bar a b).
\end{equation*}
Substituting them into Eq.~\eqref{eq:bloch-sld} gives
\[
 F^{\mathrm S}(p)=4p(1-p)\left(|b|^2+
             \frac{[\operatorname{Re}(\bar a b)]^2}{1-|a|^2}\right).
\]
At $p=0,1$ the derivative is zero. Since $4p(1-p)$ is maximized at $p=1/2$, we obtain
\[
 J_1^{\mathrm S}=\max_{p\in[0,1]}F^{\mathrm S}(p)
  =|b|^2+\frac{[\operatorname{Re}(\bar a b)]^2}{1-|a|^2}.
\]
The ancilla-free input with $p=1/2$ from Step~2 attains this value.

Subtracting this value from the asymptotic value in Step~1 gives
\[
 \JS-J_1^{\mathrm S}
 =\frac{|\bar a b|^2-[\operatorname{Re}(\bar a b)]^2}{1-|a|^2}
 =\frac{[\operatorname{Im}(\bar a b)]^2}{1-|a|^2}.
\]
The second equality decomposes $|\bar a b|^2$ into the squares of its
real and imaginary parts. This proves Eq.~\eqref{eq:qubit-gap}.
\end{proof}

For $0<|a|<1$, write the off-diagonal coefficient locally as
$a(\theta)=\eta(\theta)e^{i\phi(\theta)}$, where $\eta$ is its
magnitude and $\phi$ its phase. Substituting the evaluation-point
values $a=\eta e^{i\phi}$ and
$b=(\dot\eta+i\eta\dot\phi)e^{i\phi}$ into
Eq.~\eqref{eq:qubit-gap} gives
$\JS-J_1^{\mathrm S}=\eta^4\dot\phi^2/(1-\eta^2)$.
A parameter that locally changes only the attenuation
($\dot\phi=0$) gives one-use equality, whereas a nonzero phase
response can leave a strictly positive gap even though the RLD bound
is asymptotically attained.

\noindent\textbf{One-use and asymptotic information.}
For $|\theta|<\sqrt{3}/2$, the matrix
\begin{equation}\label{eq:qubit-example}
 A_\theta:=\begin{pmatrix}
 1&i/2+\theta\\ -i/2+\theta&1
 \end{pmatrix}
\end{equation}
defines a differentiable family of CPTP maps of the form~\eqref{eq:corr-channel}. At $\theta=0$ we have $a=i/2$ and $b=1$, so
\[
 J_1^{\mathrm S}=1<\JS=\JR=\tfrac43.
\]
Therefore $J_n^{\mathrm S}>nJ_1^{\mathrm S}$ for at least one finite $n$; otherwise the limit of $J_n^{\mathrm S}/n$ could not exceed $J_1^{\mathrm S}$. This provides an explicit instance of the gap already studied for generalized single-qubit dephasing in Zhou and Jiang~\cite[Sec.~III]{ZJ}. This example separates asymptotic RLD saturation from finite-use additivity.

This example also distinguishes the density operator in the criterion from a physical input. The certificate weights $w_1=w_2=1/2$ give $\sigma=I_2/2$ in Eq.~\eqref{eq:corr-test} and certify asymptotic saturation at $4/3$. If that same matrix is used as an input without an ancilla, Eq.~\eqref{eq:corr-channel} instead gives
\[
 \Lambda_\theta(I_2/2)=I_2/2,\qquad
 \partial_\theta\Lambda_\theta(I_2/2)=0.
\]
Its output SLD information is therefore zero, whereas the optimized one-use value is $1$.

\subsection{Optimal SLD information for any number of uses at complete dephasing}
The preceding qubit example exhibits a gain over repeating the best
one-use experiment. At complete dephasing we ask whether repeating
an optimal one-use probe is optimal for every number of uses.

At $A_0=I_d$, the output of a fixed input state $\rho$ has zero
off-diagonal entries at $\theta=0$. Their derivatives at that point
are $G_{jk}\rho_{jk}$ for $j\ne k$, by Eq.~\eqref{eq:corr-channel}.
These derivatives can carry information about the parameter.

\begin{theorem}[Optimal SLD information at complete dephasing]\label{thm:complete-dephasing}
Let $A_\theta$ be a differentiable family of $d\times d$ correlation
matrices with $A_0=I_d$, and let $G=\dot A_0=G^\dagger$.
For every integer $n\ge1$, even when ancillary systems and adaptive
processing are allowed,
\begin{equation}\label{eq:complete-dephasing-exact}
 \frac{J_n^{\mathrm S}(\Lambda;0)}{n}
 =J_1^{\mathrm S}(\Lambda;0)=\JS(\Lambda)
 =\max_{\substack{p_j\ge0\\\sum_jp_j=1}}
 4\sum_{1\le j<k\le d}|G_{jk}|^2\frac{p_jp_k}{p_j+p_k}.
\end{equation}
A summand with $p_j+p_k=0$ is defined to be zero.
An optimal one-use input can be chosen without an ancilla as
$|\psi_p\rangle=\sum_j\sqrt{p_j}|j\rangle$, where $p$ maximizes
Eq.~\eqref{eq:complete-dephasing-exact}. Repeating this probe
independently attains the optimal $n$-use output SLD information.
Using the spectral projections of the one-use SLD at the
evaluation point as a projective measurement on each output gives the same value as classical Fisher
information for the joint outcomes.
\end{theorem}

To prove this theorem, we need two ingredients. A minimax identity
first evaluates the Choi optimization, using weights $u$ on a
simplex. A repetition principle then converts equality
between the optimal one-use value and the asymptotic coefficient
into equality for every finite number of channel uses.

\begin{lemma}[Weighted coherence minimax identity]\label{lem:coherence-minimax}
Let $G=G^\dagger$ be a $d\times d$ matrix with $G_{jj}=0$, and let
$\mathcal Z=\{Z=Z^\dagger:\operatorname{diag}(Z)=0\}$ and
$\Delta_d=\{u_j\ge0:\sum_j u_j=1\}$. Then
\begin{equation}\label{eq:weighted-minimax}
 \min_{Z\in\mathcal Z}\max_j\sum_k|G_{kj}+2iZ_{kj}|^2
 =\max_{u\in\Delta_d}F_G(u),\qquad
 F_G(u):=4\sum_{j<k}|G_{jk}|^2\frac{u_ju_k}{u_j+u_k},
\end{equation}
where a term with $u_j+u_k=0$ is zero. The function $F_G$ is
continuous and concave on $\Delta_d$; both extrema are attained.
\end{lemma}
\begin{proof}
\Needspace{4\baselineskip}
\noindent\textbf{Step 1: Exchange the Choi minimization and
simplex maximization.}\par\nobreak

For $Z\in\mathcal Z$, set
$a_j(Z)=\sum_k|G_{kj}+2iZ_{kj}|^2$ and
$Q(u,Z)=\sum_j u_j a_j(Z)$. A probability-weighted average has
maximum $\max_j a_j(Z)$ at a vertex of the simplex.
We first exchange the two optimizations with an infimum over $Z$
and a supremum over $u$. Attainment of the outer extrema is
established after the exchange.
For fixed $Z$, $Q$ is continuous and affine in $u$; for fixed
$u$, it is continuous and convex in the real vector space
$\mathcal Z$. The simplex is compact and convex, and $\mathcal Z$
is convex. Compactness on the simplex side suffices for
Sion's minimax theorem~\cite[Corollary~3.5]{Sion1958}, which gives
\[
 \inf_{Z\in\mathcal Z}\sup_{u\in\Delta_d}Q(u,Z)
 =\sup_{u\in\Delta_d}\inf_{Z\in\mathcal Z}Q(u,Z).
\]

\Needspace{4\baselineskip}
\noindent\textbf{Step 2: Evaluate the infimum over each
independent pair $(j,k)$.}\par\nobreak

For $j<k$, the pair contributes
$u_j|G_{kj}+2iZ_{kj}|^2$ from column $j$ and
$u_k|G_{jk}+2iZ_{jk}|^2$ from column $k$.
Set $g=G_{jk}$ and $x=2iZ_{jk}$. Hermiticity gives
$G_{kj}=\bar g$ and $2iZ_{kj}=-\bar x$, so the sum of the two
contributions is $u_j|g-x|^2+u_k|g+x|^2$.
For $u_j+u_k>0$, completing the square yields
\[
 u_j|g-x|^2+u_k|g+x|^2
 =(u_j+u_k)\left|x-\frac{u_j-u_k}{u_j+u_k}g\right|^2
 +\frac{4u_ju_k}{u_j+u_k}|g|^2.
\]
The upper-triangular entries of $Z$ may be chosen independently,
with the lower-triangular entries fixed by Hermiticity. Thus one
matrix $Z$ makes the squared term zero for every pair with
$u_j+u_k>0$.
If both weights vanish, that pair contributes zero for any $x$.
There are no diagonal contributions. Therefore
$\inf_Z Q(u,Z)=F_G(u)$.
Each summand extends continuously to the boundary by
$4u_ju_k/(u_j+u_k)\le u_j+u_k$, so the maximum is attained.
Moreover $F_G(u)=\inf_Z Q(u,Z)$ is concave because it is a
pointwise infimum of affine functions of $u$.
Finally, the minimum over $Z$ on the left is attained. Indeed,
for every $Z\in\mathcal Z$,
\begin{equation*}
\max_j a_j(Z)\ge\frac1d\sum_{j,k}|G_{kj}+2iZ_{kj}|^2 =\frac1d\Tr[(G+2iZ)^\dagger(G+2iZ)] =\frac1d(\Tr G^2+4\Tr Z^2).
\end{equation*}
The cross term has zero trace because it is proportional to
$\Tr[G,Z]=0$. Thus the objective is coercive in the Frobenius
norm $\|Z\|_F=(\Tr Z^2)^{1/2}$. It is continuous on the closed
finite-dimensional real vector space $\mathcal Z$, so its minimum
is attained.
\end{proof}

The following lemma converts the asymptotic equality into exact
additivity for every finite number of uses, without requiring an
optimal finite-use strategy to exist.
\begin{lemma}[Finite-use repetition principle]\label{lem:repetition}
Suppose finite nonnegative numbers $a_n$ satisfy
$a_{m+n}\ge a_m+a_n$ for all positive integers $m,n$, and the finite
limit $a_\infty=\lim_{n\to\infty}a_n/n$ exists.
If $a_1=a_\infty$, then $a_n=na_1$ for every $n\ge1$.
\end{lemma}
\begin{proof}
Superadditivity gives $a_n\ge na_1$ by repeated addition.
For any fixed $n$ and each positive $k$, repetition gives
$a_{kn}\ge k a_n$; division by $kn$ and $k\to\infty$ yields
$a_n/n\le a_\infty=a_1$. Combining the inequalities proves the claim.
\end{proof}

The minimax identity now supplies the asymptotic value. To use the
repetition principle, it remains to identify the same value in the
one-use probe optimization and verify superadditivity.

\begin{proof}[Proof of Theorem~\ref{thm:complete-dephasing}]
We compute the asymptotic coefficient by optimizing over the minimax
weight $u$, and the one-use information by optimizing over the
populations $p$ of an input probe. If both optimizations maximize
the same function on $\Delta_d$, their values agree, and
Lemma~\ref{lem:repetition} gives the value for each use count.

\Needspace{4\baselineskip}
\noindent\textbf{Step 1: Prove
$\JS=\max_{u\in\Delta_d}F_G(u)$.}\par\nobreak

For $A_0=I_d$, Lemma~\ref{lem:corr-compression} gives
$C=C^+=P=WW^\dagger$, $D=WGW^\dagger$, and the feasible directions
$Y=WZW^\dagger$ with $Z\in\mathcal Z$.
By Eq.~\eqref{eq:corr-compression}, the Choi objective is diagonal,
with $j$th entry the squared norm of column $j$ of $G+2iZ$.
Theorem~\ref{thm:variational} consequently yields
\begin{equation}\label{eq:complete-dephasing-minmax}
 \JS=\min_{Z\in\mathcal Z}\max_j\sum_k|G_{kj}+2iZ_{kj}|^2.
\end{equation}
Lemma~\ref{lem:coherence-minimax} evaluates this minimum as
\[
 \JS=\max_{u\in\Delta_d}
 4\sum_{j<k}|G_{jk}|^2\frac{u_ju_k}{u_j+u_k}.
\]

\Needspace{4\baselineskip}
\noindent\textbf{Step 2: Prove
$J_1^{\mathrm S}=\max_{p\in\Delta_d}F_G(p)=\JS$
and exhibit an ancilla-free optimal probe.}\par\nobreak

A pure input with an arbitrary ancilla is
$|\Psi\rangle=\sum_j\sqrt{p_j}|j\rangle|a_j\rangle$,
where $p\in\Delta_d$ and the ancilla vectors $|a_j\rangle$
are normalized. The vectors $|j\rangle|a_j\rangle$ are orthonormal
because of the first factor. Applying Eq.~\eqref{eq:corr-channel}
to the system factor of each matrix unit gives
\[
 (\Lambda_\theta\otimes\mathrm{id})(|\Psi\rangle\langle\Psi|)
 =\sum_{j,k}\sqrt{p_jp_k}(A_\theta)_{jk}
 |j\rangle|a_j\rangle\langle k|\langle a_k|.
\]
At zero, $A_0=I_d$ gives $\rho_{jk}=\delta_{jk}p_j$ in this
orthonormal family. Differentiation gives
$\dot\rho_{jk}=\sqrt{p_jp_k}G_{jk}$. Both operators vanish on the
orthogonal complement. Equation~\eqref{eq:SLD-spectral}
gives
\begin{equation*}
F^{\mathrm S}(\rho,\dot\rho)
 =2\sum_{\substack{j\ne k\\p_j+p_k>0}}
     \frac{p_jp_k|G_{jk}|^2}{p_j+p_k} =4\sum_{\substack{j<k\\p_j+p_k>0}}|G_{jk}|^2\frac{p_jp_k}{p_j+p_k}
 =F_G(p).
\end{equation*}
For a mixed input on the channel and a memory, purify it using an
additional ancilla and trace out that purifying system after the
channel. Monotonicity of SLD information under partial trace bounds
the mixed-input information by this pure-input expression.
Conversely every $p\in\Delta_d$ is realized without an ancilla by
$|\psi_p\rangle=\sum_j\sqrt{p_j}|j\rangle$.
Thus $J_1^{\mathrm S}=\max_{p\in\Delta_d}F_G(p)$.
The variable $p$ ranges over exactly the same simplex as $u$, and
$F_G(p)$ has exactly the same formula as $F_G(u)$ in Step~1.
Therefore $J_1^{\mathrm S}=\JS$. Choose a maximizer $p_*$;
its ancilla-free probe attains that value.

\Needspace{4\baselineskip}
\noindent\textbf{Step 3: Prove
$J_n^{\mathrm S}=nJ_1^{\mathrm S}$ and attainment by
repeated local SLD measurements.}\par\nobreak

Independent execution of an $m$-use protocol and an $n$-use protocol
adds their output SLD information. Choose each protocol within
$\varepsilon$ of its supremum. Their concatenation gives
$J_{m+n}^{\mathrm S}\ge J_m^{\mathrm S}+J_n^{\mathrm S}-2\varepsilon$,
so letting $\varepsilon\downarrow0$ gives superadditivity.
The limit of $J_n^{\mathrm S}/n$ exists by
Theorem~\ref{thm:variational}, and Step~2 proves its equality with
$J_1^{\mathrm S}$. Lemma~\ref{lem:repetition} therefore gives
$J_n^{\mathrm S}=nJ_1^{\mathrm S}$ for every $n$.

To realize the one-use SLD value as classical Fisher information, take the optimal one-use
output and its derivative $\rho,\dot\rho$ from Step~2,
choose an SLD $L$, and let $P_\ell$ be its eigenprojection with
eigenvalue $\ell$. Fix this measurement at the evaluation point,
with $q_\ell=\Tr(\rho P_\ell)$ and
$\dot q_\ell=\Tr(\dot\rho P_\ell)$.
The SLD equation and $LP_\ell=\ell P_\ell$ give
$\dot q_\ell=\ell q_\ell$. Hence its classical Fisher information is
\[
 \sum_{\ell:q_\ell>0}\frac{\dot q_\ell^2}{q_\ell}
 =\sum_\ell\ell^2q_\ell=\Tr(\rho L^2)=J_1^{\mathrm S}.
\]
Zero-probability terms have zero derivative at the evaluation point.
With independent repetitions of the probe and the same measurement
on each output, the joint outcomes have classical Fisher information
$nJ_1^{\mathrm S}$ by additivity for product distributions.
\end{proof}

\subsection{A graph criterion for positive-definite correlation matrices}
The preceding result gives the optimal SLD information at complete
dephasing. We now ask when the RLD bound is attained, allowing an
arbitrary positive-definite correlation matrix $A=A_0$. We use
a directed graph to determine when the weight equations have a solution.

Even with $A$ fixed, the local estimation problem depends on how the
channel varies. For a fixed input state $\rho$, the derivative of the
off-diagonal output entry $(j,k)$ is $G_{jk}\rho_{jk}$ by
Eq.~\eqref{eq:corr-channel}. Different Hermitian, zero-diagonal
matrices $G$ thus describe different local channel families through
the same evaluation-point channel. Each such $G$ is realized by
$A_\theta=A+\theta G$ on a sufficiently small interval about zero:
the unit diagonal is preserved, and positive definiteness follows
for $|\theta|\|G\|<\lambda_{\min}(A)$.

To test saturation, write $B=GA^{-1}$. The equations in
Proposition~\ref{prop:corr} require $w_jB_{jk}=w_k\overline{B_{kj}}$
for $j\ne k$. In particular, if $w_j>0$ and $B_{jk}\ne0$,
then $w_k>0$ and $B_{kj}\ne0$. This implication motivates a
directed graph on the basis indices, with an edge $j\to k$ when
$j\ne k$ and $B_{jk}\ne0$. A strongly connected component is a
maximal set of vertices mutually reachable by directed paths.

\begin{proposition}[Graph criterion for RLD saturation]\label{prop:dephasing-graph}
Suppose $A>0$, and let $I_*$ be the maximizing index set in
Eq.~\eqref{eq:corr-R}. The RLD bound is attained asymptotically
if and only if the directed graph has a strongly connected
component $S$ with the following properties.
\begin{enumerate}
\item $S\subseteq I_*$ and no edge goes from $S$ to its complement.
\item Each edge $j\to k$ in $S$ has a reverse edge, and
      $r_{jk}:=\overline{B_{kj}}/B_{jk}$ is a positive real number.
\item The product of the ratios $r_{jk}$ along each directed
      cycle in $S$ equals one.
\end{enumerate}
For a one-vertex component the internal edge and cycle conditions are
automatic. Such a vertex must still lie in $I_*$ and have no
outgoing edge.
\end{proposition}

\begin{proof}
Suppose first that the RLD bound is attained and take weights from
Proposition~\ref{prop:corr}. Their positive support $T$ is nonempty
and contained in $I_*$. If $j\in T$ and $k\notin T$, the equation
$w_jB_{jk}=0$ rules out the edge $j\to k$.

Within $T$, both weights are positive. If $B_{jk}\ne0$, the
equation forces $B_{kj}\ne0$ and gives $r_{jk}=w_j/w_k>0$.
The ratios therefore telescope around every directed cycle in $T$.

An edge between two strongly connected components of the graph
restricted to $T$ would have a reverse edge and join those
components. Thus any such component $S$ has no edge to
$T\setminus S$, and the first paragraph rules out edges to the
complement of $T$. Consequently $S$ is also a strongly connected
component of the original graph, with no outgoing edges and with
$S\subseteq I_*$.

Conversely, take a component $S$ satisfying the three conditions.
Fix a vertex in $S$ and give it weight one. Along a directed path
from that vertex, assign the next weight by $w_k=w_j/r_{jk}$.
All these weights are positive. The reverse edge has reciprocal ratio,
since $r_{jk}$ is positive real. A closed walk decomposes into
directed cycles and therefore has ratio product one. Following
one path and returning along the reverse of another shows that
the assigned weight is independent of the path. It follows that
$w_jB_{jk}=w_k\overline{B_{kj}}$ on every edge in $S$, and both
coefficients vanish for pairs in $S$ with no edge. Set all weights
outside $S$ to zero. For $j\in S$ and $k\notin S$, $B_{jk}=0$
by the first condition, so the equation holds even if $B_{kj}\ne0$.
Its conjugate gives the equation in the reverse order, and pairs
outside $S$ have both weights zero. Normalizing the weights gives
the solution required by Proposition~\ref{prop:corr}.
\end{proof}

For $G=0$, all vertices are isolated and belong to $I_*$; both
information quantities are zero. For general $G$, the conditions
describe how saturation can fail even if $I_*$ contains several
vertices: positive weights must be consistent across all the edges
of a component that stays within $I_*$.

\paragraph{Complete dephasing.}
At the evaluation point studied in the finite-use result, the
general graph criterion reduces to a condition on weighted degrees.
When $A=I_d$, we have $B=G=G^\dagger$. All edges come in reverse
pairs and their ratios are one. We may therefore use the undirected
graph with an edge $\{j,k\}$ whenever $G_{jk}\ne0$. Giving this
edge weight $|G_{jk}|^2$ makes the weighted degree of vertex $j$
equal to $q_j=\sum_k|G_{jk}|^2$. The same weights occur in the
optimal SLD formula in Eq.~\eqref{eq:complete-dephasing-exact}.

\begin{corollary}[RLD saturation from the weighted degrees]\label{cor:weighted}
If $A=I_d$, the criterion in Eq.~\eqref{eq:corr-test}, including $w_j\ge0$ and $\sum_jw_j=1$, reduces to
\[
 (w_j-w_k)G_{jk}=0\quad(j\ne k),\qquad
 w_j=0\quad\text{if }\sum_k|G_{jk}|^2<\max_\ell\sum_k|G_{\ell k}|^2.
\]
Consequently, $\JS=\JR$ holds if and only if the graph with edges $\{j,k\}$ satisfying $G_{jk}\ne0$ has a connected component in which every vertex has weighted degree equal to the global maximum, with weights $|G_{jk}|^2$.
\end{corollary}
\begin{proof}
In the undirected description, the strongly connected components
of the original directed graph are precisely the connected components.
Each has no edges to its complement, and the ratio conditions are
automatic. Proposition~\ref{prop:dephasing-graph} therefore reduces
to the existence of a connected component contained in $I_*$.
\end{proof}

Consider two graphs on three vertices with weight one on every edge.
For the path, take $|G_{12}|=|G_{23}|=1$ and $G_{13}=0$.
Its weighted degrees are $(1,2,1)$, so its single connected
component is not contained in $I_*$. For the triangle, take
$|G_{jk}|=1$ for every $j\ne k$. Every vertex has weighted degree
two, so the RLD bound is attained. These two arrangements give
different saturation outcomes even though every present edge
has the same weight in both.

The path and triangle illustrate when the bound is attained. To
quantify the gap, consider a star, whose symmetry reduces the probe
optimization to the population of its central vertex.

\begin{corollary}[Star-graph formula and dimension-dependent RLD gap]\label{cor:star}
In Theorem~\ref{thm:complete-dephasing}, let $d=m+1$ with $m\ge1$.
Define a Hermitian $(m+1)\times(m+1)$ matrix $G$ by setting $G_{0j}=G_{j0}=1$ for $1\le j\le m$ and all other entries to zero. Its nonzero entries connect the central index $0$ to each remaining index, forming a star-shaped pattern. The resulting information values are
\begin{equation}\label{eq:star-values}
 J_n^{\mathrm S}(\Lambda;0)
 =n\frac{4m}{(1+\sqrt m)^2},\qquad
 \JR=m,\qquad
 \frac{\JR}{\JS}=\frac{(1+\sqrt m)^2}{4}.
\end{equation}
An optimal one-use input is
\begin{equation}\label{eq:star-input}
 |\psi_*\rangle:=\sqrt{p_0}|0\rangle+
   \sum_{j=1}^m\sqrt{p_j}|j\rangle,\qquad
 p_0:=\frac1{1+\sqrt m},\qquad
 p_j:=\frac1{\sqrt m(1+\sqrt m)}.
\end{equation}
In particular, the RLD-to-optimal-SLD ratio diverges as the
number of leaves grows, although each fixed-dimensional channel
family is differentiable and has finite RLD information.
\end{corollary}
\begin{proof}
The values depend only on $A_0$ and $G$ by Theorem~\ref{thm:complete-dephasing} and Eq.~\eqref{eq:corr-R}. To see that these data occur in a differentiable family, take $A_\theta=I_{m+1}+\theta G$. Let $|s\rangle=m^{-1/2}\sum_{j=1}^m|j\rangle$. We have $G|0\rangle=\sqrt m\,|s\rangle$ and $G|s\rangle=\sqrt m\,|0\rangle$, and $G$ vanishes on the orthogonal complement of their span. Thus its nonzero eigenvalues are $\pm\sqrt m$, and $A_\theta>0$ for $|\theta|<1/\sqrt m$. Its diagonal is one and $A_0=I_{m+1}$, so Theorem~\ref{thm:complete-dephasing} applies.

Let $F_G(p)$ denote the objective in Eq.~\eqref{eq:complete-dephasing-exact} at a given probability vector $p$. Lemma~\ref{lem:coherence-minimax} gives $F_G(p)=\inf_{Z\in\mathcal Z}Q(p,Z)$ and proves that this objective is concave as an infimum of affine functions of $p$. We first use symmetry and concavity to reduce the maximization to
one variable, the central population. Let $S_m$ be the permutation
group of the leaves, acting on $p$ while leaving $p_0$ fixed. The star has the same weights after any such permutation $\pi$, so $F_G(\pi p)=F_G(p)$. For $\bar p=(m!)^{-1}\sum_{\pi\in S_m}\pi p$, concavity gives
\[
 F_G(\bar p)\ge(m!)^{-1}\sum_{\pi\in S_m} F_G(\pi p)=F_G(p).
\]
The leaf entries of $\bar p$ are equal, so a maximizer can be chosen with equal leaf populations. Write $p(t)$ for these populations, with $p_0(t)=t$ and $p_j(t)=(1-t)/m$ for $1\le j\le m$. Summing the $m$ equal edge contributions in Eq.~\eqref{eq:complete-dephasing-exact} gives
\[
 F_G(p(t))=4m\frac{t(1-t)/m}{t+(1-t)/m}
     =\frac{4mt(1-t)}{1+(m-1)t},\qquad 0\le t\le1.
\]
The derivative of $F_G(p(t))$ with respect to $t$ is $4m[1-2t-(m-1)t^2]/[1+(m-1)t]^2$. Its numerator changes sign from positive to negative at $t=(1+\sqrt m)^{-1}$, including $m=1$. Substituting the maximizing $t$ gives $F_G(p(t))=4m/(1+\sqrt m)^2$ and the populations in Eq.~\eqref{eq:star-input}. Theorem~\ref{thm:complete-dephasing} gives $J_n^{\mathrm S}=nF_G(p(t))$ at this maximizer. The weighted degrees are $m$ at the center and $1$ at each leaf, so $\JR=m$. Dividing by the maximum of $F_G$ gives the ratio in Eq.~\eqref{eq:star-values}.

\end{proof}
The three-vertex path above is the case $m=2$: its exact
optimal coefficient is $8/(1+\sqrt2)^2\simeq1.373$ while its RLD
bound is $2$. The ratio in Eq.~\eqref{eq:star-values} has a
different origin from the fixed-dimensional divergence in
Corollary~\ref{cor:full-support-unbounded}: here the dimension grows
and the background Choi matrix remains rank deficient.

\subsection{The effect of phase on RLD saturation in a qutrit channel}
At complete dephasing, the magnitudes $|G_{jk}|$ suffice to decide
whether the RLD bound is attained. Do they still suffice when the
correlation matrix at the evaluation point has nonzero off-diagonal
entries? The general criterion depends on $GA^{-1}$, so we examine
what happens when the phases in $G$ vary with $A$ fixed.

We compare qutrit families with the same real background correlation
$r_0$ on each off-diagonal entry and with all three off-diagonal
magnitudes of $G$ equal to one. The parameter $\theta$ is the
quantity to be estimated; $\Phi$ specifies the family used in the
comparison. At nonzero $r_0$, varying $\Phi$ changes whether the RLD
bound is attained.

\begin{corollary}[A correlated-qutrit phase family]\label{cor:phase}
Let $\Phi\in\mathbb R$ and
\[
 A_{r_0}=(1-r_0)I_3+r_0\mathbf 1\mathbf 1^\top,\quad
 \mathbf 1=(1,1,1)^\top,\quad -\tfrac12<r_0<1,
 \qquad
 G_\Phi=\begin{pmatrix}0&1&e^{-i\Phi}\\1&0&1\\e^{i\Phi}&1&0\end{pmatrix}.
\]
For the family defined by $A_\theta:=A_{r_0}+\theta G_\Phi$ near $\theta=0$, RLD saturation holds precisely when
\begin{equation}\label{eq:phase-iff}
 r_0\sin(\Phi/2)=0.
\end{equation}
Equivalently, $r_0=0$ or $\Phi\equiv0\pmod{2\pi}$.
\end{corollary}
\Needspace{6\baselineskip}
\begin{proof}
\leavevmode\par\nobreak

\Needspace{3\baselineskip}
\noindent\textbf{Step 1: Prove that $q_1=q_3$ and $q_1-q_2=2r_0(1-\cos\Phi)/[(1-r_0)(1+2r_0)]$.}\par\nobreak

We compute the $q_j$ to find $I_*$, the indices on which the
weights in Proposition~\ref{prop:corr} may be positive.
The eigenvalue of $A_{r_0}$ on $\mathbf1$ is $1+2r_0$, and its eigenvalue on $\mathbf1^\perp$ is $1-r_0$. Both are positive in the given range. Thus $A_\theta$ remains positive definite for small $|\theta|$ and has unit diagonal, so Proposition~\ref{prop:corr} applies. The projection $P_1=\mathbf1\mathbf1^\top/3$ is onto $\operatorname{span}\{\mathbf1\}$. Inverting the two eigenvalues gives
\[
 A_{r_0}^{-1}=(1-r_0)^{-1}(I-P_1)+(1+2r_0)^{-1}P_1
 =uI+v\mathbf1\mathbf1^\top,\qquad
 u=\frac1{1-r_0},\qquad v=-\frac{r_0}{(1-r_0)(1+2r_0)}.
\]
Equation~\eqref{eq:corr-R} now gives
\[
 q_j=(G_\Phi A_{r_0}^{-1}G_\Phi)_{jj}
     =u(G_\Phi^2)_{jj}+v|(G_\Phi\mathbf1)_j|^2
     =2u+v|(G_\Phi\mathbf1)_j|^2.
\]
The last equality uses $(G_\Phi^2)_{jj}=\sum_k|(G_\Phi)_{jk}|^2=2$, since each row has two off-diagonal entries of modulus one. The row sums of $G_\Phi$ are $1+e^{-i\Phi}$, $2$, and $1+e^{i\Phi}$. Hence
\[
 q_1=q_3=\frac2{1-r_0}-\frac{2r_0(1+\cos\Phi)}{(1-r_0)(1+2r_0)},\quad
 q_2=\frac2{(1-r_0)(1+2r_0)},\quad
 q_1-q_2=\frac{2r_0(1-\cos\Phi)}{(1-r_0)(1+2r_0)}.
\]
If $r_0\ne0$ and $\Phi\not\equiv0\pmod{2\pi}$, the difference has
the sign of $r_0$. Thus $I_*=\{1,3\}$ for $r_0>0$ and
$I_*=\{2\}$ for $r_0<0$. If $r_0=0$ or
$\Phi\equiv0\pmod{2\pi}$, all three indices belong to $I_*$.

\Needspace{3\baselineskip}
\noindent\textbf{Step 2: Prove that $r_0\ne0$ and $\Phi\not\equiv0\pmod{2\pi}$ imply $\JS<\JR$.}\par\nobreak

Assume $\Phi\not\equiv0$ modulo $2\pi$, so $1-\cos\Phi>0$. The sign of $q_1-q_2$ in Step~1 is then the sign of $r_0$.

In both sign cases, the phase changes the maximizing set from all
three vertices to a proper subset. We show that every maximizing
vertex has an edge to a vertex outside that subset, ruling out
the component required by Proposition~\ref{prop:dephasing-graph}. If $r_0>0$, the maximizing set is $I_*=\{1,3\}$, so $w_2=0$. In the $(1,2)$ equation of Eq.~\eqref{eq:corr-test}, the right-hand side is zero, while the coefficient of $w_1$ is
\[
 (G_\Phi A_{r_0}^{-1})_{12}=u+v(1+e^{-i\Phi})\ne0.
\]
Indeed $|v(1+e^{-i\Phi})|\le2|v|$ and $2|v|/u=2r_0/(1+2r_0)<1$. Hence $w_1=0$. Similarly, the $(3,2)$ equation has coefficient $u+v(1+e^{i\Phi})\ne0$, and gives $w_3=0$. All three probabilities would vanish, contrary to normalization.

If $r_0<0$, then $I_*=\{2\}$, so $w_2=1$ and $w_1=w_3=0$. The $(2,1)$ equation requires $(G_\Phi A_{r_0}^{-1})_{21}=0$, since its left-hand weight is one and its right-hand weight is zero. The coefficient instead satisfies
\[
 (G_\Phi A_{r_0}^{-1})_{21}=u+2v
 =\frac1{(1-r_0)(1+2r_0)}>0
\]
because $-1/2<r_0<0$. In both cases Eq.~\eqref{eq:corr-test} has no solution, so Proposition~\ref{prop:corr} and the RLD upper bound give $\JS<\JR$.

\Needspace{3\baselineskip}
\noindent\textbf{Step 3: Prove that $r_0=0$ or $\Phi\equiv0\pmod{2\pi}$ implies $\JS=\JR$.}\par\nobreak

Step~1 gives $q_1=q_2=q_3$ in both cases, so all indices are allowed in Eq.~\eqref{eq:corr-test}. If $r_0=0$, then $A_{r_0}=I_3$. If $\Phi\equiv0\pmod{2\pi}$, then $A_{r_0}$ commutes with $G_\Phi=\mathbf1\mathbf1^\top-I_3$. In either case $G_\Phi A_{r_0}^{-1}=A_{r_0}^{-1}G_\Phi$, so $w_j=1/3$ solves the off-diagonal equations. Proposition~\ref{prop:corr} gives $\JS=\JR$. Together with Step~2, this proves Eq.~\eqref{eq:phase-iff}.
\end{proof}

For the matrices in the corollary, $G_{12}G_{23}G_{31}=e^{i\Phi}$.
We call the argument of this product the cycle phase of $G$.
For a diagonal unitary $U=\operatorname{diag}(e^{i\gamma_j})$,
composing the channel with $U$ on its output gives, in the fixed basis,
\[
 U(A_\theta\circ X)U^\dagger=(UA_\theta U^\dagger)\circ X.
\]
The resulting correlation matrix and its derivative at zero are
$(UAU^\dagger,UGU^\dagger)$.
The three phase factors in the product of the transformed entries
of $G$ cancel. In our comparison $A=A_{r_0}$ is fixed and $G_\Phi$
varies with $\Phi$. The phase condition in Eq.~\eqref{eq:phase-iff}
applies to this one-phase family. The graph criterion uses products of
edge ratios formed from $GA^{-1}$.

\begin{remark}[Environmental realization and phase meaning]\label{rem:qutrit-env}
We show how coupling the input to a finite-dimensional environment
and tracing out that environment yields the same channel
$\Lambda_\theta(X)=A_\theta\circ X$.
For sufficiently small $|\theta|$, set $A_\theta=A_{r_0}+\theta G_\Phi>0$,
choose an environment $E\cong\mathbb C^3$, and define
\[
 |e_j(\theta)\rangle=(A_\theta^{\mathsf T})^{1/2}|j\rangle_E,
 \qquad
 V_\theta|j\rangle_A=|j\rangle_B\otimes|e_j(\theta)\rangle_E.
\]
The environment vectors have inner products $\langle e_k(\theta)|e_j(\theta)\rangle=(A_\theta)_{jk}$. The output basis is orthonormal and the environment vectors have norm one because $(A_\theta)_{jj}=1$. Thus
\[
 \langle V_\theta j|V_\theta k\rangle
 =\delta_{jk}\langle e_j(\theta)|e_k(\theta)\rangle=\delta_{jk},
\]
so $V_\theta$ is an isometry. Its action on a matrix unit gives
\[
 \Tr_E[V_\theta|j\rangle\langle k|V_\theta^\dagger]
 =\langle e_k(\theta)|e_j(\theta)\rangle |j\rangle\langle k|
 =(A_\theta)_{jk}|j\rangle\langle k|.
\]
By linearity this gives the generalized dephasing channel $\Lambda_\theta(X)=A_\theta\circ X$. Locally the isometry can also be extended to a unitary interaction with a fixed pure environment.
This construction realizes the first-order response $G_\Phi$ of Corollary~\ref{cor:phase} with a finite-dimensional environment. The condition in Eq.~\eqref{eq:phase-iff} identifies when this phase pattern, together with nonzero background correlations, obstructs RLD saturation.
\end{remark}

\section{Maximally entangled inputs}\label{sec:prescribed-probes}\label{sec:entangled}\label{sec:examples}
In Section~\ref{sec:pauli}, a maximally entangled input attains the
channel RLD value in one use. We now ask when it does so for a general
channel.
Set $d_A=\dim\mathcal H_A$ and
$|\Phi\rangle=d_A^{-1/2}\sum_j|j\rangle_A|j\rangle_{A'}$,
where $A'$ is a copy of $A$. Expanding the input projector and applying
the channel to its first factor gives
\[
 (\Lambda_\theta\otimes\mathrm{id}_{A'})(|\Phi\rangle\langle\Phi|)
 =\frac{1}{d_A}\sum_{j,k}\Lambda_\theta(|j\rangle\langle k|)
   \otimes |j\rangle_{A'}\langle k|
 =\frac{C_\theta}{d_A}.
\]
The last equality identifies $A'$ with the input copy in
Eq.~\eqref{eq:choi}. The output and its derivative at zero are
$C/d_A$ and $D/d_A$.

\begin{proposition}[When the maximally entangled input reaches RLD]\label{prop:maxent}
Suppose $D=PDP$ and let $R,r$ be as in Eq.~\eqref{eq:R}. The maximally entangled input attains the channel RLD value at one use if and only if
\begin{equation}\label{eq:ME}
 [C,D]=0,\qquad R=rI_A.
\end{equation}
This is sufficient, but not necessary, for channel-level asymptotic RLD saturation.
\end{proposition}
\begin{proof}
We first compare the SLD and RLD information of the normalized Choi
state. We then compare the average eigenvalue of $R$ with its largest
eigenvalue. Set $\rho=C/d_A$ and $\dot\rho=D/d_A$. Since $D=PDP$, the derivative is supported on $\operatorname{Ran}P$, which is also the support of $\rho$. On that support $\rho$ is positive definite, and its Moore--Penrose inverse on the full space is $\rho^+=d_A C^+$. The state RLD term in Eq.~\eqref{eq:state-gap} is therefore
\[
 \Tr(\dot\rho\,\rho^+\dot\rho)
 =\Tr[(D/d_A)(d_A C^+)(D/d_A)]
 =\Tr(DC^+D)/d_A .
\]
\par\Needspace{8\baselineskip}
\[
 F^{\mathrm S}(C/d_A,D/d_A)
 \overset{(a)}{\le}\Tr(DC^+D)/d_A
 =\Tr R/d_A
 \overset{(b)}{\le}r .
\]
Inequality (a) follows from Lemma~\ref{lem:state-gap}, applied to $\rho,\dot\rho$ on $\operatorname{Ran}P$. Equality in (a) holds exactly when the restricted state and derivative commute. Both operators vanish on the orthogonal complement, so this is equivalent to $[C,D]=0$ on the whole Choi space. For (b), the average eigenvalue of $R$ equals its largest eigenvalue precisely when every eigenvalue is $r$, that is, when $R=rI_A$. The two endpoints of the chain agree exactly when both conditions in Eq.~\eqref{eq:ME} hold. In that case independent repetitions of $|\Phi\rangle$ give $\JS\ge r$, and the RLD upper bound gives $\JS=r$.
\end{proof}
The generalized Pauli example in Section~\ref{sec:pauli} satisfies both conditions in Eq.~\eqref{eq:ME}: its Choi data commute and its RLD matrix is scalar. In Section~\ref{sec:noncommuting-saturation}, the input $|1\rangle$
attains the RLD value in one use. In that example $[C,D]\ne0$, so
Eq.~\eqref{eq:ME} rules out attainment by the maximally entangled input.
This gives an asymptotically saturating channel for which the maximally
entangled input fails the one-use criterion.

\section{Proofs and explicit criteria for RLD saturation}\label{sec:positive-proof}
\subsection{Auxiliary lemmas for the general saturation criterion}
This section proves the general and positive-definite saturation
criteria stated above. We then derive a test using directional
derivatives and give explicit forms for small maximal eigenspaces.
We begin with the lemmas needed for the general criterion.

The general criterion uses three facts: separation on the largest
eigenspace, the right derivative of the largest eigenvalue, and
the orthogonal complement of the supported zero-partial-trace
directions.

The first auxiliary fact converts the absence of a positive-definite
matrix in a real linear space of Hermitian matrices into the existence
of one density operator whose trace pairing vanishes on the whole
space. This common density operator is needed in the necessity part
of the general saturation criterion.

\begin{lemma}[Semidefinite separation alternative]\label{lem:separation}
Let $E$ be a finite-dimensional nonzero Hilbert space and let
$\mathcal S$ be a real linear subspace of $\Herm(E)$. Exactly one of the
following is possible:
\begin{enumerate}[label=\textup{(\alph*)}]
\item There exists $G\in\mathcal S$ with $G>0$ on $E$.
\item There exists a density operator $\sigma\ge0$ on $E$ such that
$\Tr\sigma=1$ and $\Tr(\sigma G)=0$ for every $G\in\mathcal S$.
\end{enumerate}
\end{lemma}
\begin{proof}
\noindent\textbf{Step 1: Show that (b) excludes (a).}\par\nobreak

Suppose that a density operator $\sigma$ satisfies (b), and let
$G\in\mathcal S$ be positive definite. The smallest-eigenvalue bound
gives
\[
 \Tr(\sigma G)\ge\lambda_{\min}(G)\Tr\sigma
 =\lambda_{\min}(G)>0.
\]
This contradicts $\Tr(\sigma G)=0$ for every $G\in\mathcal S$.
Thus (a) and (b) cannot both hold.

\Needspace{4\baselineskip}
\noindent\textbf{Step 2: If (b) fails, construct a positive-definite
operator in $\mathcal S$.}\par\nobreak

For each density operator $\sigma$ on $E$, define a real linear
functional $\ell_\sigma\in\mathcal S^*$ by
\[
 \ell_\sigma(G)=\Tr(\sigma G),\qquad G\in\mathcal S.
\]
The set of density operators on $E$ is compact and convex. The map
$\sigma\mapsto\ell_\sigma$ is continuous and real affine, so its
image
\[
 \mathcal K
 =\{\ell_\sigma:\sigma\text{ is a density operator on }E\}
 \subset\mathcal S^*
\]
is also compact and convex. Failure of (b) means that no density
operator annihilates every element of $\mathcal S$. Equivalently,
the zero functional does not belong to $\mathcal K$.

Strict separation of the compact convex set $\mathcal K$ from the
origin gives a real linear functional $\Phi$ on $\mathcal S^*$
such that
\[
 \Phi(\ell_\sigma)>0
 \qquad\text{for every density operator }\sigma\text{ on }E.
\]
Because $\mathcal S$ is finite-dimensional, its canonical
identification with its double dual is onto. Hence there is a
$G\in\mathcal S$ such that
$\Phi(\ell)=\ell(G)$ for every $\ell\in\mathcal S^*$. Therefore
\[
 \Tr(\sigma G)>0
 \qquad\text{for every density operator }\sigma\text{ on }E.
\]
The minimum is attained because the set of density operators is
compact. The variational characterization of the smallest
eigenvalue then gives
\[
 \lambda_{\min}(G)
 =\min_{\sigma\text{ density on }E}\Tr(\sigma G)>0.
\]
Thus $G>0$, proving (a) when (b) fails. Together with Step~1,
this proves that exactly one of (a) and (b) holds.
\end{proof}

The next auxiliary fact determines the right directional derivative
of the largest eigenvalue at a matrix with a possibly degenerate
largest eigenvalue. We need this derivative to decide whether the
zero direction minimizes the Choi variational objective. The same
calculation also identifies a feasible direction along which the
objective becomes strictly smaller whenever the compressed linear
term has a positive-definite value.

\begin{lemma}[Quadratic spectral optimality certificate]\label{lem:spectral-certificate}
Let $V$ be a finite-dimensional real vector space, $R=R^\dagger$ an
operator on a nonzero finite-dimensional Hilbert space $K$,
$L:V\to\Herm(K)$ real linear, and $Q:V\to\Herm(K)$ satisfy
$Q(Y)\succeq0$ and $Q(tY)=t^2Q(Y)$ for real $t$.
Write $r=\lambda_{\max}(R)$,
$E=\ker(rI-R)$, and $\Pi$ for the projection onto $E$, and put
$G(Y)=\Pi L(Y)|_E$. For $M(Y)=R+L(Y)+Q(Y)$, the following are equivalent:
\begin{enumerate}[label=\textup{(\roman*)}]
\item $\lambda_{\max}M(Y)\ge r$ for all $Y\in V$;
\item a density operator $\sigma_E$ on $E$ exists such that
$\Tr_E[\sigma_EG(Y)]=0$ for \emph{every} $Y\in V$;
\item the real linear space $G(V)$ contains no positive-definite
operator on $E$.
\end{enumerate}
For every $Y\in V$, the right derivative satisfies
\begin{equation}\label{eq:positive-directional}
 \left.\frac{d}{dt}\right|_{t=0+}
 \lambda_{\max}(R+tL(Y)+t^2Q(Y))
 =\lambda_{\max}(G(Y)).
\end{equation}
If $G(Y_0)>0$, then $\lambda_{\max}M(-tY_0)<r$ for all sufficiently
small $t>0$.
\end{lemma}
\begin{proof}
We first compute the right derivative to show why a positive-definite
compression yields strict decrease in the opposite direction.
We then use separation to relate global optimality to the existence
of a common density operator.

\noindent\textbf{Step 1: Derive the right derivative in
Eq.~\eqref{eq:positive-directional}.}\par\nobreak

Fix $Y\in V$, and write $B=L(Y)$ and $T=Q(Y)$. Define
\[
 f(t):=\lambda_{\max}(R+tB+t^2T),\qquad f(0)=r.
\]
Choose a unit vector $u\in E$ that maximizes
$\langle u|B|u\rangle$ over the unit sphere of $E$. For each
$t>0$, choose a unit eigenvector $v_t$ corresponding to the largest
eigenvalue of $R+tB+t^2T$. Thus
\[
 f(t)=\langle v_t|(R+tB+t^2T)|v_t\rangle.
\]
The variational characterization of the largest eigenvalue gives
\[
 |f(t)-r|
 \le\|tB+t^2T\|_\infty
 \le t\|B\|_\infty+t^2\|T\|_\infty.
\]
It follows that
\begin{equation*}
\big|\langle v_t|R|v_t\rangle-r\big|
 \le |f(t)-r|
       +\big|f(t)-\langle v_t|R|v_t\rangle\big| \le 2t\|B\|_\infty+2t^2\|T\|_\infty
 \longrightarrow0.
\end{equation*}
Every sequence $t_k\downarrow0$ has a subsequence along which
$v_{t_k}$ converges to a unit vector $v$, by compactness of the
unit sphere. The preceding limit gives
$\langle v|R|v\rangle=r$. Since $rI-R\succeq0$, this equality
implies $(rI-R)v=0$, so $v\in E$. Thus every limit point of the
maximizing vectors $v_t$ as $t\downarrow0$ belongs to $E$.

Set $b(t)=(f(t)-r)/t$. Testing the largest eigenvalue with $u$
gives
\[
 \begin{aligned}
b(t)
 \ge \langle u|B|u\rangle+t\langle u|T|u\rangle \ge \lambda_{\max}(G(Y)),
\end{aligned}
\]
because $Ru=ru$, $T\succeq0$, and $u$ maximizes the compression
of $B$ to $E$. On the other hand, using the maximizing vector
$v_t$ gives
\[
 \begin{aligned}
b(t)
 =\frac{\langle v_t|R|v_t\rangle-r}{t}
   +\langle v_t|B|v_t\rangle
   +t\langle v_t|T|v_t\rangle \le \langle v_t|B|v_t\rangle+t\|T\|_\infty,
\end{aligned}
\]
where the inequality uses $R\preceq rI$ and $t>0$.

Choose a sequence $t_k\downarrow0$ along which
$\langle v_{t_k}|B|v_{t_k}\rangle$ approaches its upper limit.
A subsequence of the corresponding vectors converges to a unit
vector $v\in E$, as shown above. Therefore
\[
 \limsup_{t\downarrow0}\langle v_t|B|v_t\rangle
 \le\max_{\substack{w\in E\\\|w\|=1}}\langle w|B|w\rangle
 =\lambda_{\max}(G(Y)).
\]
Combining the lower and upper bounds yields
\[
 \lambda_{\max}(G(Y))
 \le\liminf_{t\downarrow0}b(t)
 \le\limsup_{t\downarrow0}b(t)
 \le\lambda_{\max}(G(Y)).
\]
The two limits coincide, proving the right-derivative formula.

\Needspace{4\baselineskip}
\noindent\textbf{Step 2: Use the right derivative to exclude
positive-definite directions and identify strict decrease along
their negatives.}\par\nobreak

Assume (i), so $\lambda_{\max}M(Z)\ge r$ for every $Z\in V$.
For any $Y\in V$ and $t>0$, the vector $-tY$ is also in $V$.
Consequently,
\[
 \lambda_{\max}(R-tL(Y)+t^2Q(Y))\ge r.
\]
The right derivative at $t=0$ is therefore nonnegative.
Applying Step~1 to the direction $-Y$ gives
\[
 0\le\lambda_{\max}(G(-Y))
 =-\lambda_{\min}(G(Y)).
\]
Thus $\lambda_{\min}(G(Y))\le0$ for every $Y\in V$, which is (iii).

Conversely, suppose that $G(Y_0)>0$ for some $Y_0\in V$.
Step~1 gives
\[
 \left.\frac{d}{dt}\right|_{t=0+}
 \lambda_{\max}(R-tL(Y_0)+t^2Q(Y_0))
 =-\lambda_{\min}(G(Y_0))<0.
\]
By the definition of a negative right derivative, there is an
$\epsilon>0$ such that
\[
 \lambda_{\max}M(-tY_0)<r
 \qquad(0<t<\epsilon).
\]
Thus a positive-definite compressed direction $G(Y_0)$ makes the
objective decrease strictly when we move from $Y=0$ along the
feasible direction $-Y_0$.

\Needspace{4\baselineskip}
\noindent\textbf{Step 3: Show that the absence of a
positive-definite compressed direction is equivalent to the
existence of one common density operator, and use it to prove
global minimality.}\par\nobreak

The image $G(V)$ is a real linear subspace of $\Herm(E)$ because
$G$ is real linear. Condition (iii) says that $G(V)$ contains no
positive-definite operator. Lemma~\ref{lem:separation}, applied to
the real linear space $G(V)$, therefore gives a density operator
$\sigma_E$ on $E$ such that
\[
 \Tr_E[\sigma_EG(Y)]=0
 \qquad\text{for every }Y\in V.
\]
This proves (ii). Conversely, if (ii) holds, then for every
positive-definite $G_0\in G(V)$ we would have
$\Tr_E(\sigma_EG_0)>0$, contradicting (ii). Thus (ii) and (iii)
are equivalent.

It remains to show that (ii) implies (i). Extend $\sigma_E$ by zero
on $E^\perp$, obtaining a density operator $\sigma$ on $K$.
Because $R$ acts as $rI_E$ on $E$, we have $\Tr(\sigma R)=r$.
For every $Y\in V$, the compression defining $G(Y)$ gives
\[
 \Tr[\sigma L(Y)]
 =\Tr_E[\sigma_EG(Y)]=0.
\]
Since $Q(Y)\succeq0$, the variational characterization of the
largest eigenvalue yields
\[
 \begin{aligned}
\lambda_{\max}M(Y)
 \ge\Tr[\sigma M(Y)] =\Tr(\sigma R)+\Tr[\sigma L(Y)]+\Tr[\sigma Q(Y)] \ge r.
\end{aligned}
\]
This proves (i) from (ii) and completes the equivalence.
\end{proof}

The final auxiliary fact converts a trace-orthogonality condition
into the operator equation required by the general saturation
criterion. Once a Hermitian operator $S$ is orthogonal to every
supported direction with zero partial trace, we use the partial-trace
map and its adjoint to construct a Hermitian operator $H$ on the
input space.

\begin{lemma}[Annihilator of supported trace-zero directions]\label{lem:supported-annihilator}
Let $P$ be an orthogonal projection on $\cH_B\otimes\cH_A$,
$\mathcal X_P=\{Y=Y^\dagger:Y=PYP\}$, and
$\cV_P=\{Y\in\mathcal X_P:\Tr_B Y=0\}$.
For a Hermitian operator $S=PSP$, the condition
\[
 \Tr(YS)=0\qquad\text{for every }Y\in\cV_P
\]
holds if and only if there exists a Hermitian operator $H$ on
$\cH_A$ such that
\[
 S=P(I_B\otimes H)P.
\]
The equation constrains $H$ only through the map
$\Phi_P(H):=P(I_B\otimes H)P$. If a nonzero Hermitian operator
$K$ satisfies $\Phi_P(K)=0$, then $H$ and $H+K$ give the same
compressed operator. Thus the lemma asserts the existence of such
an $H$; it does not assert that $H$ is unique.
\end{lemma}
\begin{proof}
We first prove that trace-orthogonality to every element of $\cV_P$
implies the existence of $H$. Equip the finite-dimensional real
spaces $\mathcal X_P$ and $\Herm(\cH_A)$ with the trace inner
product, and define the partial-trace map
\[
 T:\mathcal X_P\longrightarrow\Herm(\cH_A),\qquad
 T(Y)=\Tr_B Y.
\]
By definition, $\ker T=\cV_P$. The assumed orthogonality therefore
means that $S\in(\ker T)^\perp$. For a linear map between
finite-dimensional real inner-product spaces, the orthogonal
complement of its kernel is the range of its adjoint. Hence
\[
 S\in(\ker T)^\perp
 =\operatorname{ran}T^*,
\]
so there exists a Hermitian operator $H$ on $\cH_A$ such that
$S=T^*(H)$.

We now identify the adjoint $T^*$. For every $Y\in\mathcal X_P$,
\[
 \begin{aligned}
\Tr[Y T^*(H)]
 &\overset{(a)}{=}\Tr[(TY)H] \overset{(b)}{=}\Tr[(\Tr_B Y)H]\\
&\overset{(c)}{=}\Tr[Y(I_B\otimes H)] \overset{(d)}{=}\Tr[YP(I_B\otimes H)P].
\end{aligned}
\]
Here (a) is the definition of the adjoint with respect to the trace
inner products, (b) is the definition of $T$, (c) is the defining
property of the partial trace, and (d) uses $Y=PYP$ and cyclicity
of the trace. Both $T^*(H)$ and $P(I_B\otimes H)P$ belong to
$\mathcal X_P$. Since their trace inner products with every element
of $\mathcal X_P$ agree, the two operators are equal:
\[
 T^*(H)=P(I_B\otimes H)P.
\]
This proves the first implication.

Conversely, suppose that a Hermitian operator $H$ satisfies
$S=P(I_B\otimes H)P$. For every $Y\in\cV_P$,
\[
 \begin{aligned}
\Tr(YS)
 \overset{(a)}{=}\Tr[Y(I_B\otimes H)] \overset{(b)}{=}\Tr[(\Tr_B Y)H] \overset{(c)}{=}0.
\end{aligned}
\]
Here (a) uses $Y=PYP$ and cyclicity of the trace, (b) uses the
partial-trace identity, and (c) uses $\Tr_B Y=0$. Thus
$\Tr(YS)=0$ for every $Y\in\cV_P$, which proves the reverse
implication. The two implications establish the stated equivalence.
\end{proof}

\subsection{Proof of Theorem~\ref{thm:criterion}}\label{sec:general-proof}
\begin{proof}[\unskip\nopunct]
Under the support assumption $D=PDP$, we prove that
$\JS(\Lambda)=\JR(\Lambda)$ holds if and only if there is one pair
$(\sigma,H)$ such that $\sigma\succeq0$, $\Tr\sigma=1$,
$R(C,D)\sigma=r(C,D)\sigma$, and the supported operator equation
\eqref{eq:station} holds. The separation result, the directional-
derivative formula, and the supported annihilator established above
supply the two steps of the reduction: the spectral lemma connects
optimality to a common density operator, and the partial-trace
identity converts its trace condition to the equation involving $H$.
Fix $D=PDP$ and put $R=\Tr_B(DC^+D)$,
$r=\lambda_{\max}(R)$, and $E=\ker(rI-R)$.
The projection $\Pi$ is onto $E$ in $\cH_A$, whereas $P$ projects onto
the Choi support in $\cH_B\otimes\cH_A$.
For $Y\in\cV_P$, direct expansion gives
\begin{equation}\label{eq:expand}
 \begin{aligned}
M_{C,D}(tY)&=R+tL_P(Y)+4t^2\Tr_B(YC^+Y),\\
L_P(Y)&:=2i\Tr_B(DC^+Y-YC^+D).
\end{aligned}
\end{equation}
The quadratic term is positive semidefinite because
$YC^+Y\succeq0$. Define the compression of the linear term on $E$ by
\begin{equation}\label{eq:Gamma}
 \Gamma_{C,D}(Y)=\Pi L_P(Y)|_E.
\end{equation}
The map $Y\mapsto\Gamma_{C,D}(Y)$ is real linear.

By Theorem~\ref{thm:variational}, saturation is equivalent to
$Y=0$ minimizing $\lambda_{\max}M_{C,D}(Y)$ over $\cV_P$.
In Eq.~\eqref{eq:expand} the quadratic term is positive semidefinite
and homogeneous of degree two. Lemma~\ref{lem:spectral-certificate}
therefore identifies optimality of zero with the existence of a
density operator $\sigma_E$ on $E$ such that
\[
 \Tr_E[\sigma_E\Gamma_{C,D}(Y)]=0\qquad\text{for every }Y\in\cV_P.
\]
We translate this condition into Eq.~\eqref{eq:station} in both
directions. Write
\[
 S_\sigma^+:=2i[(I_B\otimes\sigma)DC^+
                  -C^+D(I_B\otimes\sigma)].
\]

\Needspace{4\baselineskip}
\noindent\textbf{Step 1: Establish the trace pairing
$\Tr_E[\sigma_E\Gamma_{C,D}(Y)]=\Tr[YP S_\sigma^+P]$.}\par\nobreak

Let $E=\ker(rI-R)$ and let $\iota:E\hookrightarrow\cH_A$ be the
inclusion. For a density operator $\sigma_E$ on $E$, define its
zero extension $\sigma=\iota\sigma_E\iota^\dagger$ on $\cH_A$.
The extension is positive, and the rectangular trace identity gives
\[
 \Tr_{\cH_A}\sigma
 =\Tr_E(\sigma_E\iota^\dagger\iota)
 =\Tr_E\sigma_E=1.
\]
Since $R\iota=r\iota$, it also satisfies $R\sigma=r\sigma$.
Conversely, $R\sigma=r\sigma$ places the range of $\sigma$ in
$E$. Hermiticity then gives $\sigma=\Pi\sigma\Pi$, so $\sigma$
is the zero extension of its restriction to $E$.
For $Y\in\cV_P$, the definition of the compression in
Eq.~\eqref{eq:Gamma} gives
$\Gamma_{C,D}(Y)=\iota^\dagger L_P(Y)\iota$. Using also
$\sigma=\iota\sigma_E\iota^\dagger$, we obtain the following
equation chain:
\begin{align}
\Tr_E[\sigma_E\Gamma_{C,D}(Y)]
 &\overset{(a)}{=}\Tr_E[\sigma_E\iota^\dagger
                         L_P(Y)\iota] \overset{(b)}{=}\Tr_{\cH_A}[\iota\sigma_E\iota^\dagger
                         L_P(Y)] \overset{(c)}{=}\Tr_{\cH_A}[\sigma L_P(Y)]\notag\\
&\overset{(d)}{=}2i\Tr_{\cH_A}\!\left[
       \sigma\Tr_B(DC^+Y-YC^+D)\right]\notag\\
&\overset{(e)}{=}2i\Tr_{\cH_B\otimes\cH_A}\!\left[
       (I_B\otimes\sigma)(DC^+Y-YC^+D)\right]\notag\\
&\overset{(f)}{=}2i\Tr\!\left[
       Y(I_B\otimes\sigma)DC^+
       -YC^+D(I_B\otimes\sigma)\right] \overset{(g)}{=}\Tr(YS_\sigma^+) \overset{(h)}{=}\Tr(YP S_\sigma^+P). \label{eq:pair}
\end{align}
Here (a) uses the definition of $\Gamma_{C,D}$ as the compression
of $L_P$ to $E$, (b) uses the rectangular trace identity for the inclusion
$\iota:E\hookrightarrow\cH_A$ and its adjoint, (c) uses the definition of the
zero extension $\sigma=\iota\sigma_E\iota^\dagger$, (d) uses
the definition of $L_P(Y)$, (e) is the defining property of the
partial trace, (f) uses cyclicity of the trace, (g) uses the
definition of $S_\sigma^+$, and (h) uses $Y=PYP$ and cyclicity.
Thus Eq.~\eqref{eq:pair} holds for every density operator $\sigma_E$
on $E$ and every feasible $Y$.

\Needspace{4\baselineskip}
\noindent\textbf{Step 2: Prove that the existence of $(\sigma,H)$
implies $\JS=\JR$.}\par\nobreak

Assume that $\sigma,H$ satisfy the conditions in the theorem.
Because $R\sigma=r\sigma$, the range of $\sigma$ is contained in $E$.
Thus $\sigma$ is the zero extension of a density operator $\sigma_E$
on $E$. By Eq.~\eqref{eq:pair} and the operator equation,
\[
 \begin{aligned}
\Tr[\sigma L_P(Y)]
 &\overset{(a)}{=}\Tr[YP S_\sigma^+P] \overset{(b)}{=}\Tr[YP(I_B\otimes H)P]\\
&\overset{(c)}{=}\Tr[Y(I_B\otimes H)] \overset{(d)}{=}\Tr[(\Tr_B Y)H] \overset{(e)}{=}0.
\end{aligned}
\]
Here (a) is Eq.~\eqref{eq:pair}, (b) is the operator equation
assumed in the theorem, (c) uses $Y=PYP$ and cyclicity of the trace,
(d) is the defining property of the partial trace, and (e) uses
$\Tr_B Y=0$. Thus this same $\sigma$ annihilates the linear term
for every feasible direction.

For every feasible $Y$, the variational characterization of
the largest eigenvalue and Eq.~\eqref{eq:expand} now give
\[
\begin{aligned}
 \lambda_{\max}M_{C,D}(Y)&\ge \Tr[\sigma M_{C,D}(Y)]\\
 &=r+4\Tr[\sigma\Tr_B(YC^+Y)]\ge r .
\end{aligned}
\]
In the equality we used $\Tr(\sigma R)=r$ and
$\Tr[\sigma L_P(Y)]=0$. The last inequality follows because
$YC^+Y\succeq0$, partial trace preserves positivity, and
$\sigma\succeq0$. The feasible point $Y=0$ attains the lower bound
because $M_{C,D}(0)=R$. The variational formula
\eqref{eq:var} therefore gives $\JS=r=\JR$.

\Needspace{4\baselineskip}
\noindent\textbf{Step 3: Prove that $\JS=\JR$ implies the existence
of $(\sigma,H)$.}\par\nobreak

By saturation and the spectral lemma, there is a density operator
$\sigma_E$ whose trace pairing vanishes for all $Y\in\cV_P$.
Its zero extension $\sigma$ has the state properties checked in
Step~1, and Eq.~\eqref{eq:pair} gives
\[
 \Tr[YP S_\sigma^+P]=\Tr_E[\sigma_E\Gamma_{C,D}(Y)]=0
 \qquad(Y\in\cV_P).
\]
The operator $P S_\sigma^+P$ is Hermitian because the two terms
inside the brackets in $S_\sigma^+$ are adjoints of one another.
It is supported on $P$ by construction. Lemma~\ref{lem:supported-annihilator} gives a Hermitian
$H$ with $PS_\sigma^+P=P(I_B\otimes H)P$. This is
Eq.~\eqref{eq:station} for the same $\sigma$.
\end{proof}

\subsection{Proof of Theorem~\ref{thm:positive}}
\begin{proof}[\unskip\nopunct]
Theorem~\ref{thm:criterion} gives a saturation criterion on the
support of the Choi matrix. We show how it simplifies when $C$
has full rank. We first check that the general theorem applies,
so the asymptotic coefficient exists. We then show that its
conditions on a pair $(\sigma,H)$ reduce to
Eqs.~\eqref{eq:positive-state} and~\eqref{eq:L} for the same pair.

The channel family in Theorem~\ref{thm:positive} is finite-dimensional,
differentiable, and CPTP. Since $C>0$, its support projection is
$P=I_{BA}$, so $D=PDP$. Thus Theorem~\ref{thm:criterion} applies
and ensures that the limit defining $\JS$ exists.

Also $C^+=C^{-1}$, so the RLD matrix and its largest eigenvalue in
the two theorems coincide. The positivity, normalization, and
eigenspace conditions on $\sigma$ are consequently identical.
Substituting $P=I_{BA}$ and $C^+=C^{-1}$ into the left and right
sides of Eq.~\eqref{eq:station} gives, respectively,
\[
 2i[(I_B\otimes\sigma)DC^{-1}-C^{-1}D(I_B\otimes\sigma)]
 \qquad\text{and}\qquad I_B\otimes H.
\]
These are the two sides of Eq.~\eqref{eq:L}, so a pair satisfies
either system exactly when it satisfies the other.

For sufficiency, a pair satisfying the positive-definite conditions
therefore satisfies the general conditions, and
Theorem~\ref{thm:criterion} gives $\JS=\JR$.
For necessity, if $\JS=\JR$, the pair provided by
Theorem~\ref{thm:criterion} satisfies the positive-definite conditions
by the same correspondence.
\end{proof}

\Needspace{17\baselineskip}
\subsection{Directional-derivative characterization of RLD saturation}\label{sec:first-order}
The proof of Theorem~\ref{thm:criterion} identifies saturation
with optimality at $Y=0$. For each feasible direction $Y$, we
examine the right derivative at $t=0$ of
$t\mapsto\lambda_{\max}M_{C,D}(tY)$. Here $t$ varies the
optimization variable along $Y$. The derivative is
$\lambda_{\max}(\Gamma_{C,D}(Y))$, where
$\Gamma_{C,D}(Y)$ is the coefficient of $t$ in $M_{C,D}(tY)$
compressed to the largest eigenspace $E$ of $R$.
A positive-definite compression gives a negative right derivative
along $-Y$. The positive-semidefinite quadratic term makes the
absence of a direction with negative right derivative sufficient
for global optimality.

\begin{proposition}[Directional-derivative criteria for RLD saturation]\label{prop:first-order}
Under the assumptions of Theorem~\ref{thm:criterion}, the following
are equivalent:
\begin{enumerate}[label=\textup{(\roman*)}]
\item $\JS(\Lambda)=\JR(\Lambda)$.
\item The real linear space $\{\Gamma_{C,D}(Y):Y\in\cV_P\}$
contains no positive-definite matrix on $E$.
\item $\lambda_{\min}(\Gamma_{C,D}(Y))\le0$ for every $Y\in\cV_P$.
\end{enumerate}
If these conditions fail, choose $Y_0\in\cV_P$ with
$\Gamma_{C,D}(Y_0)>0$. Then, for all sufficiently small $t>0$,
\begin{equation}\label{eq:strict}
 \lambda_{\max}M_{C,D}(-tY_0)<r(C,D).
\end{equation}
\end{proposition}

\Needspace{6\baselineskip}
\begin{proof}[Proof of Proposition~\ref{prop:first-order}]
Set $f(Y)=\lambda_{\max}M_{C,D}(Y)$ on $\cV_P$.
The variational formula \eqref{eq:var} and the identity
$M_{C,D}(0)=R(C,D)$ give
\[
 \min_{Y\in\cV_P}f(Y)=\JS,\qquad f(0)=\JR .
\]
Thus (i) is equivalent to global optimality of $Y=0$.
In Lemma~\ref{lem:spectral-certificate}, take the real space of
directions to be $\cV_P$, the linear term to be $L_P$, and the
quadratic term to be $Y\mapsto4\Tr_B(YC^+Y)$ as in
Eq.~\eqref{eq:expand}. This term is positive semidefinite and
homogeneous of degree two. The compression of $L_P$ to $E$ is $\Gamma_{C,D}$, so the lemma gives the equivalence of
(i) and (ii). A Hermitian matrix on $E$
is positive definite precisely when its smallest eigenvalue is
positive, which gives the equivalence of (ii) and (iii).

Finally, suppose that these conditions fail. Then some
$Y_0\in\cV_P$ satisfies $\Gamma_{C,D}(Y_0)>0$. The
right-derivative formula in the spectral optimality lemma gives
\[
 \begin{aligned}
\left.\frac{d}{dt}\right|_{t=0+}
 \lambda_{\max}M_{C,D}(-tY_0)
 \overset{(a)}{=}\lambda_{\max}(-\Gamma_{C,D}(Y_0)) \overset{(b)}{=}-\lambda_{\min}(\Gamma_{C,D}(Y_0)) <0.
\end{aligned}
\]
Here (a) applies the directional-derivative formula to $-Y_0$
and uses $\Gamma_{C,D}(-Y_0)=-\Gamma_{C,D}(Y_0)$ by real linearity.
Equality (b) uses the identity
$\lambda_{\max}(-A)=-\lambda_{\min}(A)$ for Hermitian $A$.
The negative right derivative implies that
$\lambda_{\max}M_{C,D}(-tY_0)<r(C,D)$ for all sufficiently
small $t>0$, which is Eq.~\eqref{eq:strict}.
\end{proof}

\subsection{Finite-dimensional tests for RLD saturation}
We now turn the test over all feasible directions into a test using the images of a basis. Under $D=PDP$, Proposition~\ref{prop:first-order} characterizes
RLD saturation by the absence of positive-definite operators in the
real linear space
\[
 \Gamma_{C,D}(\cV_P)=\{\Gamma_{C,D}(Y):Y\in\cV_P\}
 \subseteq\Herm(E).
\]
Since $\cV_P$ is a finite-dimensional real vector space,
choose a real basis $Y_1,\ldots,Y_m$. For any feasible $Y$ we have
$Y=\sum_{\mu=1}^m a_\mu Y_\mu$ with $a_\mu\in\mathbb R$, and
real-linearity gives
\[
 \Gamma_{C,D}(Y)=\sum_{\mu=1}^m
 a_\mu\Gamma_{C,D}(Y_\mu).
\]
Consequently, the saturation criterion asks whether the real span
of the finitely many Hermitian matrices
$\Gamma_{C,D}(Y_1),\ldots,\Gamma_{C,D}(Y_m)$ contains a
positive-definite operator on $E$. In the one- and two-dimensional
cases for $E$, the following corollaries turn this question into
explicit scalar and real linear-algebra tests, respectively.
\begin{corollary}[Simple largest RLD eigenvalue]\label{cor:simple}
Suppose $\dim E=1$, and let $|v\rangle$ be a unit vector spanning $E$.
Then $\JS=\JR$ if and only if
\begin{equation}\label{eq:simple}
 \langle v|\Tr_B[2i(DC^+Y-YC^+D)]|v\rangle=0
 \quad\text{for all }Y\in\cV_P.
\end{equation}
To decide whether RLD saturation holds, it is enough to verify this
scalar equality for the elements of any real basis of $\cV_P$.
\end{corollary}
\begin{proof}
Define the real-valued functional
\[
 g(Y):=\langle v|\Gamma_{C,D}(Y)|v\rangle,
 \qquad Y\in\cV_P.
\]
By the definition of $\Gamma_{C,D}$ in Eq.~\eqref{eq:Gamma},
$g(Y)$ is exactly the left-hand side of Eq.~\eqref{eq:simple}.
Since $\dim E=1$, $\Gamma_{C,D}(Y)$ has only one matrix entry,
namely $g(Y)$, in the orthonormal basis consisting of $v$.
If $g(Y)\ne0$ for some feasible $Y$, then $g(Y)>0$ or
$g(-Y)>0$, because $\cV_P$ is a real vector space and
$g(-Y)=-g(Y)$. Thus a positive-definite operator belongs to
$\Gamma_{C,D}(\cV_P)$ whenever $g$ is not identically zero.
Conversely, if $g(Y)=0$ for all $Y$, the image contains no
positive-definite operator. Proposition~\ref{prop:first-order}
therefore shows that RLD saturation is equivalent to
$g(Y)=0$ for every feasible $Y$.

Now take a real basis $Y_1,\ldots,Y_m$ of $\cV_P$.
For any $Y=\sum_{\mu=1}^m a_\mu Y_\mu$ with real $a_\mu$,
real-linearity of $\Gamma_{C,D}$ gives
\[
 g(Y)=\sum_{\mu=1}^m a_\mu g(Y_\mu).
\]
Hence $g$ vanishes on all of $\cV_P$ if and only if
$g(Y_\mu)=0$ for every basis element. These finitely many
scalar equalities are sufficient to test the RLD saturation
condition in the one-dimensional case.
\end{proof}

When $E$ is two-dimensional, a density operator on $E$ can be
described by three real coordinates in the Pauli basis. This
turns the trace conditions into linear equations with a unit-ball
constraint on those coordinates.

\begin{corollary}[A direct test when the largest RLD eigenvalue is double]\label{cor:double}
Suppose $\dim E=2$. Choose a real basis $Y_1,\ldots,Y_m$ of
$\cV_P$ and Pauli matrices $\tau_1,\tau_2,\tau_3$ on $E$.
For each basis element $Y_\mu$, expand the Hermitian matrix
$\Gamma_{C,D}(Y_\mu)$ in the Pauli basis as
\begin{equation}\label{eq:Ab}
 \begin{gathered}
\Gamma_{C,D}(Y_\mu)=b_\mu I_E+\sum_{j=1}^3 T_{\mu j}\tau_j,\qquad b_\mu=\tfrac12\Tr_E\Gamma_{C,D}(Y_\mu),\qquad
 T_{\mu j}=\tfrac12\Tr_E[\tau_j\Gamma_{C,D}(Y_\mu)].
\end{gathered}
\end{equation}
The coefficients are real because the trace of a product of two
Hermitian operators is real.
\Needspace{7\baselineskip}
Define $b=(b_\mu)\in\mathbb R^m$
and $T=(T_{\mu j})\in\mathbb R^{m\times3}$. Then RLD saturation
holds if and only if the real linear equation $Tu=-b$ has a
solution $u\in\mathbb R^3$ with $\|u\|_2\le1$.
Equivalently, the following two conditions are both necessary
and sufficient for RLD saturation:
\begin{equation}\label{eq:pseudo}
 b\in\operatorname{ran}T,\qquad \|T^+b\|_2\le1.
\end{equation}
The first condition ensures that $Tu=-b$ is solvable; the second
ensures that at least one solution lies in the Euclidean unit ball.
Here $T^+$ is the Moore--Penrose inverse of the real matrix $T$.
\end{corollary}
\begin{proof}
We first express the saturation criterion as a feasibility condition
for a density operator on $E$. Proposition~\ref{prop:first-order}
shows that saturation is equivalent to the absence of
positive-definite matrices in $\Gamma_{C,D}(\cV_P)$.
By Lemma~\ref{lem:separation}, this is equivalent to the existence
of one density operator $\sigma$ on $E$ satisfying
\[
 \Tr_E[\sigma\Gamma_{C,D}(Y)]=0
 \qquad\text{for every }Y\in\cV_P.
\]
Since the pairing is real linear in $Y$, it is enough to impose
these equations for $Y=Y_\mu$, $1\le\mu\le m$.

Every trace-one Hermitian operator on the two-dimensional space $E$
can be written uniquely as
\[
 \sigma=\frac12\left(I_E+\sum_{j=1}^3u_j\tau_j\right),
 \qquad u=(u_1,u_2,u_3)\in\mathbb R^3.
\]
The Pauli relations yield
\[
 \left(\sum_{j=1}^3u_j\tau_j\right)^2
 =\|u\|_2^2 I_E.
\]
Thus the eigenvalues of $\sigma$ are
$(1+\|u\|_2)/2$ and $(1-\|u\|_2)/2$.
Consequently, $\sigma$ is a density operator precisely when
$\|u\|_2\le1$.

For each basis element, Eq.~\eqref{eq:Ab} and the identities
$\Tr_E\tau_j=0$ and $\Tr_E(\tau_j\tau_k)=2\delta_{jk}$ give
\[
 \begin{aligned}
\Tr_E[\sigma\Gamma_{C,D}(Y_\mu)]
 =\frac12\Tr_E\!\left[
  \left(I_E+\sum_j u_j\tau_j\right)
  \left(b_\mu I_E+\sum_k T_{\mu k}\tau_k\right)\right] =b_\mu+\sum_{j=1}^3T_{\mu j}u_j.
\end{aligned}
\]
The trace pairing therefore vanishes on every real basis element
if and only if $Tu=-b$.
This equation is solvable if and only if $b\in\operatorname{ran}T$.
When this condition holds, all solutions are
\[
 u=-T^+b+z,\qquad z\in\ker T.
\]
The Moore--Penrose solution $-T^+b$ is orthogonal to $\ker T$,
so
\[
 \|u\|_2^2=\|T^+b\|_2^2+\|z\|_2^2.
\]
It follows that $-T^+b$ is a solution of minimum Euclidean norm,
and a solution with $\|u\|_2\le1$ exists if and only if
$\|T^+b\|_2\le1$. Together with the solvability condition,
this proves the two conditions in Eq.~\eqref{eq:pseudo} and hence
the stated necessary and sufficient criterion for RLD saturation.
\end{proof}

Given $C,D$, the calculation is as follows. Compute $P,C^+$ from Eq.~\eqref{eq:pseudoinverse} and $R(C,D)$ from Eq.~\eqref{eq:R}. Find its largest-eigenvalue space $E$ and the projection $\Pi$ onto that space. Choose a real basis of the Hermitian solutions of $Y=PYP$ and $\Tr_B Y=0$, and compute their images under Eq.~\eqref{eq:Gamma}. Use Corollary~\ref{cor:simple} or~\ref{cor:double} when applicable; in larger dimension test whether the real span of these images contains a positive-definite matrix. Approximate degeneracy in a numerical calculation must be distinguished from exact eigenvalue multiplicity.

\section{Conclusions}
For finite-dimensional quantum channels with finite RLD information,
we have given necessary and sufficient Choi-space conditions for the
RLD upper bound to equal the optimized asymptotic SLD coefficient.
The auxiliary density operator $\sigma$ must be supported on the
largest-eigenvalue space of $R(C,D)$. It must also satisfy the linear
operator equation with a Hermitian $H$. For rank-deficient Choi matrices
with $D=PDP$, that equation is restricted to the Choi support. The formulation allows degeneracy of the largest
RLD eigenvalue and pointwise treatment of rank-changing channel
families. 
Zhou and Jiang~\cite{ZJ} established the optimized asymptotic SLD coefficient and its attainability in the regime of at most linear information growth. Ko\l{}ody\'nski and Demkowicz-Dobrza\'nski~\cite{KD2013} related the RLD value to a particular Kraus representation. These criteria determine when that representation is optimal using only Choi data, and Proposition~\ref{prop:first-order} provides a feasible direction that lowers the objective when the criterion fails.

For generalized dephasing channels, Proposition~\ref{prop:corr} tests saturation using linear equations in the certificate weights $w_j=\sigma_{jj}$. At complete dephasing, Theorem~\ref{thm:complete-dephasing} gives $J_n^{\mathrm S}/n=J_1^{\mathrm S}=\JS$ for every $n\ge1$, attained by repeating an ancilla-free pure probe and its local SLD measurement. Proposition~\ref{prop:dephasing-graph} gives a graph criterion
for saturation for every positive-definite correlation matrix at
the evaluation point. At complete dephasing it reduces to the
weighted-degree criterion of Corollary~\ref{cor:weighted}.

The ratio $\JR/\JS$ diverges as the number of leaves in the star graph grows. For the qubit replacer family, it diverges at fixed dimension as the smallest Choi eigenvalue tends to zero. The qutrit example gives a phase-dependent obstruction to saturation within an explicitly realized channel family. Proposition~\ref{prop:maxent} gives a separate test for a prescribed maximally entangled input, and the rank-changing example distinguishes pointwise information from the nearby-parameter limit.

\appendix
\section{Affine tangent replacement}\label{app:tangent}
When $C$ is rank deficient, the original family may not admit a differentiable Kraus representation with constant rank near zero. We first replace it by a family with the same channel and derivative at zero. The comparison is made protocol by protocol before taking any supremum or asymptotic limit.

Define the replacement on input operators by
\begin{equation}\label{eq:replacement-map}
 \widehat\Lambda_\theta(X):=\Lambda_0(X)+\theta\dot\Lambda_0(X)
 =\Tr_A[(C+\theta D)(I_B\otimes X^{\mathsf T})].
\end{equation}
\begin{lemma}[Replacing a channel family by its affine tangent]\label{lem:tangent}
If Eq.~\eqref{eq:supp} holds, then for some $\epsilon>0$ the maps in Eq.~\eqref{eq:replacement-map} are CPTP for $|\theta|<\epsilon$. Their Choi matrices are
\begin{equation}\label{eq:affine}
 \widehat C_\theta=C+\theta D,
\end{equation}
with constant rank $\operatorname{rank}C$. They satisfy $\widehat\Lambda_0=\Lambda_0$ and $\dot{\widehat\Lambda}_0=\dot\Lambda_0$. For every positive integer $n$,
\begin{equation}\label{eq:finite-eq}
 J_n^{\mathrm S}(\Lambda;0)=J_n^{\mathrm S}(\widehat\Lambda;0),
\end{equation}
both for parallel strategies and for the full class of causally ordered adaptive strategies. Their pointwise channel RLD values also coincide.
For every fixed protocol with output state $\rho_n(0)$ and support
projection $P_n$, its output derivative satisfies
$\dot\rho_n(0)=P_n\dot\rho_n(0)P_n$.
\end{lemma}
\Needspace{6\baselineskip}
\begin{proof}
\leavevmode\par\nobreak

\Needspace{3\baselineskip}
\noindent\textbf{Step 1: Prove that $\widehat C_\theta=C+\theta D$ defines CPTP maps of constant Choi rank for small $|\theta|$.}\par\nobreak

Let $c>0$ be the smallest eigenvalue of $C$ on $\operatorname{Ran}P$. Since $D=PDP$, the matrix $C+\theta D$ is zero on the orthogonal complement and satisfies
\[
 (C+\theta D)|_{\operatorname{Ran}P}
 \ge(c-|\theta|\|D\|_\infty)I_{\operatorname{Ran}P}.
\]
It is positive definite on $\operatorname{Ran}P$ for sufficiently small $|\theta|$. Also $\Tr_B(C+\theta D)=I_A$. Equations~\eqref{eq:choi} and~\eqref{eq:choi-inverse} identify it as the Choi matrix of the CPTP map in Eq.~\eqref{eq:replacement-map}, with rank $\operatorname{rank}C$.

\Needspace{3\baselineskip}
\noindent\textbf{Step 2: Prove Eq.~\eqref{eq:finite-eq} and equality of the channel RLD values.}\par\nobreak

For a fixed parameter-independent protocol, write $\mathcal F(\mathcal M_1,\ldots,\mathcal M_n)$ for the output when $\mathcal M_j$ is inserted in slot $j$. The input state and the operations between slots are fixed. Composition with those linear operations and with the identity on the memory makes $\mathcal F$ linear in each slot separately. For the original family, the product rule gives
\begin{align*}
\rho_n(0)=\mathcal F(\Lambda_0,\ldots,\Lambda_0),\qquad \dot\rho_n(0)=\left.\frac{d}{d\theta}\mathcal F(\Lambda_\theta,\ldots,\Lambda_\theta)\right|_{\theta=0}
              =\sum_{j=1}^n X_j,\\
X_j:=\mathcal F(\Lambda_0,\ldots,\dot\Lambda_0,\ldots,\Lambda_0),
\end{align*}
where $\dot\Lambda_0$ is in slot $j$. The same formulas hold for the replacement, with $\widehat\Lambda_0=\Lambda_0$ and $\dot{\widehat\Lambda}_0=\dot\Lambda_0$. This proves equality of the output states and derivatives for each protocol. The SLD information in Eq.~\eqref{eq:SLD-state} depends only on that pair. Taking the supremum in each strategy class separately gives Eq.~\eqref{eq:finite-eq}. The channel RLD values also agree by Eq.~\eqref{eq:R}, since the Choi data $C,D$ coincide.

\Needspace{4\baselineskip}
\noindent\textbf{Step 3: Establish support of the output derivative.}\par\nobreak

Let $P_n$ be the support projection of the common output state
$\rho_n(0)$ for a fixed protocol. The channel support condition
implies that the output derivative annihilates every vector in
$\ker\rho_n(0)$. To prove this, we bound the derivative above and
below by multiples of $\rho_n(0)$ using completely positive maps
inserted into each slot.

Let $c>0$ be the smallest eigenvalue of $C$ on $\operatorname{Ran}P$, and set $a=\|D\|_\infty/c$. On $\operatorname{Ran}P$, $aC\ge \|D\|_\infty P$ bounds both $D$ and $-D$ from above. On its orthogonal complement both $C$ and $D$ vanish, so $aC\pm D\succeq0$ on the whole Choi space. The maps $a\Lambda_0\pm\dot\Lambda_0$ with these Choi matrices are completely positive. Insert one of them in slot $j$. Complete positivity ensures positivity even with the memory attached, and the fixed operations preserve positivity through the other slots. The resulting output can be unnormalized. Multilinearity and then summation over $j$ give
\begin{equation*}
a\rho_n(0)\pm X_j
 =\mathcal F(\Lambda_0,\ldots,a\Lambda_0\pm\dot\Lambda_0,\ldots,\Lambda_0)\succeq0,\qquad na\rho_n(0)\pm\dot\rho_n(0)=\sum_{j=1}^n(a\rho_n(0)\pm X_j)\succeq0 .
\end{equation*}
If $v\in\ker\rho_n(0)$, the last two positive semidefinite operators have opposite quadratic forms on $v$. Both forms must be zero, and a positive semidefinite operator annihilates every vector on which its quadratic form is zero. Subtracting the two resulting vector equations gives $\dot\rho_n(0)v=0$. Hermiticity then gives
\begin{equation}\label{eq:output-tangent-support}
 \dot\rho_n(0)=P_n\dot\rho_n(0)P_n.
\end{equation}
\end{proof}

\section{From the Kraus optimization to the Choi formula}\label{app:choi-translation}
\subsection{Supported Kraus--Choi gauge correspondence}
This appendix derives the Choi variational formula from the
asymptotic Kraus optimization. The first step is to identify
the Kraus gauge and its objective with variables and operators
on the fixed Choi support supplied by the affine replacement.

The constant-rank affine replacement in Lemma~\ref{lem:tangent} allows
us to use a differentiable Choi factor whose range is the same at all
nearby parameters. The following lemma carries out the gauge-to-Choi
translation once on that support. Full Choi support requires no
separate derivation.

\begin{lemma}[Supported Kraus--Choi gauge correspondence]\label{lem:gauge-correspondence}
Let $P$ be a projection on $\cH_B\otimes\cH_A$, and let
$Q_\theta:\mathbb C^s\to\operatorname{Ran}P$ be differentiable near
zero, with $Q_\theta Q_\theta^\dagger=C_\theta$ and each
$Q_\theta$ bijective onto $\operatorname{Ran}P$. Write
$Q=Q_0$, $C=QQ^\dagger$, $D=\dot Q Q^\dagger+Q\dot Q^\dagger$,
and assume $\Tr_B D=0$. For Hermitian $h$ on $\mathbb C^s$, put
$\dot Q_h=\dot Q-iQh$, $A_h=\dot Q_hQ^\dagger$, and define
\begin{equation}\label{eq:Ah}
 A_h=D/2-iY,\qquad
 Y(h)=\tfrac i2(\dot Q Q^\dagger-Q\dot Q^\dagger)+QhQ^\dagger.
\end{equation}
Then $h\mapsto Y(h)$ is an affine bijection from $\Herm(\mathbb C^s)$
onto $\{Y=Y^\dagger=PYP\}$. For the Kraus operators obtained by
reshaping the columns of $Q$ and $\dot Q_h$ into operators, define
$\beta_h=i\sum_\ell K_\ell^\dagger\dot K_{h,\ell}$ and
$\alpha_h=\sum_\ell\dot K_{h,\ell}^\dagger\dot K_{h,\ell}$.
Then
\[
 \beta_h^{\mathsf T}=\Tr_B Y,\qquad
 M_{C,D}(Y)=4\alpha_h^{\mathsf T}.
\]
Consequently $\beta_h=0$ is equivalent to $Y\in\cV_P$, and
$\lambda_{\max}M_{C,D}(Y)=4\|\alpha_h\|_\infty$ for corresponding
feasible variables.
\end{lemma}
\begin{proof}
\Needspace{4\baselineskip}
\noindent\textbf{Step 1: Establish the affine bijection
$h\longleftrightarrow Y$.}\par\nobreak

Because $P Q_\theta=Q_\theta$ near zero, both $Q$ and $\dot Q$
have range contained in $\operatorname{Ran}P$. Taking the Hermitian
and anti-Hermitian parts of
$A_h=(\dot Q-iQh)Q^\dagger$ proves Eq.~\eqref{eq:Ah} and $Y=PYP$.
The first term of $Y(h)$ is fixed. Since $Q:\mathbb C^s\to
\operatorname{Ran}P$ is bijective, conjugation $h\mapsto QhQ^\dagger$
is a bijection between the Hermitian operators on these spaces:
for any supported Hermitian $Y$, the inverse is
\[
 h=Q^{-1}\left[Y-\tfrac i2(\dot Q Q^\dagger-Q\dot Q^\dagger)\right]
       (Q^{-1})^\dagger,
\]
where $Q^{-1}:\operatorname{Ran}P\to\mathbb C^s$ is the inverse of $Q$.
This inverse formula produces a Hermitian $h$ for every supported
Hermitian $Y$, while the displayed expression for $Y(h)$ is
single-valued. Hence $h\mapsto Y(h)$ is the claimed affine
bijection, completing Step~1.

\Needspace{4\baselineskip}
\noindent\textbf{Step 2: Identify the constraint
$\beta_h^{\mathsf T}=\Tr_B Y$.}\par\nobreak

With the vectorization $|U\rangle\rangle=\sum_{b,j}U_{bj}|bj\rangle$,
\begin{equation}\label{eq:transpose}
 [\Tr_B|U\rangle\rangle\langle\langle V|]_{jk}
 =\sum_bU_{bj}\overline{V_{bk}}=(V^\dagger U)_{kj}.
\end{equation}
Applying it to each column gives
$\beta_h^{\mathsf T}=i\Tr_B A_h$. By Eq.~\eqref{eq:Ah},
\[
 i\Tr_B A_h=\tfrac i2\Tr_B D+\Tr_B Y=\Tr_B Y.
\]
The final equality uses trace preservation $\Tr_B D=0$.
In particular $Y=0$ determines a feasible Hermitian gauge.

\Needspace{4\baselineskip}
\noindent\textbf{Step 3: Prove
$M_{C,D}(Y)=4\alpha_h^{\mathsf T}$.}\par\nobreak

As $Q$ is a bijective factor of $C$ on its support,
$Q^\dagger C^+Q=I_s$. To verify this without treating $Q$ as a
square matrix, write $Q=VB$ with $V:\mathbb C^s\to\operatorname{Ran}P$
an isometry and $B$ invertible. Then
$C^+=V(BB^\dagger)^{-1}V^\dagger$, and
$Q^\dagger C^+Q=B^\dagger(BB^\dagger)^{-1}B=I_s$.
Since $\dot Q_h$ is supported,
$A_hC^+Q=\dot Q_hQ^\dagger C^+Q=\dot Q_h$. It follows that
$\dot Q_h\dot Q_h^\dagger=A_hC^+A_h^\dagger$.
Using $2A_h=D-2iY$ and Eq.~\eqref{eq:transpose},
\begin{equation}\label{eq:objective-translation}
 M_{C,D}(Y)=4\Tr_B(A_hC^+A_h^\dagger)
 =4\Tr_B(\dot Q_h\dot Q_h^\dagger)
 =4\alpha_h^{\mathsf T}.
\end{equation}
Because $\alpha_h\succeq0$, transposition preserves eigenvalues,
which proves the objective identity and the lemma.
\end{proof}

The next bound is used to turn the Choi-space infimum into a minimum.
It also applies to rank-deficient Choi matrices, because the optimization
variable and the derivative are supported.
\begin{lemma}[Coercivity of the supported Choi objective]\label{lem:coercivity}
Assume $D=PDP$, and let $\mu>0$ be the smallest positive eigenvalue
of $C^+$ and $d_A=\dim\cH_A$. For every $Y\in\cV_P$,
\begin{equation}\label{eq:coercivity}
 \lambda_{\max}M_{C,D}(Y)
 \ge\frac{\mu}{d_A}(\Tr D^2+4\Tr Y^2).
\end{equation}
Consequently the minimum of $\lambda_{\max}M_{C,D}(Y)$ over
$\cV_P$ is attained.
\end{lemma}
\begin{proof}
For $X=D+2iY$ we have $PX=X$ and $C^+\succeq\mu P$, so
$X^\dagger C^+X\succeq\mu X^\dagger X$. By the largest-eigenvalue
bound on the average eigenvalue and the trace-preserving property of
partial trace,
\[
 \lambda_{\max}M_{C,D}(Y)
 \ge\frac{\Tr M_{C,D}(Y)}{d_A}
 \ge\frac\mu{d_A}\Tr(X^\dagger X).
\]
Expanding $X^\dagger X$ and cancelling the traces of the commutator
terms gives $\Tr(X^\dagger X)=\Tr D^2+4\Tr Y^2$.
Thus the sublevel set at the feasible point $Y=0$ is bounded;
it is closed by continuity of the objective on the closed
finite-dimensional space $\cV_P$. Its compactness yields an attained
minimum, also minimal over all $\cV_P$.
\end{proof}

\subsection{Derivation of the general variational formula}\label{subsec:tangent-variational}
We use the preceding gauge correspondence to translate the
asymptotic Kraus formula for the affine replacement into the
Choi formula for the original family. To specify the formula
and its hypothesis, let $\mathcal N$ be a channel family represented by a differentiable
Kraus column $K_\theta$, with $K=K_0$. In the stacked Kraus convention, a Hermitian gauge $h_{\mathrm{ZJ}}$
changes $iK^\dagger\dot K$ to
$iK^\dagger\dot K+K^\dagger h_{\mathrm{ZJ}}K$.
Thus a gauge with $\beta_h=0$ exists exactly when
$iK^\dagger\dot K$ belongs to the real linear space
$\{K^\dagger X K:X=X^\dagger\}$. Under this hypothesis,
Zhou and Jiang~\cite[Theorem~2, Eq.~(12)]{ZJ} give
\begin{equation}\label{eq:ZJ-SQL}
 \JS(\mathcal N)=4\min_{h=h^\dagger:\,\beta_h=0}
                  \|\alpha_h\|_\infty.
\end{equation}
Their result establishes existence of the asymptotic coefficient in
the applicable constant-rank setting and equality of the parallel and
causally ordered adaptive asymptotic coefficients. Our gauge acts on
the right of the Choi factor: the corresponding vertically stacked
Kraus gauge is $h^{\mathsf T}$, as explained in
Subsection~\ref{app:stationarity}.

\begin{proof}[Proof of Theorem~\ref{thm:variational}]
We construct a fixed-support Kraus factor for the affine
replacement, translate its constrained minimization using the
gauge correspondence, and transfer the resulting coefficient
through the finite-use identities.
\Needspace{4\baselineskip}
\noindent\textbf{Step 1: Obtain a factor
$Q_\theta Q_\theta^\dagger=C+\theta D$ of fixed range.}\par\nobreak

Lemma~\ref{lem:tangent} replaces the given family by the CPTP family
$\widehat\Lambda_\theta$ with Choi matrix
$\widehat C_\theta=C+\theta D$ and the same finite-use pointwise
information. Let $s=\operatorname{rank}C$ and choose an isometry
$V:\mathbb C^s\to\operatorname{Ran}P$.
The factor
\[
 Q_\theta=V(V^\dagger\widehat C_\theta V)^{1/2}
\]
is differentiable and maps $\mathbb C^s$ bijectively onto
$\operatorname{Ran}P$ for every sufficiently small $|\theta|$:
positivity and invertibility on the support are established in
Lemma~\ref{lem:tangent}. Moreover, multiplying the factors shows
$Q_\theta Q_\theta^\dagger=\widehat C_\theta=C+\theta D$.

\Needspace{4\baselineskip}
\noindent\textbf{Step 2: Convert the constrained Kraus minimization
into the supported Choi minimization.}\par\nobreak

Apply Lemma~\ref{lem:gauge-correspondence} to this factor.
Its affine bijection matches the constraint $\beta_h=0$ precisely
with $Y\in\cV_P$, and matches each Kraus objective with
$\lambda_{\max}M_{C,D}(Y)=4\|\alpha_h\|_\infty$.
It also supplies a gauge with $\beta_h=0$ by choosing $Y=0$.
Therefore Eq.~\eqref{eq:ZJ-SQL} applies with $\mathcal N=\widehat\Lambda$.
Lemma~\ref{lem:coercivity} establishes attainment of the Choi
minimum. These results give
\[
 \JS(\widehat\Lambda)=
 \min_{Y\in\cV_P}\lambda_{\max}M_{C,D}(Y).
\]

\Needspace{4\baselineskip}
\noindent\textbf{Step 3: Transfer the asymptotic coefficient
to the original family.}\par\nobreak

By Eq.~\eqref{eq:finite-eq}, the original family and its affine
replacement have identical $J_n^{\mathrm S}$ for \emph{each} $n$,
separately in the parallel and adaptive strategy classes.
Thus their limits agree, exist, and have the parallel/adaptive
equality furnished by Zhou and Jiang for the replacement.
Substituting $Y=0$ in Eq.~\eqref{eq:M} yields
$M_{C,D}(0)=\Tr_B(DC^+D)=R(C,D)$.
This proves Eq.~\eqref{eq:var}, including the rank-changing case.
The finite-use bound for the original family is given separately
in Eq.~\eqref{eq:finite-rld-bound}.
\end{proof}

\subsection{The RLD representation and Kraus stationarity}\label{app:stationarity}
The preceding derivation translates the entire minimization.
Here we identify the particular Kraus representation corresponding
to $Y=0$. This representation gives the RLD value. We relate its
stationarity condition to the Choi criterion in the main text.
\noindent\textbf{The RLD gauge on the Choi support.}
Assume $D=PDP$, and take the factor
$Q:\mathbb C^s\to\operatorname{Ran}P$ of the constant-rank affine
replacement from Appendix~\ref{app:tangent}. Here $Q$ is a bijection from $\mathbb C^s$ onto $\operatorname{Ran}P$.
By Lemma~\ref{lem:gauge-correspondence}, $Y=0$ corresponds to a unique
Hermitian gauge $h_0$ with $A_{h_0}=D/2$. Since
$Q^\dagger C^+Q=I_s$ and the range of $\dot Q_{h_0}$ lies in
$\operatorname{Ran}P$, we can recover the derivative from
$A_{h_0}=\dot Q_{h_0}Q^\dagger$ as
$\dot Q_{h_0}=A_{h_0}C^+Q$. Therefore
\begin{equation}\label{eq:RLD-gauge}
 \dot Q_{h_0}Q^\dagger=D/2,\qquad
 \dot Q_{h_0}=(D/2)C^+Q,\qquad
 4\alpha_{h_0}^{\mathsf T}=R(C,D).
\end{equation}
The last identity follows from Eq.~\eqref{eq:objective-translation}
with $Y=0$. In general $h_0\ne0$: the condition $Y=0$ selects a
Kraus derivative after a gauge change of the chosen constant-rank
Choi factor.

\medskip
\noindent\textbf{Full-support comparison with the RLD Kraus representation.}
Now suppose in addition that $C>0$. Then $P=I_{BA}$,
$s=\dim(\cH_B\otimes\cH_A)$, $Q$ is square and invertible on the
entire Choi space, and $C^+=C^{-1}$. Equation~\eqref{eq:RLD-gauge} then gives
\[
 \dot Q_{h_0}=(D/2)C^{-1}Q,\qquad
 Q^{-1}\dot Q_{h_0}
 =\tfrac12Q^{-1}D(Q^\dagger)^{-1}.
\]
The matrix on the right is Hermitian since $D=D^\dagger$.
Thus the derivative of this full-support Kraus representation is a
Hermitian linear combination of its Kraus operators. This is the
representation used to obtain the RLD value in
Ko\l{}ody\'nski and Demkowicz-Dobrza\'nski~\cite[Appendix~C,
Eq.~(C.2)]{KD2013}. 

\medskip
\noindent\textbf{Stationarity in supported Kraus variables.}
The stationarity comparison can be made on the support without the
full-support inverse formula. Stack the Kraus operators as
$K=(K_1,\ldots,K_s)^{\mathsf T}$ and their reference derivatives in
the same order. The factor $Q$ is unchanged by $e^{-i\theta h}$ at
$\theta=0$, while the derivative of its $\ell$th column changes by
$-i\sum_jK_jh_{j\ell}$. Thus in the vertical stack the gauge matrix
of Zhou and Jiang is $h_{\mathrm{ZJ}}=h^{\mathsf T}$, and the derivative
is $\dot K-ih_{\mathrm{ZJ}}K$. In particular,
$\alpha_h=(\dot K-ih_{\mathrm{ZJ}}K)^\dagger
(\dot K-ih_{\mathrm{ZJ}}K)$; write
$h_{\mathrm{ZJ},0}=h_0^{\mathsf T}$.

For a Hermitian variation $\delta h_{\mathrm{ZJ}}$, set
\begin{equation}\label{eq:gauge-variation}
 Z=Q(\delta h_{\mathrm{ZJ}})^{\mathsf T}Q^\dagger.
\end{equation}
It is Hermitian and satisfies $Z=PZP$. Since $Q$ is bijective
\emph{onto} $\operatorname{Ran}P$, every supported Hermitian $Z$ is
obtained in this way. Equation~\eqref{eq:transpose} gives
\[
 (K^\dagger\delta h_{\mathrm{ZJ}}K)^{\mathsf T}=\Tr_B Z.
\]
Hence the constraint $K^\dagger\delta h_{\mathrm{ZJ}}K=0$
in Zhou and Jiang~\cite[Appendix~F, Eq.~(F4)]{ZJ} is equivalent to
$Z\in\cV_P$, including for rank-deficient $C$.

Set $\rho_{\mathrm{ZJ}}:=\sigma^{\mathsf T}$. Equation~\eqref{eq:RLD-gauge}
gives $4\alpha_{h_0}^{\mathsf T}=R$. Transposing $R\sigma=r\sigma$
gives $\rho_{\mathrm{ZJ}}(4\alpha_{h_0})=r\rho_{\mathrm{ZJ}}$.
Taking adjoints gives $4\alpha_{h_0}\rho_{\mathrm{ZJ}}=r\rho_{\mathrm{ZJ}}$,
since both factors are Hermitian and $r$ is real. Both operations
are reversible.
Thus $\rho_{\mathrm{ZJ}}$ has the corresponding top-eigenspace
support. Set
\[
 B_0:=\dot K-ih_{\mathrm{ZJ},0}K,\qquad
 \delta B:=-i\delta h_{\mathrm{ZJ}}K.
\]
For a constrained variation, $tZ$ is feasible for every real $t$.
Lemma~\ref{lem:gauge-correspondence} then gives $Y=tZ$ for the
varied gauge, so that
$M_{C,D}(tZ)=4[(B_0+t\delta B)^\dagger
(B_0+t\delta B)]^{\mathsf T}$.
Differentiating at $t=0$, using the linear term $L_P(Z)$ in
Eq.~\eqref{eq:expand} and
$\Tr(\sigma X^{\mathsf T})=\Tr(\sigma^{\mathsf T}X)$, yields
\begin{equation}\label{eq:F4-correspondence}
\begin{aligned}
\Tr[\sigma\,2i\Tr_B(DC^+Z-ZC^+D)]
 &\overset{(a)}{=}4\Tr\!\left[\rho_{\mathrm{ZJ}}
       (\delta B^\dagger B_0+B_0^\dagger\delta B)\right]\\
 &\overset{(b)}{=}8\operatorname{Re}\Tr[\rho_{\mathrm{ZJ}}\delta B^\dagger B_0].
\end{aligned}
\end{equation}
Equality (a) is the derivative of the supported Choi--Kraus objective
identity; (b) uses the Hermiticity of $\rho_{\mathrm{ZJ}}$.
Since $\delta B^\dagger=iK^\dagger\delta h_{\mathrm{ZJ}}$,
the derivative is eight times the real-part pairing in Eq.~(F4)
of Zhou and Jiang, evaluated at the RLD gauge. It vanishes
exactly when that pairing vanishes. Their condition is
formulated at an optimal gauge; here we ask whether the particular
RLD gauge $h_0$ is optimal, equivalently whether one density operator
$\sigma$ with $R\sigma=r\sigma$ makes this pairing vanish for
\emph{every} $Z\in\cV_P$.

To convert that condition to a Choi-space operator equation, set
\[
 S_\sigma^+:=2i[(I_B\otimes\sigma)DC^+-C^+D(I_B\otimes\sigma)].
\]
Equation~\eqref{eq:pair} gives
$\Tr[\sigma L_P(Z)]=\Tr(ZPS_\sigma^+P)$. By
Lemma~\ref{lem:supported-annihilator}, its vanishing for all
$Z\in\cV_P$ is equivalent to
$PS_\sigma^+P=P(I_B\otimes H)P$ for some Hermitian $H$, precisely
Eq.~\eqref{eq:station}. For full support this is Eq.~\eqref{eq:L}.
When the Choi support is a proper subspace the Kraus comparison uses the constant-rank
replacement family, not an assumed differentiable constant-rank
factorization of the original family. Equation~\eqref{eq:finite-eq}
identifies the original and replacement families' finite-use
pointwise SLD information and therefore transfers the criterion.

\section{Kraus-factor construction for measurement--preparation channels}\label{app:mp-kraus}
To bound the optimal asymptotic SLD coefficient in
Proposition~\ref{prop:mp}, we construct a Kraus representation for
which the gauge constraint holds at $h=0$. The first identity
below ensures that this gauge is feasible, and the second gives
the objective value used for the upper bound.

\begin{lemma}[Kraus representation for fixed-basis measurement and preparation]
\label{lem:mp-kraus-factor}
Consider the channel family in Eq.~\eqref{eq:mp}. Assume each
$\rho_{x,\theta}$ is differentiable at zero and $\rho_x=\rho_{x,0}>0$.
Let $f_x=F^{\mathrm S}(\rho_x,\dot\rho_{x,0})$.
There exists a Kraus representation $\{K_\ell(\theta)\}$ of the family
near zero, differentiable at zero, such that
\[
 \sum_\ell K_\ell^\dagger\dot K_\ell=0,\qquad
 4\sum_\ell\dot K_\ell^\dagger\dot K_\ell
 =\sum_x f_x|x\rangle\langle x|.
\]
Here $K_\ell=K_\ell(0)$ and $\dot K_\ell$ is its derivative at zero.
\end{lemma}

\begin{proof}
For each output state we choose a factor whose derivative is determined
by its SLD $L_x$. This choice makes the first Kraus sum vanish and
sets the diagonal entries of the second sum.

\Needspace{3\baselineskip}
\noindent\textbf{Step 1: Construct $V_{x,\theta}V_{x,\theta}^\dagger=\rho_{x,\theta}$ with $\dot V_x=L_xV_x/2$.}\par\nobreak

Since each $\rho_{x,0}>0$ and the state family is differentiable at zero, its positive square root is differentiable there. Choose a reference factor $W_{x,\theta}=\rho_{x,\theta}^{1/2}$ and put $W_x=\rho_x^{1/2}$. We choose a differentiable unitary $U_{x,\theta}$ with $U_{x,0}=I$ and set $V_{x,\theta}=W_{x,\theta}U_{x,\theta}$. Multiplication by a unitary preserves the state factorization. The product rule at zero gives $\dot V_x=\dot W_x+W_x\dot U_x(0)$ and $V_x=W_x$. To obtain $\dot V_x=L_xW_x/2$, we need $\dot U_x(0)=\Omega_x$, where
\[
 \Omega_x=W_x^{-1}(L_xW_x/2-\dot W_x).
\]
The SLD equation $D_x=(\rho_xL_x+L_x\rho_x)/2$ and the derivative $D_x=\dot W_xW_x+W_x\dot W_x$ of $\rho_x=W_x^2$ give
\[
 W_x(\Omega_x+\Omega_x^\dagger)W_x
 =\tfrac12(L_x\rho_x+\rho_xL_x)
  -(\dot W_xW_x+W_x\dot W_x)=0.
\]
As $W_x>0$ is invertible, $\Omega_x$ is anti-Hermitian. Hence the differentiable unitary $U_{x,\theta}=\exp(\theta\Omega_x)$ achieves
\[
 \dot V_x=\dot W_x+W_x\Omega_x=L_xV_x/2,
 \qquad V_{x,\theta}V_{x,\theta}^\dagger=\rho_{x,\theta}.
\]

\Needspace{3\baselineskip}
\noindent\textbf{Step 2: Verify $\sum_{x,a}K_{x,a}(\theta)XK_{x,a}(\theta)^\dagger=\Lambda_\theta(X)$.}\par\nobreak

Let $v_{x,a}(\theta)$ denote the columns of $V_{x,\theta}$ and set $K_{x,a}(\theta)=|v_{x,a}(\theta)\rangle\langle x|$. For every $X$,
\[
 \sum_a K_{x,a}(\theta)XK_{x,a}(\theta)^\dagger
 =\langle x|X|x\rangle V_{x,\theta}V_{x,\theta}^\dagger
 =\langle x|X|x\rangle\rho_{x,\theta}.
\]
Summing over $x$ gives
$\sum_{x,a}K_{x,a}(\theta)XK_{x,a}(\theta)^\dagger
=\Lambda_\theta(X)$, as defined in Eq.~\eqref{eq:mp}.

\Needspace{3\baselineskip}
\noindent\textbf{Step 3: Evaluate the two Kraus sums.}\par\nobreak

By Step~2, the coefficient of $|x\rangle\langle x|$ in $\sum_{x,a}K_{x,a}^\dagger\dot K_{x,a}$ is $\sum_a\langle v_{x,a}|\dot v_{x,a}\rangle$. Using $\dot v_{x,a}=L_xv_{x,a}/2$ from Step~1 and $\sum_a|v_{x,a}\rangle\langle v_{x,a}|=V_xV_x^\dagger=\rho_x$, we get
\begin{align*}
 \sum_a\langle v_{x,a}|\dot v_{x,a}\rangle
 &=\tfrac12\Tr(L_x V_xV_x^\dagger)=\tfrac12\Tr(\rho_x L_x)=0,\\
 4\sum_a\langle\dot v_{x,a}|\dot v_{x,a}\rangle
 &=\sum_a\langle v_{x,a}|L_x^2|v_{x,a}\rangle
 =\Tr(L_x^2 V_xV_x^\dagger)=f_x .
\end{align*}
The zero follows by taking the trace of the SLD equation for $D_x$, since $\Tr D_x=0$. The second line is the coefficient of $|x\rangle\langle x|$ in $4\sum_{x,a}\dot K_{x,a}^\dagger\dot K_{x,a}$.
Summing the contributions over $x$ gives both identities in the lemma.
\end{proof}

\section{Earlier results and examples}\label{app:comparisons}
\subsection{Selected earlier results}
Table~\ref{tab:prior-work} lists the results used from the literature and their roles in the analysis. Table~\ref{tab:examples} summarizes the channel examples.

\begingroup
\footnotesize
\setlength{\tabcolsep}{3pt}
\renewcommand{\arraystretch}{1.12}
\begin{longtable}{@{}L{.225\textwidth}L{.17\textwidth}L{.30\textwidth}L{.235\textwidth}@{}}
\caption{Selected previous results and their roles.}\label{tab:prior-work}\\
\toprule
Reference & Focus & Established result & Role in this paper \\
\midrule
\endfirsthead
\multicolumn{4}{c}{\tablename\ \thetable\ (continued)}\\
\toprule
Reference & Focus & Established result & Role in this paper \\
\midrule
\endhead
\midrule
\multicolumn{4}{r}{Continued on next page}\\
\endfoot
\bottomrule
\endlastfoot
\multicolumn{4}{@{}l}{\textbf{A. One-use information and channel bounds}}\\
Fujiwara \& Imai (2003)~\cite{FI2003} & Pauli channels & Entangled probe optimal & Known example. \\
Fujiwara \& Imai (2008)~\cite{FI2008} & One-use SLD & Kraus minimization & Precursor to the Kraus optimization. \\
Hayashi (2011)~\cite{Hayashi} & Channel RLD & Choi formula & Choi upper bound tested here. \\
Demkowicz-Dobrza\'nski et al. (2012)~\cite{DDKG2012} & Noisy metrology & Channel-extension bound & Repeated-use upper bounds. \\
Ko\l{}ody\'nski \& Demkowicz-Dobrza\'nski (2013)~\cite{KD2013} & Channel-extension and RLD bounds & Kraus representation giving RLD & Feasible point $Y=0$. \\[3pt]
\multicolumn{4}{@{}l}{\textbf{B. Asymptotic coefficient and tightness}}\\
Zhou \& Jiang (2021)~\cite{ZJ} & Asymptotic SLD & Kraus optimum, stationarity, and parallel attainment & Choi formulation and a support-restricted criterion for the RLD gauge. \\
Katariya \& Wilde (2021)~\cite{KW} & Adaptive RLD & Multiparameter RLD bound & Adaptive bound for several parameters. \\
Hayashi \& Ouyang (2026)~\cite{HayashiOuyang2026} & General processes & RLD-based additive bounds & Indefinite causal order. \\
Huang \& Wilde (2026)~\cite{HuangWilde2026} & General processes & Additive RLD inequality & Indefinite causal order. \\
Albarelli \& Demkowicz-Dobrza\'nski (2022)~\cite{AD2022} & Purification & RLD as one choice & Tightness context. \\[3pt]
\multicolumn{4}{@{}l}{\textbf{C. Global scaling and model-specific context}}\\
Hayashi et al. (2022)~\cite{HLY2022} & Global scaling & Phase noise $O(n^{-1})$ threshold & Different regime. \\
\v{S}afr\'anek (2017)~\cite{Safranek2017} & Rank changes & Discontinuity of SLD information & Pointwise convention. \\
Horodecki et al. (2003)~\cite{HSR2003} & Measure--prepare & Entanglement-breaking class & Model context. \\
Singh \& Datta (2022)~\cite{SinghDatta2022} & Dephasing & Generalized dephasing channels & Correlation-matrix parametrization. \\
Huang et al. (2024)~\cite{HLW2024} & Bosonic dephasing & Exact sensing limits & Different model. \\
\end{longtable}
\endgroup

\subsection{Examples and their conclusions}
Table~\ref{tab:example-guide} in the Introduction groups the examples
by the mathematical question they address. Table~\ref{tab:examples}
lists the channel families, quantities tested, and corresponding conclusions.

\begingroup
\footnotesize
\setlength{\tabcolsep}{3pt}
\renewcommand{\arraystretch}{1.12}
\begin{longtable}{@{}L{.24\textwidth}L{.23\textwidth}L{.26\textwidth}L{.21\textwidth}@{}}
\caption{Examples, mathematical questions, and conclusions.}\label{tab:examples}\\
\toprule
Example & Setting & Tested quantity & Conclusion \\
\midrule
\endfirsthead
\multicolumn{4}{c}{\tablename\ \thetable\ (continued)}\\
\toprule
Example & Setting & Tested quantity & Conclusion \\
\midrule
\endhead
\midrule
\multicolumn{4}{r}{Continued on next page}\\
\endfoot
\bottomrule
\endlastfoot
\multicolumn{4}{@{}l}{\textbf{A. Pauli and measurement--preparation channels}}\\
Generalized Pauli~\cite{FI2003} & Weyl-error probabilities & One-use optimum & Maximally entangled. \\
Measure--prepare: equality & $[C,D]\ne0$ & RLD saturation & Attained. \\
Measure--prepare: strict gap & Selected noncommuting block & RLD versus SLD & $1/3$ versus $1/4$. \\
Qubit replacer channel & Full-rank Choi matrices approaching rank deficiency & $\JR/\JS$ & Unbounded in fixed dimension. \\
Rank-changing replacer channel & Choi rank varies near $0$ & Pointwise and nearby SLD & Different values. \\[3pt]
\multicolumn{4}{@{}l}{\textbf{B. Structured dephasing and finite uses}}\\
Qubit dephasing~\cite{ZJ} & Magnitude and phase & $J_1^{\mathrm S}$ and $\JS$ & $J_1^{\mathrm S}$ can be smaller. \\
Generalized dephasing channel~\cite{SinghDatta2022} & Entrywise multiplication by a correlation matrix, $A_0>0$ & RLD saturation & Linear equations in the weights. \\
Complete-dephasing point & $A_0=I$; derivative $G$ & All finite uses & Exact additivity. \\
Star graph & Unit edge weights of $G$ & $\JR/\JS$ & Star: $\sim m/4$. \\
Correlated-qutrit phase family & Cycle phase $\Phi$ & RLD saturation & $r_0\sin(\Phi/2)=0$. \\[3pt]
\multicolumn{4}{@{}l}{\textbf{C. Maximally entangled inputs}}\\*
Prescribed maximally entangled input & Fixed one-use probe & Attainment of channel RLD & Criterion in Eq.~\eqref{eq:ME}. \\

\end{longtable}
\endgroup

\section*{Use of generative AI tools}
The author used ChatGPT and Microsoft Copilot 
to assist with literature searches, exploration of
possible approaches to the research problem, and revision of the
exposition. 
The manuscript was written under
the author's direction, and the author is solely responsible for all mathematical
content, proofs, references, and conclusions.

\section*{Funding}
The author was supported in part by the General R\&D Projects of 1+1+1
CUHK-CUHK(SZ)-GDST Joint Collaboration Fund (Grant No.~GRDP2025-022) and
the Guangdong Provincial Quantum Science Strategic Initiative
(Grant No.~GDZX2505003).

\end{document}